\documentclass[prx,aps,amsmath,amssymb,nofootinbib,epsfig,showpacs,twocolumn,eqsecnum,superscriptaddress]{revtex4-1}

\usepackage{amsmath,cleveref,amssymb,amsfonts,graphicx,multirow,color,bm,textcomp,mathtools}
\usepackage{paralist}
\usepackage{srcltx}
\usepackage[all,cmtip]{xy}
\usepackage{mathrsfs}
\usepackage{srcltx,soul,subfigure}
\usepackage{threeparttable,dsfont,bbm}

\crefname{equation}{Eq.}{Eqs.}
\crefname{figure}{Fig.}{Figs.}

\newcommand{\bra}[1]{\langle #1 |}
\newcommand{\ket}[1]{| #1 \rangle}

\newcommand{\bee}{\begin{eqnarray}}
\newcommand{\ee}{\end{eqnarray}}
\newcommand{\bma}{\begin{pmatrix}}
\newcommand{\ema}{\end{pmatrix}}
\newcommand{\balig}{\begin{align}}
\newcommand{\ealig}{\end{align}}
\newcommand{\peq}{P_{N-1}^{\rm{eq}}}

\newcommand{\ba}{\begin{align}}
\newcommand{\ea}{\end{align}}

\newcommand{\ignore}[1]{}

\newcommand{\bI}{\mathbbm{1}}

\usepackage{dcolumn}
\newcolumntype{C}[1]{>{\centering\let\newline\\\arraybackslash\hspace{0pt}}m{#1}}

\begin{document}

\title{Conductance of a superconducting Coulomb blockaded Majorana nanowire}

\author{Ching-Kai Chiu}
\affiliation{
Condensed Matter Theory Center and Joint Quantum Institute and Station Q Maryland, Department of Physics, University of Maryland, College Park, MD 20742, USA}

\author{Jay D. Sau}
\affiliation{
Condensed Matter Theory Center and Joint Quantum Institute and Station Q Maryland, Department of Physics, University of Maryland, College Park, MD 20742, USA}

\author{S. Das Sarma}
\affiliation{
Condensed Matter Theory Center and Joint Quantum Institute and Station Q Maryland, Department of Physics, University of Maryland, College Park, MD 20742, USA}

\begin{abstract}
In the presence of an applied magnetic field introducing Zeeman spin splitting, a superconducting (SC) proximitized one-dimensional (1D) nanowire with spin-orbit coupling can pass through a topological quantum phase transition developing zero-energy topological Majorana bound states (MBSs) on the wire ends. One of the promising experimental platforms in this context is a Coulomb blockaded island, where by measuring the two-terminal conductance one can in principle investigate the MBS properties. Here, we theoretically study the tunneling transport of a single electron across the superconducting Coulomb blockaded nanowire at finite temperature in order to obtain the generic conductance equation. By considering all possible scenarios where only MBSs are present at the ends of the nanowire, we compute the nanowire conductance as a function of the magnetic field, the temperature, and the gate voltage. 
In the simplest 1D topological SC model, the oscillations of the conductance peak spacings (OCPSs) arising from the Majorana overlap from the two wire ends manifest an increasing oscillation amplitude with increasing magnetic field (in disagreement with a recent experimental observation). We develop a generalized finite temperature master equation theory including not only multiple subbands in the nanowire, but also the possibility of ordinary Andreev bound states in the non-topological regime.  Inclusion of all four effects (temperature, multiple subbands, Andreev bound states, and MBSs) provides a complete picture of the tunneling transport properties of the Coulomb blockaded nanowire. Based on this complete theory, we indeed obtain OCPSs whose amplitudes decrease with increasing magnetic field in qualitative agreement with recent experimental results, but this happens only for rather high temperatures with multisubband occupancy and the simultaneous presence of both Andreev bound states and MBSs in the system.  Thus, the experimentally observed OCPSs manifesting decreasing amplitude with increasing magnetic field can be explained in our theory only if the experimental magnetic field range encompasses both the non-topological and the topological regimes so that both Andreev bound states and Majorana bound states are contributing to these oscillations as well as the applicable electron temperature in the nanowire is rather high.  A particularly significant aspect of our theory is that in such a high-temperature Coulomb blockaded nanowire, the OCPSs no longer have a one-to-one correspondence with the nanowire quasiparticle energy spectrum as is generic in the low temperature unblockaded situation.  This implies that the OCPSs cannot be used to conclude about the low energy spectrum so that no statement can be made about the so-called ``topological protection" based on such OCPSs. In particular, the length dependence of the oscillation peak in such a situation is nongeneric and does not directly contain useful information about the Majorana splitting energy, reflecting only the physics of Andreev bound states in the finite size nanowires used in the experiment.
\end{abstract}

\date{\rm\today}
\maketitle

\section{Introduction} \label{section I}

	The integer quantum Hall effect~\cite{Klitzing} ushered in the era of topological systems and phenomena in condensed matter physics, although the fact that the precise quantization of the Hall conductance in two dimensions (2D) is indeed a direct manifestation of topological robustness took several years to be appreciated.  Of course, the 2016 Nobel Prize~\cite{Thouless:1982rz,Haldane1988} in physics has made this fact rather universally celebrated.
The classification of topological insulators and superconductors~\cite{Kitaev2009,Schnyder2008} provides guidance to look for topological systems and materials. The essential signature of the topological phase is the presence of stable gapless (zero energy) states on the boundary of the system with the bulk having a robust energy gap. In fact, this can be construed as an equivalent definition of a topological phase for the quantum Hall effect, the bulk gap corresponds to the cyclotron gap imposed  by the external magnetic field whereas the boundary gapless states are the edge states confined to the 1D edge of the 2D layer.  Indeed , the symmetry-based topological classification scheme~\cite{hasan:rmp,review_TIb,RevModPhys.88.035005} is restricted to insulators (e.g. 2D quantum Hall states, 2D quantum spin Hall insulators, 3D topological insulators) and superconductors simply because these are the systems with bulk energy gaps separating ground states from excited states.  But, obviously, very special constraints are necessary to ensure that an insulator or a superconductor would have gapless or zero energy boundary states, and this is why the subject of topological systems has become active only in recent years because of deep theoretical advances in spite of insulators and superconductors having been known for more than hundred years.  Insulators and superconductors all have bulk gaps, but only the ones having robust boundary gapless states are called topological (whereas those not having such special boundary states are called nontopological or trivial).

The topological quantum phase transition (TQPT) between topological and trivial phases is also a subject of great current interest, and such a transition can only happen through the vanishing of the bulk gap at the TQPT. Recently, there has been an enormous inter-disciplinary interest in one particular type of zero energy boundary states associated with both 1D and 2D topological superconductor systems.  These are the so-called Majorana (zero-energy) bound states (MBSs)~\cite{elliott_franz_review,Kitaev2001}, which form the central theme of the current work.  In particular, these MBSs are strange quantum objects obeying anyonic non-Abelian braiding statistics~\cite{Sarma:2015aa,RMP_braiding,Ivanov_braiding}, which can be used for fault-tolerant quantum computation, thus bringing together physicists, mathematicians, computer scientists, electrical engineers, and materials scientists in an effort to build such a quantum computer.  Much popular interest has recently focused on the possibility of Majorana-based topological quantum computation since Microsoft Corporation has just announced a large commercial effort to build such a computer.
An MBS, which is its own anti-particle, has zero energy protected by particle-hole symmetry stemming from superconductivity. Furthermore, these MBSs can be realized on the ends of 1D topological superconductors~\cite{Kitaev2001} and at point defects of 2D topological superconductors~\cite{Fu:2008fk}. The most promising experimental platform for the realization of MBSs is SC-proximitized semiconductor nanowire with strong spin-orbit coupling~\cite{Roman_SC_semi,Gil_Majorana_wire,PhysRevB.82.214509} in an applied magnetic field to create a Zeeman spin splitting. 
The precise theoretical prediction~\cite{Roman_SC_semi,Gil_Majorana_wire} is that a spin-orbit-coupled semiconductor nanowire (e.g. InSb, InAs) with superconducting proximity effect induced by a nearby regular superconductor (e.g. Nb, Al) would become a topological superconductor with zero-energy (i.e. mid-gap) MBSs localized at the wire ends provided the Zeeman spin splitting induced by the applied magnetic field is large enough to overcome the induced SC gap.  Here, the applied magnetic field is the tuning parameter inducing the TQPT with MBSs appearing as localized zero energy bound states at the wire ends at high enough magnetic field values. One technique to experimentally look for MBSs in the nanowires is to study the tunneling conductance~\cite{PhysRevB.82.214509,PhysRevB.63.144531}, which should manifest a quantized zero bias peak associated with the MBS in the topological phase. The recent experimental observation of such a predicted zero-bias conductance peak in nanowires above a critical applied magnetic field has been a milestone to hint at the possible existence of the MBSs in the nanowire-superconductor hybrid structure \cite{Mourik_zero_bias,Das:2012aa,ballistic_M,doi:10.1021/nl303758w,PhysRevB.87.241401,PhysRevLett.110.126406,2016arXiv161004555C}. These experimental observations of zero bias conductance peaks in nanowires in a finite external magnetic field by multiple independent laboratories using different semiconductor (InSb or InAs)-superconductor (NbTiN or Al) combinations have created a great deal of excitement because of the implication that the conductance peak is providing strong evidence for the existence of  topological Majorana modes in these nanowires.  But the non-Abelian nature of these possible MBSs localized in nanowires still remains to be demonstrated experimentally, and this area is one of the most active current research areas in all of physics.


A key development in this subject is the recent experimental paper by Albrecht et al.~\cite{Albrecht:2016aa} reporting the apparent observation of exponential ``topological protection'' in Majorana nanowires, which, if validated and understood, would be a singular landmark in the field.  In particular, the most straightforward interpretation of the observation of Albrecht et al.~\cite{Albrecht:2016aa} is that the two MBSs localized at the two ends of the nanowire in their system are sufficiently far apart so that their wavefunction overlap is exponentially small. This exponential weakness is reflected in the MBS splitting oscillation showing an exponential decrease with increasing wire length as predicted theoretically~\cite{PhysRevLett.103.107001,PhysRevB.85.165124,PhysRevB.86.220506,PhysRevLett.109.166403}. This would imply that each MBS can now be thought of as an independent topological entity obeying non-Abelian statistics and hence suitable for use in topological quantum computation.  Unfortunately, however, such a straightforward interpretation seems inapplicable to the experiment of Ref.~\cite{Albrecht:2016aa}  since the magnetic field dependence of the MBS overlap seems to disagree with the theoretical predictions \cite{PhysRevB.86.220506,Tudor_oscillation_damp} in spite of the length dependence manifesting the predicted theoretical exponential behavior.  The current work is aimed at an understanding of the Albrecht experiment~\cite{Albrecht:2016aa}, which because of its singular importance (i.e.~``topological protection") must be taken extremely seriously.

The conundrum here is the following.  The theoretical exponential behavior~\cite{PhysRevB.86.220506}, as manifesting in the oscillations of the conductance peak spacings (OCPSs) as a function of the applied magnetic field or the wire length, reflects an $e^{-L/\xi}$ dependence in the overlap between the MBSs localized at the two wire ends, where $L$ and $\xi$ are respectively the wire length (or more precisely, the separation between the two MBSs) and the Majorana localization length (or more precisely, the nanowire superconducting coherence length).  Since the coherence length increases with increasing magnetic field B~\cite{PhysRevB.86.220506}, the Majorana oscillations in the wire length and magnetic field are intimately coupled.  An observation of the exponential decrease of oscillations in the wire length (as reported in Ref.~\cite{Albrecht:2016aa}) must therefore automatically come with an increase in the oscillation amplitude as a function of magnetic field since $L/\xi$ decreases with increasing magnetic field at fixed $L$.  Seeing one without the other does not make any sense from the perspective of the minimal theory~\cite{PhysRevB.86.220506}.  The experimental situation in Ref.~\cite{Albrecht:2016aa} is actually worse since the oscillation amplitude seems to decrease (instead of increasing) with increasing magnetic field (while at the same time, the amplitude decreases with increasing L), which is completely the opposite of the predicted theoretical behavior.  A resolution of this conundrum in Ref.~\cite{Albrecht:2016aa} is obviously of key importance in the context of the ``exponential protection" claim in terms of wire length.  The hope for understanding the puzzling results reported in Ref.~\cite{Albrecht:2016aa} (i.e. apparent exponential decrease in MBS overlap as a function of wire length along with a decreasing overlap as a function of increasing magnetic field) lies in the fact that the experimental system in Ref.~\cite{Albrecht:2016aa} is not consistent with the minimal model of a nanowire coupled to a SC as considered in most MBS theories, but presents a more complex situation (discussed below) involving a Coulomb blockade in the nanowire.  Our goal in the current work is to generalize Majorana theories to include Coulomb blockade to see if the results of Ref.~\cite{Albrecht:2016aa} can be explained and understood.

Coulomb blockade at the basic level means that the system is small enough so that the Coulomb charging energy for putting electrons into the system is significant in affecting the experimental behavior.  
We must therefore incorporate Coulomb blockade in the Majorana nanowire theory to develop an appropriate model for the situation studied Ref.~\cite{Albrecht:2016aa}.  This new model inspired by Ref.~\cite{Albrecht:2016aa} includes a spin-orbit coupled proximitized SC nanowire in the presence of non-SC quantum dots at wire ends under an applied magnetic field. The study of Coulomb blockade in quantum dots (without any superconductivity) has a long history~\cite{RevModPhys.64.849}. 
Coulomb blockade was first discovered experimentally by Dolan and Fulton in small metallic tunnel junctions manifesting charging oscillations associated with the finite Coulomb energy in small systems~\cite{PhysRevLett.59.109}. Subsequently, Fulton et al.\ observed the interplay of superconductivity and charging energy in small SC tunnel junctions in 1989~\cite{PhysRevLett.63.1307}. 
The periodic Coulomb oscillation of conductance peaks in semiconductor quantum dots was first observed also in 1989~\cite{PhysRevLett.62.583} and supported by followup experiments~\cite{PhysRevB.40.5871,PhysRevLett.65.771}. The phenomenon was explained by single-electron tunneling in the presence of Coulomb blockade physics~\cite{early_Blockade_glazman,generic_blockade,PhysRevLett.63.1893}. The single-electron Coulomb blockaded normal tunneling through small dots (with small capacitance) exhibits $1e$ periodicity of the conductance peak oscillation as the gate voltage varies. As the dot becomes superconducting and the SC gap is greater than the charging energy, $1e$ tunneling becomes blocked since single electrons cannot tunnel through SC gaps. It is, however, possible for small SC dots to manifest $2e$ Cooper pair tunneling~\cite{PhysRevLett.70.4138} leading to $2e$ periodicity in the tunneling conductance oscillation as a function of gate voltage as observed in Refs.~\cite{PhysRevLett.70.1862,PhysRevLett.69.1997}. It  was also observed there can be a transition from $2e$ periodicity to $1e$ periodicity by increasing the temperature to suppress superconductivity~\cite{PhysRevLett.72.3234}. In principle, the possibility exists in superconducting tunnel junctions at finite temperatures for both $2e$ and $1e$ charge oscillations to occur as a function of gate voltage since transport could take place through sub-gap and above-gap states. On the other hand Coulomb blockaded transport through zero energy MBSs at the wire ends in a 1D topological SC should manifest 1e charge oscillations since nonlocal resonant tunneling of single electrons are allowed through the MBSs~\cite{PhysRevB.84.165440,PhysRevB.92.020511,PhysRevLett.104.056402,conductance_coulomb_blockade_roman}.

In addition to the experimental work by Albrecht et al.~\cite{Albrecht:2016aa}, which motivates our work, there have been a few recent theoretical papers~\cite{PhysRevB.84.165440,PhysRevB.92.020511,PhysRevLett.104.056402,conductance_coulomb_blockade_roman} on Coulomb blockaded Majorana nanowires, but our work extends the theory to the realistic situation of finite temperature and multi-subband occupancy along with the inclusion of both ordinary Andreev bound states and Majorana states in contributing to transport so that a meaningful comparison to the important results of Ref.~\cite{Albrecht:2016aa} could be carried out. We first briefly discuss the qualitative expectations for tunneling transport in a Coulomb blockaded Majorana nanowire with increasing magnetic field so that the system evolves from being a non-topological ordinary (proximitized) SC at low field to being a topological SC at high fields, and then perhaps becoming a gapless SC at very high magnetic fields.
At low magnetic field, the oscillation has $2e$ periodicity since the superconducting gap is larger than the charging energy. For higher magnetic field, while still being in the non-topological phase, the superconducting gap decreases eventually becoming smaller than the charging energy, and therefore, each conductance peak starts to split into two. Eventually, at high enough magnetic field the system goes through the TQPT, and in the presence of zero energy MBSs the conductance peak is expected to exhibit $1e$ periodicity since transport is now occurring through resonant tunneling through both MBSs.  The possibility that at very high magnetic field, the 1e metallic periodicity may arise simply because the nanowire has developed a zero SC gap cannot be ruled out. 
In Ref.~\cite{Albrecht:2016aa}, modifications to this strictly 1e resonant conduction due to MBS are claimed to be arising from the presence of MBS overlap from the two ends, and this modification leads to a measurement of the Majorana energy splitting, leading to the topological protection observed in Ref.~\cite{Albrecht:2016aa}.
Thus, it is crucial to understand how MBS overlap affects the resonant tunneling in the topological regime so that the topological protection claim of Ref.~\cite{Albrecht:2016aa} can be validated.  This is particularly critical given that Ref.~\cite{Albrecht:2016aa} finds an exponential MBS splitting in the wire length, but inconsistent results as a function of magnetic field in the same wire as discussed above and also below.	

%
%
%
%
%
%
%
%

The major problem in the Majorana interpretation is the inconsistency in the magnetic field behavior between the experimental observation and the theoretical prediction in the $1e$ periodic region. The oscillation amplitude of the even and odd conductance peak spacings observed in the experiment~\cite{Albrecht:2016aa} \emph{always} decreases as the magnetic field increases. 
At low enough temperatures the conductance peak spacing should reflect the positive and negative energies of the two superconducting states close to zero energy~\cite{conductance_coulomb_blockade_roman} (i.e.~the conductance should be an approximate map of the underlying mid-gap energy spectra near zero energy). The spectrum of a simple 1D SC proximitized semiconductor nanowire~\cite{Tudor_oscillation_damp} shows the oscillation amplitude of the two energy levels close to zero energy becoming larger as the magnetic field increases corresponding to an increasing MBS overlap due to the SC gap suppression with increasing field. Thus, if the Coulomb blockaded nanowire in the experiment possessing the MBSs is described by the 1D semiconductor model, the theoretical prediction and the experimental observation have a serious inconsistency in the oscillation amplitude. The goal of this manuscript is to resolve this inconsistency by considering possible scenarios in the presence or absence of the MBSs (i.e.~by probing both the topological and the non-topological regimes of the nanowire). Several mechanisms could possibly lead to this observed damped oscillation with increasing field, such as finite temperature, small bulk gap, multi-subband contribution, and the presence of Andreev bound states in the topologically trivial region. 
To understand the experimental results, in particular the $1e$ periodicity region, which arises presumably from the MBSs localized in the nanowire, we develop a transport theory for a superconducting Coluomb blockaded nanowire in the presence of spin-orbit coupling and Zeeman splitting including effects of finite temperature, multi-subband occupancy, and Andreev bound states.
Surprisingly, such a generic transport equation has not been derived in the literature in spite of the fact that Beenakker developed the non-equilibrium transport theory for a non-SC metallic Coulomb blockaded quantum dot 25 years ago~\cite{generic_blockade}. We find that such a transport theory description 
enables us to obtain the non-local conductance contributions resulting from both conventional Andreev bound state~\cite{PhysRevB.92.020511} and topological MBSs~\cite{PhysRevB.92.020511,PhysRevLett.104.056402,conductance_coulomb_blockade_roman}. Such contributions lead to qualitative distinctions between the conductance peak structure and the low-lying energy spectra even in the tunneling limit. This will lead us to reevaluate the simple
interpretation of Majorana oscillations measured from Coulomb-blockaded resonances allowing an understanding of the experimental results, which are much more nuanced and subtle than the simple exponential protection in length scenario envisioned in Ref.~\cite{Albrecht:2016aa}.  Our theory obviously thus has wide-ranging consequences in the current search for non-Abelian Majorana modes in Coulomb blockaded nanowires in the quest toward building a topological quantum computer. We emphasize that the Coulomb blockade may arise simply from the superconductor-semiconductor hybrid structure being small in size without there being any explicit quantum dots being present in the system.

	The rest of this manuscript is organized as follows. In \cref{section II}, we derive the generic tunneling transport equation for a \emph{superconducting} Coulomb blockaded nanowire at finite temperature in the $1e$ periodicity region with the tunneling leads \emph{weakly} coupled to the nanowire. In \cref{section III}, we describe the nanowire Hamiltonian along with the description for the leads, which are relevant for studying the system. 
In \cref{section IV}, we numerically compute the OCPSs at different temperatures by using a simplified version of the general transport equations. We also present in \cref{section IV} our main results in the context of the experimental findings in Ref.~\cite{Albrecht:2016aa}, carefully searching for situations where the OCPSs could have decreasing amplitude with increasing magnetic field as observed experimentally. By considering several scenarios leading to the bulk gap shrinking as the magnetic field increases (as observed in Ref.~\cite{Albrecht:2016aa}), we compare the conductance  peak spacings with the experimental observation. 
By introducing Andreev bound states in the non-topological (or trivial) region in the 1D superconducting nanowire model, we also look for an alternative explanation for the observed OCPSs in the absence of the MBSs. In the last part of \cref{section IV}, we combine all of the aforementioned physics to find the best scenario that explains the conductance peak spacing oscillation in the experimental observation. We consider the length dependence of the conductance peak spacings in sec. V, concluding in sec. VI with a summary of our results and a discussion of the open questions.
	

\section{Transport formalism for Coulomb blockaded superconducting nanowires} \label{section II}

\begin{figure}[t!]
\begin{center}
\includegraphics[clip,width=0.75\columnwidth]{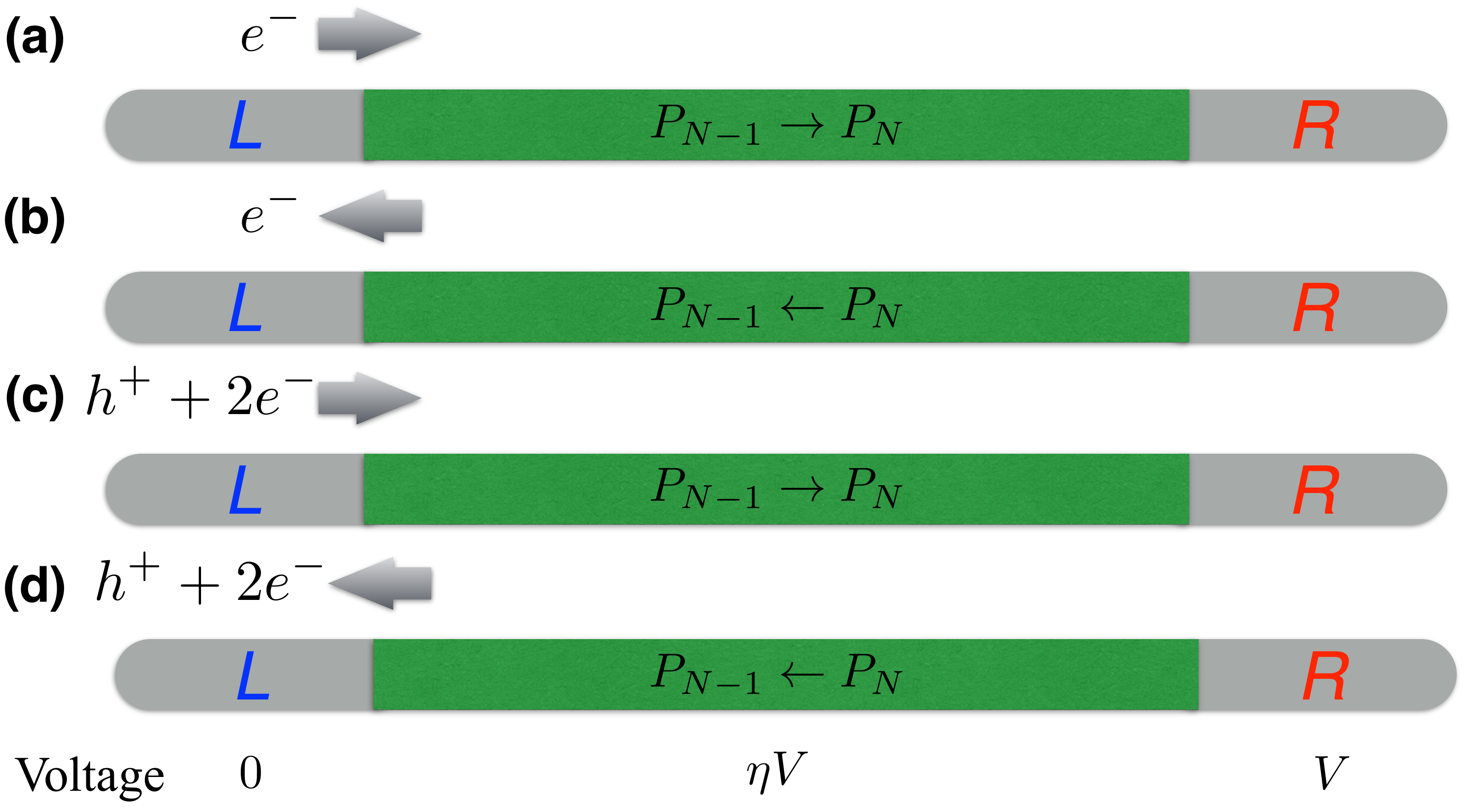}
\end{center}
  \caption{ An electron, a hole, or a Cooper pair move to(from) the nanowire from(to) the left lead. (a) an electron with energy $\epsilon^l_p$ moves to the nanowire. (b) an electron moves from the energy level $E_p$ of the nanowire to the left lead. (c) a hole (+2e Cooper pair) with energy $\tilde{\epsilon}^l_p$ moves to the nanowire. (d) a hole (+2e Cooper pair) moves from the energy level $-E_p$ of the nanowire to the left lead. } 
  \label{transport_schematic}
\end{figure}

In this section we develop a superconducting generalization of the Coulomb blockaded transport theory originally considered for metallic islands~\cite{generic_blockade}. Consider the two ends of a SC quantum dot or nanowire that are \emph{weakly} coupled with the left lead with zero voltage and the right lead with a small applied voltage $V$. The voltage on the nanowire is described by $\eta V$ with the indeterminate factor $\eta$ dropping out of our final equations. 
We label $E_p$ in ascending order as the energy levels of the quasi-particle of the nanowire with respect to the BCS ground state. The possibility of adding hole excitations, which are electron excitations combined with the loss of a Cooper pair to the SC (i.e.~the Andreev process) is what distinguishes 
the SC case from the metallic dot considered by Beenakker~\cite{generic_blockade}. We define $P_N(\{n_i\})$ to indicate the probability of the fermion configuration $\{n_i=0,1\}$ in the nanowire with the total particle number $N$. Since moving in and out by the Cooper pairs does not cost any energy in a SC, $N=\sum_i n_i + 2N_c$, where $N_c$ is the additional number of the Cooper pairs away from the charge neutral point as the gate voltage of the nanowire is zero. To simplify the problem, we assume that the charging energy $E_c$ is large enough so that only the two lowest energy levels $U(N)$ and $U(N-1)$ of the electrostatic energy 
\begin{align}
U(N)=&E_c (N-n_g)^2,  \label{staticenergy}
\end{align} 
are involved with the energy levels of the other electron numbers being too high to be included, where $n_g$ is proportional to the gate voltage of the nanowire. As the gate voltage is zero ($n_g=0$), at the charge neutral point $N=0$ the electrostatic energy reaches its minimum. More energy levels are straightforward to include theoretically, but the results become nontransparent and complicated with no new qualitative insight. We further assume that one-electron transfer process at the left and right junctions are incoherent and independent (the conductance measured  in Ref.~\cite{Albrecht:2016aa}, which is less than $e^2/h$, indicates the process is incoherent) and the tunneling rates $\Gamma_p^{l,r}$ and $\Lambda_p^{l,r}$ are much less than temperature ($\Gamma_p^{l,r}, \Lambda_p^{l,r} \ll T$), where labels $\Gamma_p^{l,r}$ and $\Lambda_p^{l,r}$ indicate the tunneling rates of the left($l$) and right($r$) for quasiparticle with energy $E_p$ and quasihole with energy $-E_p$ respectively (see \cref{tunneling}).  
The current flowing from the left lead to the nanowire is given by (see \cref{transport_schematic} for the actual tunneling processes)
\begin{small}
\begin{align}
I= &-e \sum_{p}\sum_{\{n_i\}}  \nonumber \\
&\Big \{ \Gamma_p^l \big [ P_{N-1}(\{n_i\})\delta_{n_p,0} f(\epsilon^l_p) - P_N(\{n_i\})\delta_{n_p,1}(1-f(\epsilon^l_p)) \big ] \nonumber \\
&+\Lambda_p^l \big [ P_{N-1}(\{n_i\})\delta_{n_p,1} f(\tilde{\epsilon}^{l}_{p}) - P_N(\{n_i\})\delta_{n_p,0}(1-f(\tilde{\epsilon}^{l}_{p})) \big ] \Big \}\label{current}
\end{align}
\end{small}
\noindent where $\epsilon^l_p= E_p + \Delta U +\eta eV$, $\tilde{\epsilon}^l_p=- E_p + \Delta U +\eta eV $, 
the Fermi-Dirac distribution $f(\epsilon)=1/(1+e^{\beta \epsilon})$ and the difference between the electrostatic energies with the different particle numbers $\Delta U =U(N)-U(N-1)$. The first term describes that an electron with energy $\epsilon^l_p$ at the left lead moves to the energy level $E_p$ of the nanowire and its electrostatic energy is changed to $U(N)$ from $U(N-1)$ (see \cref{transport_schematic} (a)); the second term describes that an electron in the energy level $E_p$ of the nanowire moves to the energy level $\epsilon^l_p$ at the left lead and the electrostatic energy of the nanowire changes to $U(N-1)$ from $U(N)$ (see \cref{transport_schematic} (b)). Since the nanowire is superconducting, the last two terms, which are not included in Beenakker's paper [\onlinecite{generic_blockade}], represent the movement of a hole. The third term represents the hole ($+2e$ Cooper pair) with energy $\tilde{\epsilon}^l_p$ moving from the left lead to the energy level $-E_p$ of the SC nanowire (see \cref{transport_schematic} (c)); likewise, the fourth term represents the hole ($+2e$ Cooper pair) moving from the energy level $-E_p$ of the SC nanowire to the energy level $\tilde{\epsilon}^l_p$ of the left lead (see \cref{transport_schematic} (d)).

Since the nanowire is in a stationary state, the probability of the electron configuration should be time-independent. That is, since the flow in and out of the two leads are equal, the stationary equations for particle number $N$ and $N-1$ are given by 
\begin{small}
\begin{align}
0=&\frac{\partial }{\partial t} P_N(\{n_i \})\nonumber \\
= &-\sum_p P_N(\{n_i \}) \delta_{n_p,1} \Big \{ \Gamma_p^l (1- f(\epsilon_p^l))+ \Gamma_p^r (1- f(\epsilon_p^r))\Big \} \nonumber \\
&-\sum_p P_N(\{n_i \}) \delta_{n_p,0} \Big \{ \Lambda_p^l (1- f(\tilde{\epsilon}_p^l))+ \Lambda_p^r (1- f(\tilde{\epsilon}_p^r))\Big \} \nonumber \\
&+\sum_p P_{N-1} (n_p=0) \delta_{n_p,1} \Big \{ \Gamma_p^l  f(\epsilon_p^l)+ \Gamma_p^r  f(\epsilon_p^r)\Big \} \nonumber \\
&+\sum_p P_{N-1} (n_p=1) \delta_{n_p,0} \Big \{ \Lambda_p^l  f(\tilde{\epsilon}_p^l)+ \Lambda_p^r  f(\tilde{\epsilon}_p^r)\Big \},  \label{PN}
\end{align}
\begin{align}
0=&\frac{\partial }{\partial t} P_{N-1}(\{n_i \}) \nonumber \\
=&-\sum_p P_{N-1}(\{n_i \}) \delta_{n_p,0} \Big \{ \Gamma_p^l f(\epsilon_p^l)+ \Gamma_p^r f(\epsilon_p^r)\Big \} \nonumber \\
&-\sum_p P_{N-1}(\{n_i \}) \delta_{n_p,1} \Big \{ \Lambda_p^l  f(\tilde{\epsilon}_p^l)+ \Lambda_p^r f(\tilde{\epsilon}_p^r)\Big \} \nonumber \\
&+\sum_p P_{N} (n_p=1) \delta_{n_p,0} \Big \{ \Gamma_p^l  (1-f(\epsilon_p^l))+ \Gamma_p^r (1- f(\epsilon_p^r))\Big \} \nonumber \\
&+\sum_p P_{N} (n_p=0) \delta_{n_p,1} \Big \{ \Lambda_p^l (1- f(\tilde{\epsilon}_p^l))+ \Lambda_p^r  (1-f(\tilde{\epsilon}_p^r))\Big \}, \label{stationary}
\end{align}
\end{small}
\noindent where $\epsilon_p^r=E_p+\Delta U- (1-\eta) e V$, and $\tilde{\epsilon}_p^r=-E_p +\Delta U-(1-\eta) eV $. 
Signs ``+/-" in the front of the summations indicate an electron moving in and out in the nanowire respectively. 
At the applied voltage $V=0$, $P_N(\{n_i\})$ reaches equilibrium described by detailed balance as 
\begin{align}
P_N^{\rm{eq}}(n_p=1)e^{\beta \epsilon_p}=&P_{N-1}^{\rm{eq}}(n_p=0), \nonumber \\
 P_N^{\rm{eq}}(n_p=0)e^{\beta  \tilde{\epsilon}_p}=&P_{N-1}^{\rm{eq}}(n_p=1), \label{P relation}
\end{align}
where $\epsilon_p= E_p + \Delta U$ and $\tilde{\epsilon}_p= -E_p + \Delta U$, since in the equilibrium the probability distribution is described by the Gibbs distribution 
\begin{align}
P_{N}^{\rm{eq}}(\{ n_i\})=Z^{-1} \exp \Big [ -\beta (\sum_i E_i n_i + U(N)) \Big ]
\end{align}
where $Z$ is the partition function. We note that in the presence of Cooper pairs the total electron number $N$ is not always $\sum_i n_i$. When a small applied voltage $V$ is turned on, the probability distribution in the nanowire can be expanded to the linear order around equilibrium 
\begin{align}
P_{N}(\{ n_i\})=P_{N}^{\rm{eq}}(\{ n_i\})(1+\beta e V \Phi_N(\{n_i\})), \label{PN expression}
\end{align} 
where $\Phi_N(\{n_i\})$ is an unknown functional to be obtained by solving the transport master equations to be described below. 
\begin{widetext}
By using the two equations above, we expand the current $I$ (\ref{current}) to the linear order of $V$
\begin{small}
\begin{align}
&\frac{dI}{dV}=-\beta e^2 \sum_{p} \sum_{\{n_i\}} \nonumber \\
 &\bigg \{ \delta_{n_p,0} P_{N-1}^{\rm{eq}} (\{ n_i \})f(\epsilon_p ) \Gamma_p^l \Big [ \Phi_{N-1} (\{n_i \}) -\eta f(-\epsilon_p)\Big ] 
-\delta_{n_p,1} P_{N}^{\rm{eq}} (\{ n_i \})  \Gamma_p^l \Big [ \Phi_{N} (\{n_i \}) \big ( 1- f(\epsilon_p ) \big )+\eta f(\epsilon_p)f(-\epsilon_p)\Big ] \nonumber \\
&+\delta_{n_p,1} P_{N-1}^{\rm{eq}} (\{ n_i \})f(\tilde{\epsilon}_p )  \Lambda_p^l \Big [ \Phi_{N-1} (\{n_i \}) -\eta f(-\tilde{\epsilon}_p)\Big ] 
-\delta_{n_p,0} P_{N}^{\rm{eq}} (\{ n_i \}) \Lambda_p^l \Big [ \Phi_{N} (\{n_i \}) \big ( 1- f(\tilde{\epsilon}_p ) \big )+\eta f(\tilde{\epsilon}_p) f(-\tilde{\epsilon}_p)\Big ] 
\bigg \}
\end{align}
\end{small}
Again using the identity of the Fermi-Dirac distribution $1-f(\epsilon)=f(\epsilon)e^{\beta \epsilon}$ and \cref{P relation}, we have

\begin{align} 
\frac{dI}{dV}=& \beta e^2 \sum_{p} \sum_{\{n_i\}} \bigg \{ \delta_{n_p,0} P_{N-1}^{\rm{eq}} (\{ n_i \})f(\epsilon_p ) \Gamma_p^l \Big [ \Phi_{N} (n_p=1,\{n_{i\neq p} \}) -\Phi_{N-1} (n_p=0,\{n_{i\neq p} \}) +\eta \Big ]  \nonumber \\
&+\delta_{n_p,1} P_{N-1}^{\rm{eq}} (\{ n_i \})f(\tilde{\epsilon}_p ) \Lambda_p^l \Big [ \Phi_{N} (n_p=0,\{n_{i\neq p} \}) -\Phi_{N-1} (n_p=1,\{n_{i\neq p} \}) +\eta \Big ] \bigg \}  \label{right current} 
\end{align}

To solve for the conductance $dI/dV$, we can keep the linear terms in $V$ in the stationary equations (\cref{PN,stationary}) by using \cref{PN expression} 
\begin{small}
\begin{align}
0 & = \sum_p  P_{N-1}^{\rm{eq}}(n_p=0,\{n_{i\neq p}\}) f(\epsilon_p) (\Gamma_p^l+\Gamma_p^r) \Big ( \Phi_N(n_p=1,\{n_{i\neq p}\})-\Phi_{N-1}(n_p=0,\{n_{i\neq p}\})+\eta - \frac{\Gamma^r_p}{\Gamma_p^l+\Gamma_p^r}\big ) \delta_{n_p,1} \nonumber  \\
  &+ \sum_p  P_{N-1}^{\rm{eq}}(n_p=1,\{n_{i\neq p}\}) f(\epsilon_p) (\Lambda_p^l+\Lambda_p^r) \Big ( \Phi_N(n_p=0,\{n_{i\neq p}\})-\Phi_{N-1}(n_p=1,\{n_{i\neq p}\})+\eta - \frac{\Lambda^r_p}{\Lambda_p^l+\Lambda_p^r}\big ) \delta_{n_p,0}   \label{vPN} \\
  0 & = \sum_p  P_{N-1}^{\rm{eq}}(n_p=0,\{n_{i\neq p}\}) f(\epsilon_p) (\Gamma_p^l+\Gamma_p^r) \Big ( \Phi_N(n_p=1,\{n_{i\neq p}\})-\Phi_{N-1}(n_p=0,\{n_{i\neq p}\})+\eta - \frac{\Gamma^r_p}{\Gamma_p^l+\Gamma_p^r}\big ) \delta_{n_p,0} \nonumber  \\
  &+ \sum_p  P_{N-1}^{\rm{eq}}(n_p=1,\{n_{i\neq p}\}) f(\epsilon_p) (\Lambda_p^l+\Lambda_p^r) \Big ( \Phi_N(n_p=0,\{n_{i\neq p}\})-\Phi_{N-1}(n_p=1,\{n_{i\neq p}\})+\eta - \frac{\Lambda^r_p}{\Lambda_p^l+\Lambda_p^r}\big ) \delta_{n_p,1}  \label{vPN1}
\end{align}
\end{small} 
\end{widetext}
If the nanowire has $n_{\rm{max}}$ energy levels, then a set of $2^{n_{\rm{max}}}$ stationary equations determine $\Phi_N$ and $\Phi_{N-1}$. One of the stationary equations is redundant due to the charge conservation 
\bee
\frac{\partial } {\partial t}\Big ( \sum_{\{ n_i\}} P_N {\{ n_i\}} + \sum_{\{ n_i\}} P_{N-1} {\{ n_i\}} \Big )=0
\ee
Although  the total number of the $\Phi_N$ and $\Phi_{N-1}$ is $2^{n_{\rm{max}}}$, one of $\Phi_N$'s can be set to zero with no loss of generality and $\eta$ can be neglected. There are then $2^{n_{\rm{max}}}-1$ variables to be determined, where $n_{\rm{max}}$ is the number of nanowire energy levels participating in the transport process. For an arbitrary $n_{\rm{max}}$, it is difficult to analytically solve for $dI/dV$ except perhaps under some special conditions. Later, in computing the conductance of the superconducting nanowire, we will numerically solve \cref{vPN,vPN1} for generic cases (e.g.~the presence of Andreev bound states) and use the analytical solution for special cases, which are discussed in the next subsection.  

	The physical parameters affecting the conductance are finite temperature $T$, all of the energy levels $E_p$'s less than and close to temperature, the charging energy $E_c$, and the tunneling rates $\Gamma_p^{l,r}, \Lambda_p^{l,r}$. In the next section, we consider the realistic 1D SC proximitized semiconductor model and use its physical parameters to compute the Coulomb blockaded conductance.


\subsection{The analytic solution for fixed tunneling ratio}
\begin{figure}[t!]
\begin{center}
\includegraphics[clip,width=0.75\columnwidth]{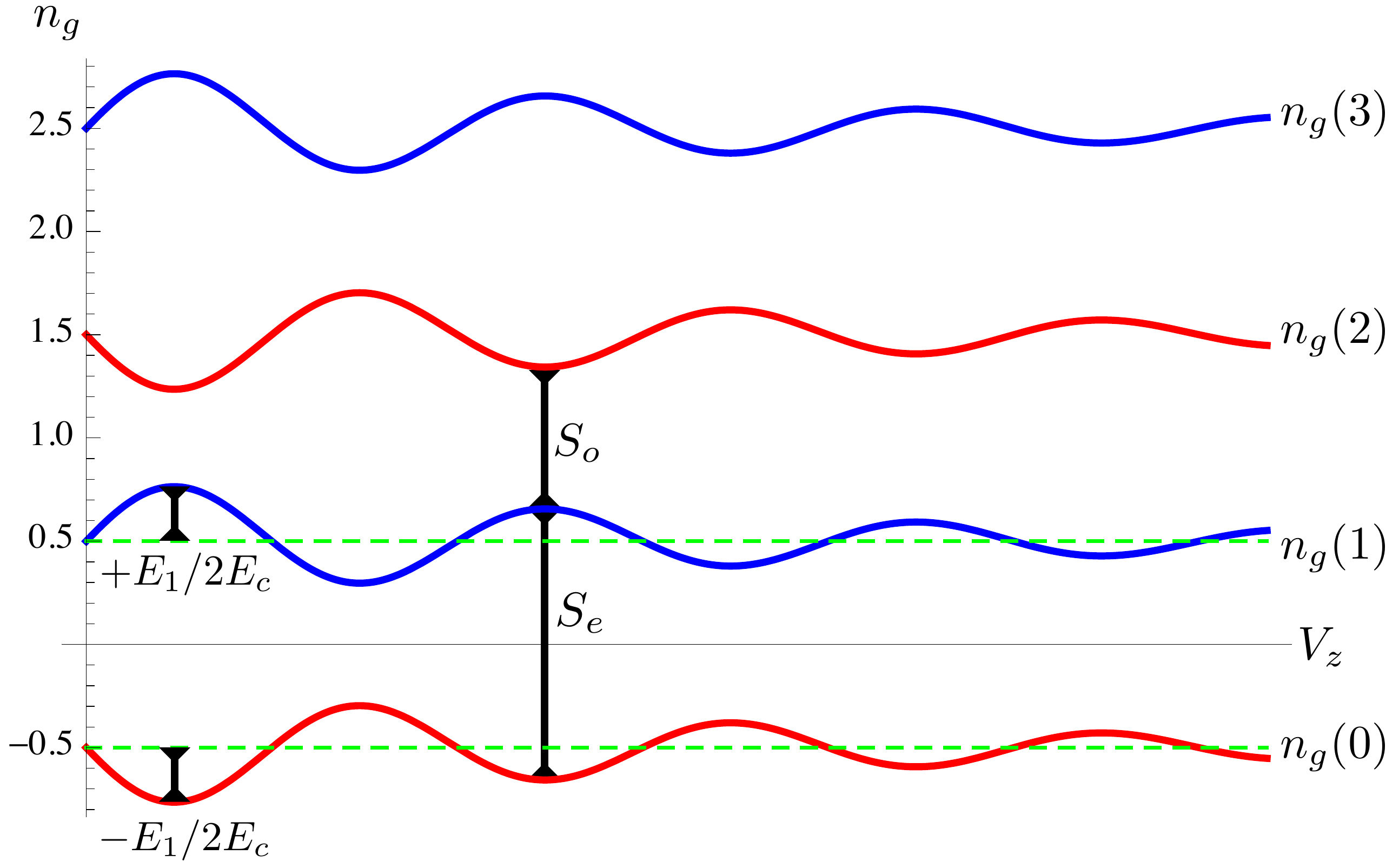}
\end{center}
  \caption{The conductance peaks as a function of Zeeman splitting $V_z$ and the particle number $N$. The red(blue) line represent the conductance peak for even(odd) $N$. The main physical quantities we study in the following are the conductance peak spacings ($S_e$ and $S_o$), which are given by the difference of the two closest conductance peaks.} 
  \label{ng_plot}
\end{figure}

Obtaining the analytic expression for the current (\ref{right current}) for an arbitrary number of energy levels $p$ is complicated. In the special situation when the tunneling ratios are fixed $\{ \Gamma_p^l/\Gamma_p^r= \Lambda_p^l/\Lambda_p^r=\rm{const}\}$ for all energy level $p$, the stationary current conditions (\cref{PN,stationary}) can be analytically solved, and the expression for the current can be explicitly written down. If the two ends of the nanowire are almost identical and localized states are absent, then $\{ \Gamma_p^l=\Gamma_p^r$, $\Lambda_p^l=\Lambda_p^r\}$ fulfills this special condition. 
First, we assume that these stationary equations still hold as the summation over $p$ is removed (later, we will go back to check if this assumption is valid.)
\begin{align}
&P_{N}(n_p=1)\Big [ \Gamma^l_p(1- f(\epsilon^l_p))+\Gamma_p^r(1- f(\epsilon^r_p)) \Big ]  \nonumber \\
= & P_{N-1}(n_p=0)\Big [ \Gamma^l_p f(\epsilon^l_p )+ \Gamma_p^r  f(\epsilon^r_p) \Big ], \label{particle} 
\end{align}
\begin{align}
&P_{N}(n_p=0)\Big [ \Lambda^l_p(1- f(\tilde{\epsilon}^l_p))+\Lambda_p^r(1- f(\tilde{\epsilon}^r_p)) \Big ] \nonumber \\
=& P_{N-1}(n_p=1)\Big [ \Lambda^l_p f(\tilde{\epsilon}^l_p )+ \Lambda_p^r  f(\tilde{\epsilon}^r_p) \Big ]. \label{hole}
\end{align}
We expand the lead voltage $V$ to the linear order by using \cref{PN expression}
\begin{widetext}
\begin{small}
\begin{align}
&\beta e P_{N}^{\rm{eq}}(n_p=1) \bigg \{  \Phi_N(n_p=1,\{n_{i\neq p}\}) (\Gamma_p^l + \Gamma_p^r)(1- f(\epsilon_p)) +     f(\epsilon_p) f(-\epsilon_p)  \Big [   \eta  \Gamma_p^l - (1-\eta) \Gamma_p^r  \Big]   \bigg \} \nonumber \\
=&\beta e P_{N-1}^{\rm{eq}}(n_p=0) \bigg \{  \Phi_{N-1}(n_p=0,\{n_{i\neq p}\}) (\Gamma_p^l + \Gamma_p^r)f(\epsilon_p) -     f(\epsilon_p) f(-\epsilon_p)  \Big [   \eta  \Gamma_p^l - (1-\eta) \Gamma_p^r  \Big]   \bigg \}  \\
&\beta e P_{N}^{\rm{eq}}(n_p=0) \bigg \{  \Phi_N(n_p=0,\{n_{i\neq p}\}) (\Lambda_p^l + \Lambda_p^r)(1- f(\tilde{\epsilon}_p)) +     f(\tilde{\epsilon}_p) f(-\tilde{\epsilon}_p)  \Big [   \eta  \Lambda_p^l - (1-\eta) \Lambda_p^r  \Big]   \bigg \} \nonumber \\
=&\beta e P_{N-1}^{\rm{eq}}(n_p=1) \bigg \{  \Phi_{N-1}(n_p=1,\{n_{i\neq p}\}) (\Lambda_p^l + \Lambda_p^r)f(\tilde{\epsilon}_p) -     f(\tilde{\epsilon}_p) f(-\tilde{\epsilon}_p)  \Big [   \eta  \Lambda_p^l - (1-\eta) \Lambda_p^r  \Big]   \bigg \} 
\end{align}
\end{small}
After the simplification 
 \begin{align}
 0= 
\Phi_N(n_p=1,\{n_{i\neq p}\})-\Phi_{N-1}(n_p=0,\{n_{i\neq p}\})+\eta - \frac{\Gamma^r_p}{\Gamma_p^l+\Gamma_p^r} \label{eq const1}, \\
0= 
  \Phi_N(n_p=0,\{n_{i\neq p}\})-\Phi_{N-1}(n_p=1,\{n_{i\neq p}\})+\eta - \frac{\Lambda^r_p}{\Lambda_p^l+\Lambda_p^r} \label{eq const2},
 \end{align}
we use the condition $\{ \Gamma_p^l/\Gamma_p^r= \Lambda_p^l/\Lambda_p^r=\rm{const}\}$ and let $\alpha\equiv \frac{\Gamma^r_p}{\Gamma_p^l+\Gamma_p^r}=\frac{\Lambda^r_p}{\Lambda_p^l+\Lambda_p^r}$. \Cref{eq const1,eq const2} have the relevant solutions; hence, dropping $\sum_p$ in \cref{PN,stationary} is legitimate.  Again using the identity $1-f(\epsilon)=f(\epsilon)e^{\beta \epsilon}$ and \cref{P relation}, we obtain the expression for the tunneling conductance from \cref{right current}
\begin{align} 
\frac{dI}{dV}
 =&\beta e^2 \sum_{p} \sum_{\{n_i\}} \bigg \{ \delta_{n_p,0} P_{N-1}^{\rm{eq}} (\{ n_i \})f(\epsilon_p ) \frac{\Gamma^l_p\Gamma_p^r}{\Gamma^l_p+\Gamma_p^r}  
 +\delta_{n_p,1} P_{N-1}^{\rm{eq}} (\{ n_i \})f(\tilde{\epsilon}_p ) \frac{\Lambda^l_p\Lambda_p^r}{\Lambda^l_p+\Lambda_p^r}  
\bigg \} \nonumber
\\
 =&\beta \alpha e^2 \sum_{p} \sum_{\{n_i\}} \bigg \{ \delta_{n_p,0} P_{N-1}^{\rm{eq}} (\{ n_i \})f(\epsilon_p ) \Gamma_p^l
 +\delta_{n_p,1} P_{N-1}^{\rm{eq}} (\{ n_i \})f(\tilde{\epsilon}_p )  \Lambda_p^l
\bigg \}  \label{current part}
\end{align}
\end{widetext}
Since the fermion parity is the only conserved quantity and the particle number is not conserved, the physics is not altered by the transformation $N\rightarrow N+2$ and $n_g \rightarrow n_g+2$ in \cref{staticenergy};  hence, only even and odd $N$'s lead to distinct conductances, which can be explicitly written down for numerical calculations. When $N$ is even, the conductance in the explicit form is given by  
\begin{equation} \label{oddc}
\frac{dI}{dV}=\frac{e^2\alpha}{kT}\frac{\sum_{p} \big [ A_{\rm{odd}}(\epsilon_p ) + B_{\rm{even}}(\tilde{\epsilon}_p)\big ] }{F_{\rm{even}}e^{-\beta \Delta U}+F_{\rm{odd}}}  
\end{equation}
where
\begin{small}
\begin{align*}
A_{\rm{even/odd}}(\epsilon_p ) & = f(\epsilon_p)  F_{\rm{even/odd}}(E_p)\Gamma_p^l \\
B_{\rm{even/odd}}(\epsilon_p )&  = f(\tilde{\epsilon}_p)  e^{-\beta E_p}F_{\rm{even/odd}}(E_p)\Lambda_p^l \\
F_{\rm{even/odd}}&=\sum_{\sum_i n_i=\rm{even/odd}} e^{-\beta \sum_i n_iE_i} \\
F_{\rm{even/odd}}(E_p)&=\sum_{\sum_{i\neq p} n_i=\rm{even/odd}} e^{-\beta \sum_i n_iE_i}
\end{align*}
\end{small}
Likewise, when $N$ is odd, the conductance is given by 
\begin{equation} \label{evenc}
\frac{dI}{dV}=\frac{e^2\alpha}{kT}\frac{\sum_{p}\big [ A_{\rm{even}}(\epsilon_p ) + B_{\rm{odd}}(\tilde{\epsilon}_p)\big ]}{F_{\rm{even}}+F_{\rm{odd}}e^{-\beta \Delta U}} . 
\end{equation}
We further consider the low temperature limit $T\ll | E_{i>1} |$ (still $T \gg \Gamma_p^{l,r},\ \Lambda_p^{l,r}$). The conductance can then be written in the simple form, which is consistent with [\onlinecite{conductance_coulomb_blockade_roman}].
For even $N$, the low-temperature conductance can be simply written as 
\begin{equation}
\frac{dI}{dV}=\frac{e^2}{kT}\frac{\Lambda^l_1 \Lambda^r_1}{\Lambda^l_1+\Lambda^r_1}\frac{1}{4 \cosh^2 (\beta \tilde{\epsilon}_1)} \label{lowT_conductance_odd}
\end{equation}
When $N-1$ is even and $N$ odd, we have the following low-temperature limit 
\begin{equation}
\frac{dI}{dV}=\frac{e^2}{kT}\frac{\Gamma^l_1 \Gamma^r_1}{\Gamma^l_1+\Gamma^r_1}\frac{1}{4 \cosh^2 (\beta \epsilon_1)} \label{lowT_conductance_even}
\end{equation}
The conductance reaches the maximum as $\tilde{\epsilon}_1=0$ or $\epsilon_1=0$, and the broadening of the conductance peak is proportional to $T$.  Hence, at low temperatures the conductance peaks for even and odd $N$ are located at $n_g(N_e)=N_e-1/2-E_1/2E_c$ and  $n_g(N_o)=N_o-1/2+E_1/2E_c$ respectively, where $N_e(N_o)$ indicates even(odd) $N$. The key quantities studied in the experiment \cite{Albrecht:2016aa} are the even and odd $1e$ oscillation peak spacings 
\begin{align}
S_o=&n_g(N_o+1)-n_g(N_o)=1-E_1/E_c, \label{SO} \\
S_e=&n_g(N_e+1)-n_g(N_e)=1+E_1/E_c, \label{SE}
\end{align}
which are the differences between the two closest peaks (see fig.~\ref{ng_plot}). Except for odd and even $N$, the spacings should be independent of $N$. In the experiment, the multiple conductance peaks are measured as the gate voltage of the nanowire varies in a wide region. Since the peaks might fluctuate for different $N$, the even and odd peak spacings are averaged over multiple $N$'s to suppress non-universal effects. All information about the underlying physics is extracted from these conductance peak spacings in Ref.~\cite{Albrecht:2016aa}.

\subsection{The  analytic solution for 2-level systems (tunneling from localized states)}

 Now we consider only
2 energy levels in the nanowire when $N$ is even. This may apply at very low temperatures where only the lowest two energy levels near zero energy are operational in transport processes. The total number of the occupied quasiparticle $n_1+n_2$ is $0$ or $2$. The four stationary equations can be explicitly written as \\
\begin{small}
\cref{vPN}: $\partial P_N(1,1)/\partial t=0 $
\begin{align}
0=&\peq(0,1)f(\epsilon_1)\Gamma_1[\Phi_N(1,1)-\Phi_{N-1}(0,1)+\eta - \gamma^r_1] \nonumber \\
+&\peq(1,0)f(\epsilon_2)\Gamma_2[\Phi_N(1,1)-\Phi_{N-1}(1,0)+\eta - \gamma^r_2],
\end{align}
\cref{vPN}: $\partial P_N(0,0)/\partial t=0 $
\begin{align}
0=&\peq(1,0)f(\tilde{\epsilon}_1)\Lambda_1[\Phi_N(0,0)-\Phi_{N-1}(1,0)+\eta - \lambda^r_1] \nonumber \\
+&\peq(0,1)f(\tilde{\epsilon}_2)\Lambda_2[\Phi_N(0,0)-\Phi_{N-1}(0,1)+\eta - \lambda^r_2],
\end{align}
\cref{vPN1}: $\partial P_{N-1}(0,1)/\partial t=0 $
\begin{align}
0=&\peq(0,1)f(\epsilon_1)\Gamma_1[\Phi_N(1,1)-\Phi_{N-1}(0,1)+\eta - \gamma^r_1] \nonumber \\
+&\peq(0,1)f(\tilde{\epsilon}_2)\Lambda_2[\Phi_N(0,0)-\Phi_{N-1}(0,1)+\eta - \lambda^r_2],
\end{align}
\cref{vPN1}: $\partial P_{N-1}(1,0)/\partial t=0 $
\begin{align}
0=&\peq(1,0)f(\tilde{\epsilon}_1)\Lambda_1[\Phi_N(0,0)-\Phi_{N-1}(1,0)+\eta - \lambda^r_1] \nonumber \\
+&\peq(1,0)f(\epsilon_2)\Gamma_2[\Phi_N(1,1)-\Phi_{N-1}(1,0)+\eta - \gamma^r_2],
\end{align}
\end{small} \noindent
where $\Gamma_i=\Gamma^l_i+\Gamma^r_i$. By solving 3 of these 4 stationary equations, we obtain the explicit expression for the conductance
\begin{widetext}
\begin{small}
\begin{align}
\frac{dI}{dV}=&\beta e^2 \bigg \{ \peq(0,1) f(\epsilon_1)\frac{\Gamma^l_1\Gamma_1^r}{\Gamma_1} +\peq(1,0) f(\epsilon_2)\frac{\Gamma^l_2\Gamma_2^r}{\Gamma_2}+\peq (1,0) f(
\tilde{\epsilon}_1)\frac{\Lambda^l_1\Lambda_1^r}{\Lambda_1}+\peq(0,1) f(\tilde{\epsilon}_2)\frac{\Lambda^l_2\Lambda_2^r}{\Lambda_2} \nonumber \\
&- (\gamma^l_1-\gamma_2^l+\lambda^l_1-\lambda_2^l)(\gamma_1^r-\gamma_2^r+\lambda_1^r-\lambda_2^r) \nonumber \\
&\times \Big( \frac{1}{\peq (0,1) f(\epsilon_1)\Gamma_1}+\frac{1}{\peq (1,0) f(\epsilon_2)\Gamma_2}+\frac{1}{\peq (1,0) f(\tilde{\epsilon}_1)\Lambda_1}+\frac{1}{\peq (0,1) f(\tilde{\epsilon}_2)\Lambda_2} \Big )^{-1} \bigg \}, \label{odd_even}
\end{align}
\end{small}
where $\gamma^{l,r}_i=\Gamma^{l,r}_i/\Gamma_i$ and $\lambda^{l,r}_i=\Lambda^{l,r}_i/\Lambda_i$. 
Similarly, when $N-1$ is even and $N$ is odd, the expression for the conductance is given by
\begin{small}
\begin{align}
\frac{dI}{dV}=&\beta e^2 \bigg \{ \peq(0,0) f(\epsilon_1)\frac{\Gamma^l_1\Gamma_1^r}{\Gamma_1} +\peq(0,0) f(\epsilon_2)\frac{\Gamma^l_2\Gamma_2^r}{\Gamma_2}+\peq (1,1) f(
\tilde{\epsilon}_1)\frac{\Lambda^l_1\Lambda_1^r}{\Lambda_1}+\peq(1,1) f(\tilde{\epsilon}_2)\frac{\Lambda^l_2\Lambda_2^r}{\Lambda_2} \nonumber \\
&- (\gamma^l_1-\gamma_2^l+\lambda^l_1-\lambda_2^l)(\gamma_1^r-\gamma_2^r+\lambda_1^r-\lambda_2^r) \nonumber \\
&\times \Big( \frac{1}{\peq (0,0) f(\epsilon_1)\Gamma_1}+\frac{1}{\peq (0,0) f(\epsilon_2)\Gamma_2}+\frac{1}{\peq (1,1) f(\tilde{\epsilon}_1)\Lambda_1}+\frac{1}{\peq (1,1) f(\tilde{\epsilon}_2)\Lambda_2} \Big )^{-1} \bigg \}, \label{even_odd}
\end{align}
\end{small}
As shown in the next section, the tunneling rates $\Gamma_p^l$, $\Gamma_p^r$, $\Lambda_p^l$, and $\Lambda_p^r$ depend on the wavefunctions of the quasiparticles and quasiholes on the ends of the nanowire. Since the tunneling ratios are the same constant $( \Gamma_p^l/\Gamma_p^r= \Lambda_p^l/\Lambda_p^r=\rm{const})$ in our approximation, the last terms vanish; the conductance equations are reduced to \cref{current part}. Because the MBSs are located on the two ends, the conductance peaks stem from the first two terms of \cref{odd_even,even_odd}.

Consider two MBSs, which possess the hybridization energy $E_1$ due to wavefunction overlap, located on the left and the middle of the nanowire and assume that the first excited state with energy $E_2$ is delocalized. Since $\Gamma_1^r, \Lambda_1^r \sim 0$ in the absence of the Majorana on the right end, the first and second terms vanish in $dI/dV$. The tunneling ratios are different constants so that  the last term survives.
However, an electron still can propagate through the MBSs in the presence of the last term of the conductance equations. The physical meaning of the last term mixing with the two-energy-level rates is that after encountering the MBS near the left lead, an electron moves out to the right lead through the extended states in the nanowire.

These transport equations can capture $1e$ tunneling of the conventional (i.e. non-topological) Andreev bound states. Consider two Andreev bound states localized on the two ends separately and the other energy levels much higher than the temperature; the localization leads to $\Gamma_1^r=\Lambda_1^r=\Gamma_2^l=\Lambda_2^l=0$. The conductance does not vanish in this limit and is given by:
\begin{align}
\frac{dI}{dV}=&4\beta e^2
\Big( \frac{1}{\peq (0,1) f(\epsilon_1)\Gamma_1}+\frac{1}{\peq (1,0) f(\epsilon_2)\Gamma_2}+\frac{1}{\peq (1,0) f(\tilde{\epsilon}_1)\Lambda_1}+\frac{1}{\peq (0,1) f(\tilde{\epsilon}_2)\Lambda_2} \Big )^{-1}.  \label{two_Andreev}
\end{align}
\end{widetext}
This result is consistent with the observation in an earlier work~\cite{PhysRevB.92.020511} that the so-called ``teleportation" phenomenon typically associated with MBSs in the Coulomb blockade regime~\cite{PhysRevLett.104.056402}, can also occur for non-topological Andreev bound states~\cite{PhysRevB.92.020511}.

To see the clear meaning of \cref{two_Andreev} we consider a special case that the system preserves reflection symmetry and the particle and hole tunneling rates are identical. Hence, the notations can be simplified to $E\equiv E_1=E_2$, $\Gamma\equiv \Gamma_1=\Gamma_2=\Lambda_1=\Lambda_2$. The conductance in Eq.\ 2.32 can be rewritten as
\begin{equation*}
\frac{dI}{dV}=8 \beta e^2 \Gamma [2+e^{\beta \epsilon}+e^{\beta \tilde{\epsilon}}]^{-1}[2+e^{-\beta \epsilon}+e^{-\beta \tilde{\epsilon}}]^{-1},
\end{equation*}
where $\epsilon=E+\Delta U$ and $\tilde{\epsilon}=-E+\Delta U$. As $\Delta U=0$, the conductance reaches to the maximum with the value 
\begin{equation*}
\frac{dI}{dV}=8 \beta e^2 \Gamma [2+e^{\beta E}+e^{-\beta E}]^{-2}.
\end{equation*}
The location of the peak is independent of the energy of the two states. Furthermore, the conductance peak corresponds to $n_g=N+1/2$. We note that this case is one of the extreme limits. It does not imply the conductance peak is always independent of the energy of the states.



\section{Tunneling and nanowire Hamiltonians} \label{section III}

The energy levels $E_p$ and the tunneling rates $\Gamma_p^l$, $\Gamma_p^r$, $\Lambda_p^l$, and $\Lambda_p^r$ are the necessary microscopic input parameters in order to perform the conductance calculations for the SC proximitized semiconductor nanowire. We assume that the superconducting order parameter $\Delta$ in the semiconductor nanowire is proximity induced through contact with a regular metallic superconductor and study the BdG Hamiltonian of the 1D model described by reasonable physical parameters. We note that the energy scales of our results depend on our assumptions about these physical parameters, in particular the SC gap, the Coulomb blockade energy,  the nanowire effective mass and g-factor, the chemical potential, the spin-orbit coupling, various hopping amplitudes, and the actual confinement potential in the nanowire.  Most, if not all, of these parameters are unknown for the real experimental systems. Therefore, one should not attach special significance to our absolute numbers, particularly the precise temperature scales defining our high and low temperature regimes.

%
%

\subsection{Tunneling rate}\label{tunneling}

The tunneling rates $\Gamma_p^l$, $\Gamma_p^r$, $\Lambda_p^l$, and $\Lambda_p^r$ are related to the wavefunction with energy $E_p$ near the leads. We first write 
the BdG Hamiltonian of the superconducting nanowire in real space 
\begin{align}
\hat{H}_{\rm{nano}}= C^\dagger 
\bma 
H_o & i \Delta   \bI \otimes \sigma_y \\
-i \Delta^*  \bI \otimes \sigma_y  & -H_o^* \\
\ema
C,
\end{align} 
where the annihilation operator including all of the lattice sites is written as $C=(\ldots,c_{x\uparrow},c_{x\downarrow},c_{x\uparrow}^\dagger,c_{x\downarrow}^\dagger,\ldots)^T$. By diagonalizing $H_{\rm{nano}}$, the Hamiltonian can be rewritten in the diagonal form 
\bee
\hat{H}_{\rm{nano}}= \sum_{p} \Big ( E_p a^\dagger_p a_p - E_p a_p a_p^\dagger \Big ),\quad E_p\geq 0.
\ee
The quasiparticle and quasihole for the energy level $p$ are given by 
\begin{align}
a_p^\dagger=& \sum_{x, \alpha=\uparrow,\downarrow} \Big ( u_{p,x\alpha}  c^\dagger_{x\alpha}+ v_{p,x\alpha}  c_{x\alpha} \Big ) \\
a_p=& \sum_{x, \alpha=\uparrow,\downarrow} \Big (  v_{p,x\alpha}^* c^\dagger_{x\alpha}+ u_{p,x\alpha}^*   c_{x\alpha} \Big )
\end{align}
The normalization leads to $\sum_{x,\alpha} (|u_{p,x\alpha}|^2+|v_{p,x\alpha}|^2)=1$. The electron creation and annihilation operators are written in terms of quasiparticles and holes
\begin{align}
c_{i\alpha}^\dagger=&  \sum_{p} \Big ( u_{p,x\alpha}^* a^\dagger_p + v_{p,x\alpha} a_p\Big ) \\
c_{i\alpha}=&  \sum_{p} \Big ( v_{p,x\alpha}^* a^\dagger_p + u_{p,x\alpha} a_p\Big ) 
\end{align}
Now we determine the tunneling rates $\Gamma_p^l$, $\Gamma_p^r$, $\Lambda_p^l$, and $\Lambda_p^r$ by assuming $kT,\ \Delta E \gg \Gamma,\ \Lambda$, where $\Delta E$ is the energy level separation. According to Fermi's golden rule, the tunneling rates are proportional to 
\bee
|\bra{f} H_t \ket{i}|^2,
\ee
where $\ket{f}$ and $\ket{i}$ are the initial and final states respectively and $H_t$ is the tunneling part of the Hamiltonian. First, consider the tunneling between the left lead and the superconducting nanowire; the tunneling Hamiltonian can be written as
\bee
H_t^l= \sum_{p,\alpha} t_p \bigg ( L_{\epsilon_p}^\dagger c_{1,\alpha} + c_{1,\alpha} ^\dagger  L_{\epsilon_p} \bigg),
\ee
where $c_{1,\alpha}, c_{1,\alpha}^\dagger$ are the annihilation and creation operators located at the left end of the nanowire. As the electron moves from the left lead to the energy level $p$ in the nanowire, the initial and final states are given by
\bee
\ket{i}_p=L^\dagger_{\epsilon_p} \ket{O_L} \ket{BCS},\quad \ket{f}_p=a^\dagger_p \ket{O_L} \ket{BCS}
\ee
where $\ket{O_L}$ and $\ket{BCS}$ are the normalized wavefunction in the left lead and the superconducting nanowire respectively and $L^\dagger_{\epsilon_p}$ is the electron creation operator with energy $\epsilon_p$. Hence, 
\begin{small}
\begin{align}
\bra{f} H_t \ket{i}_p&= t_p \sum_{\alpha} \bra{BCS} \bra{O_L} a_p \big (  L_{\epsilon_p}^\dagger c_{1,\alpha} + c_{1,\alpha} ^\dagger  L_{\epsilon_p}  \big) L^\dagger_{\epsilon_p} \ket{O_L} \ket{BCS} \nonumber \\
&= t_p \bra{BCS} \bra{O_L} a_p \big (   \sum_{\alpha} u_{p,1\alpha}^* a^\dagger_p   \big)  \ket{O_L} \ket{BCS} \nonumber \\
&= t_p \sum_{\alpha} u_{p,1\alpha}
\end{align}
\end{small}
We can obtain the tunneling rate 
\begin{align}
\Gamma_p^l=t_p'^2 |\bra{f} H_t \ket{i}_p|^2= t_p'^2 |\sum_{\alpha} u_{p,1\alpha}|^2,
\end{align}
where $t_p'$ absorbs all of the constants from Fermi's golden rule. Similarly, for an electron moving from the nanowire to the left end the tunneling rate is identical. On the other hand, for a hole movement, by following a similar derivation, we have the tunneling rate 
\bee
\Lambda_p^l = w_p'^2 |\sum_{\alpha} v_{p,1\alpha}|^2
\ee
We assume the undetermined tunneling constants $t_p'=w'_p=1$ ($\Gamma_p^{l,r}, \Lambda_p^{l,r} \ll T$ still holds) for convenience leading to 
\begin{align}
\Gamma_p^l= |\sum_{\alpha} u_{p,1\alpha}|^2,\ \Gamma_p^r= |\sum_{\alpha} u_{p,L\alpha}|^2, \nonumber \\
\Lambda_p^l = |\sum_{\alpha} v_{p,1\alpha}|^2,\  \Lambda_p^r = |\sum_{\alpha} v_{p,L \alpha}|^2,
\end{align}
where $L$ is the length of the nanowire (We intentionally do not provide the unit of $\Gamma_p^l$ and $\Lambda_p^l$ since the details of the experimental setups are unknown and non-universal. Our focus is on the OCPS, which is universal and is not necessarily determined by the exact value of the conductance). 


\subsection{The Hamiltonian for 1D superconducting nanowire }

\begin{figure}[t!]
\begin{center}
\includegraphics[clip,width=0.80\columnwidth]{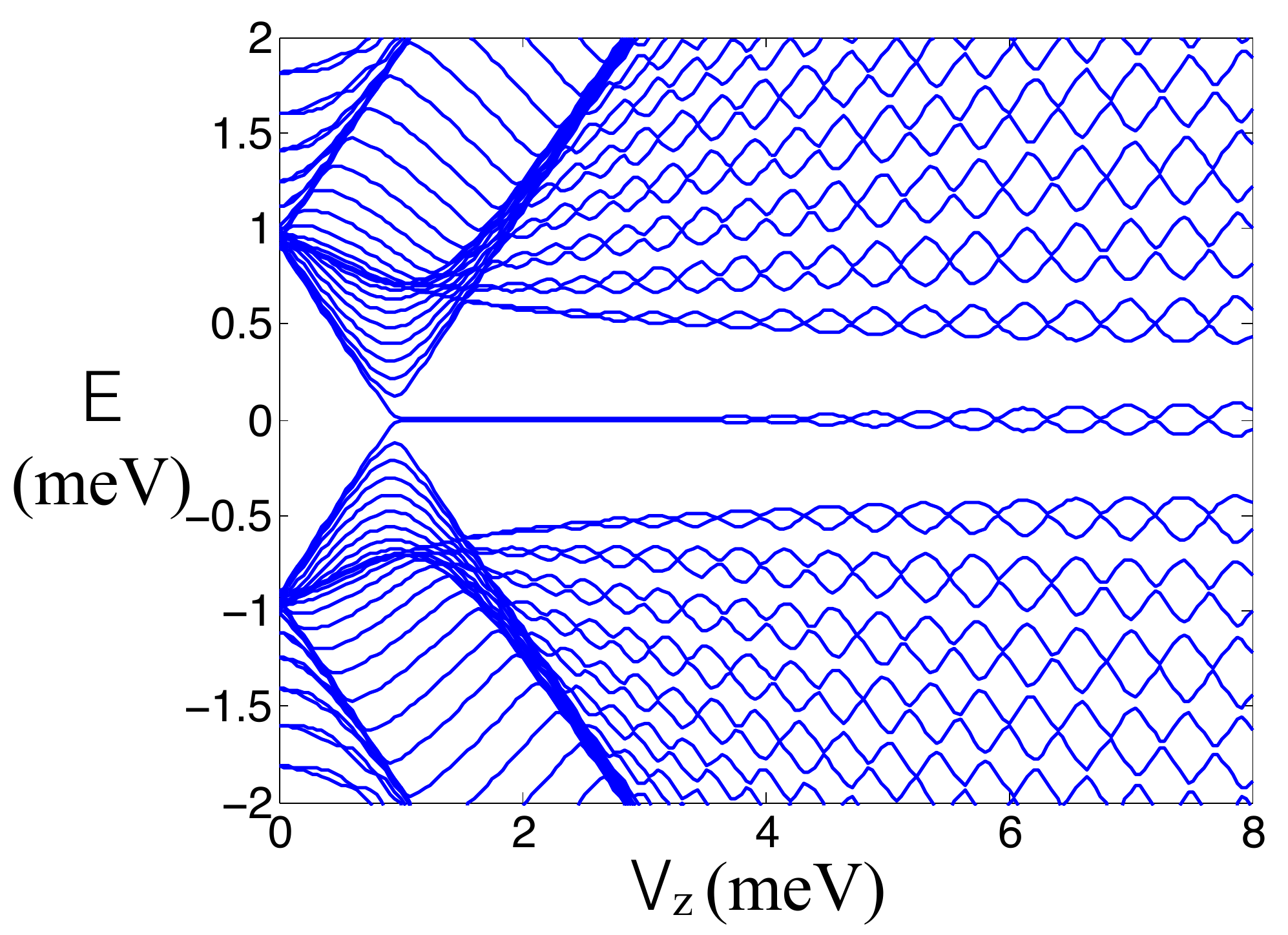}
\end{center}
  \caption{The superconducting nanowire spectrum for $t=6$meV, $\Delta=0.9$meV, $\alpha=1.2$meV, $\mu=0.2$meV, $V_y=0$meV, $L=80$. The TQPT point is located at $V_z=\sqrt{\Delta^2+\mu^2}=0.922$meV. After the TQPT point, the oscillation of Majorana hybridization energy grows as the magnetic field increases.  }\label{spectrum_L80key}
\end{figure}

 The 1D SC proximitized semiconductor nanowire with spin-orbit coupling in the presence of a field-induced Zeeman spin splitting~\cite{Gil_Majorana_wire,Roman_SC_semi} can be described in momentum space as
\begin{widetext}
\bee
H_{\rm{BdG}}(k)=\big [ 2t( 1- \cos ka ) - \mu \big]\tau_z\sigma_0 + \Delta \tau_y \sigma_y + V_z \tau_z \sigma_x + V_y \tau_0 \sigma_y + 2\alpha \sin ka \tau_z \sigma_y 
\ee
Using the open boundary condition, the Hamiltonian can be written in the following form suitable for numerical calculations  
\bee
\hat{H}_{\rm{BdG}}=\sum_x \bigg \{ C^\dagger_x \Big [ \big ( 2t - \mu \big )\tau_z\sigma_0 + \Delta \tau_y \sigma_y + V_z \tau_z \sigma_x + V_y \tau_0 \sigma_y \Big ] C_x + \Big [ C_{x+a}^\dagger  (-t\tau_z\sigma_0 + \alpha i\tau_z \sigma_y )C_{x} +h.c. \Big ] \bigg\},  \label{nanowire Hamiltonian}
\ee
\end{widetext}
where $C_i=(c_{\uparrow i}, c_{\downarrow i},c_{\uparrow i}^\dagger, c_{\downarrow i}^\dagger)$.  In the following calculations (unless specified otherwise), our choice of the physical parameters is based on \cite{parameter_1DMajorana} with slight differences: hopping strength $t=6$meV, spin-orbit coupling $\alpha=1.2$meV, superconducting order parameter $\Delta=0.9$meV, the chemical potential $\mu=0.2$meV, and the Zeeman splitting $V_z=1.2B$meV, $V_y=0$, where $B$ is the magnetic field in unit of Tesla \cite{ballistic_M}. We consider the number of lattice sites to be $L=80$ (we vary $L$ later in presenting our results) in the unit of the lattice constant $a=10$nm so the length of the nanowire is 0.8 $\mu$m and show the wire spectrum in fig.~\ref{spectrum_L80key}. 
These parameters are representative for the experimental system used in Ref.~\cite{Albrecht:2016aa}, and changing these parameters to other reasonable values for currently used semiconductor-superconductor hybrid systems does not change any of our qualitative conclusions. 
We have made no attempt to obtain quantitative agreement with experiments since our goal here is to understand the findings of Ref.~\cite{Albrecht:2016aa} qualitatively, in particular to see if the intrinsic inconsistency between the length and field dependence of the quoted oscillation amplitude in Ref.~\cite{Albrecht:2016aa} can be explained by some mechanism.  There are sufficient numbers of unknown parameters (e.g. the chemical potential and the nanowire confinement potential) in the problem (even in the clean limit without invoking any disorder) rendering quantitative comparisons meaningless.

By performing exact-diagonalization of the Hamiltonian, the eigenenergies of the nanowire are given by $\pm \varepsilon_i$, where $i$ is a positive integer and $\varepsilon_i$ is positive due to particle-hole symmetry in the BdG Hamiltonian. The relation between the quasi-particle and the quasi-hole can be described by  
\begin{equation}
a_{\pm \varepsilon_i}^\dagger= a_{\mp \varepsilon_i}
\end{equation}
Since varying the magnetic field through the nanowire is an adiabatic process, the fermion parity of the BCS wavefunction is conserved. In the absence of the magnetic field, we start with the BCS ground state obeying
\begin{equation}
a_{\varepsilon_i}\ket{\rm{BCS}}=0 \label{even BCS}, 
\end{equation}
and define the energy levels as $E_p=\varepsilon_p$ for all $p$ for the transport calculation. 
As the magnetic field increases, the lowest positive energy of the quasiparticle $\varepsilon_1$ reaches zero at the TQPT. Due to the Majorana wavefunction overlap in the finite wire, the quasiparticle $a_1$, which is the hybridization of the two MBSs, adiabatically evolves to become a quasiparticle with negative energy $-\varepsilon_1$. After this energy level crossing at zero energy, the BCS ground state evolves to the first excited state obeying 
\begin{eqnarray} 
    a_{\varepsilon_{i>1}}\ket{{\rm{BCS}_e}}=0,\ a_{\varepsilon_{1}}^\dagger\ket{\rm{BCS}_e}=0  
\end{eqnarray}
Due to the parity fermionic conversion, we still use this state as the basis state (not the ground state) to compute the conductance; the energy levels have to be redefined $E_1=-\varepsilon_1$, $E_p=\varepsilon_p$ for all $p>1$. As the magnetic field keeps increasing, after the next zero energy crossing, the BCS state goes back to the original definition in \cref{even BCS} until the third level crossing and so on. In the following conductance calculation, the basis BCS states are determined by the level crossings as the magnetic field varies. We carry out our numerical calculations following the above prescription and present our results in the next section.

\begin{figure}[t!]
\begin{center}
\includegraphics[clip,width=0.99\columnwidth]{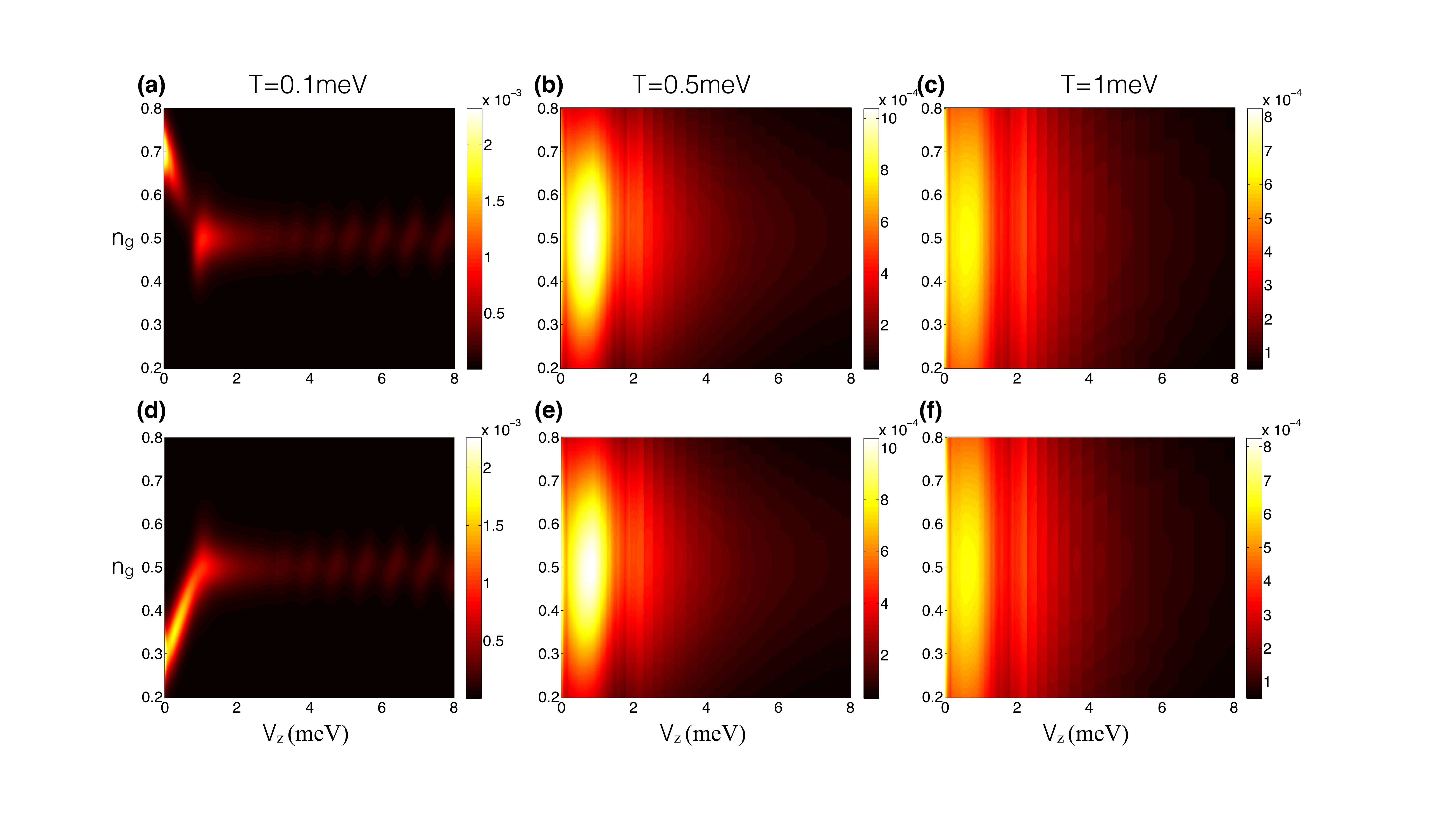}
\end{center}
  \caption{The conductance (arbitrary unit) at three different temperatures ($T=0.1 $(a,d)$,\ 0.5 $(b,e)$,\ 1.0$ (c,f)meV) as $n_g$ and $V_z$ vary. We shift the main conductance peak at $0.5$ by defining $n_g'=n_g-N+0.5$. The top three subfigures are for odd $N$ and the bottom three subfigures are for even $N$. High temperature broadens the conductance peak.} \label{conductance_finite_T}
\end{figure}

\section{Numerical results for conductance}\label{section IV}

\begin{figure*}[htp!]
\begin{center}
\includegraphics[clip,width=1.8\columnwidth]{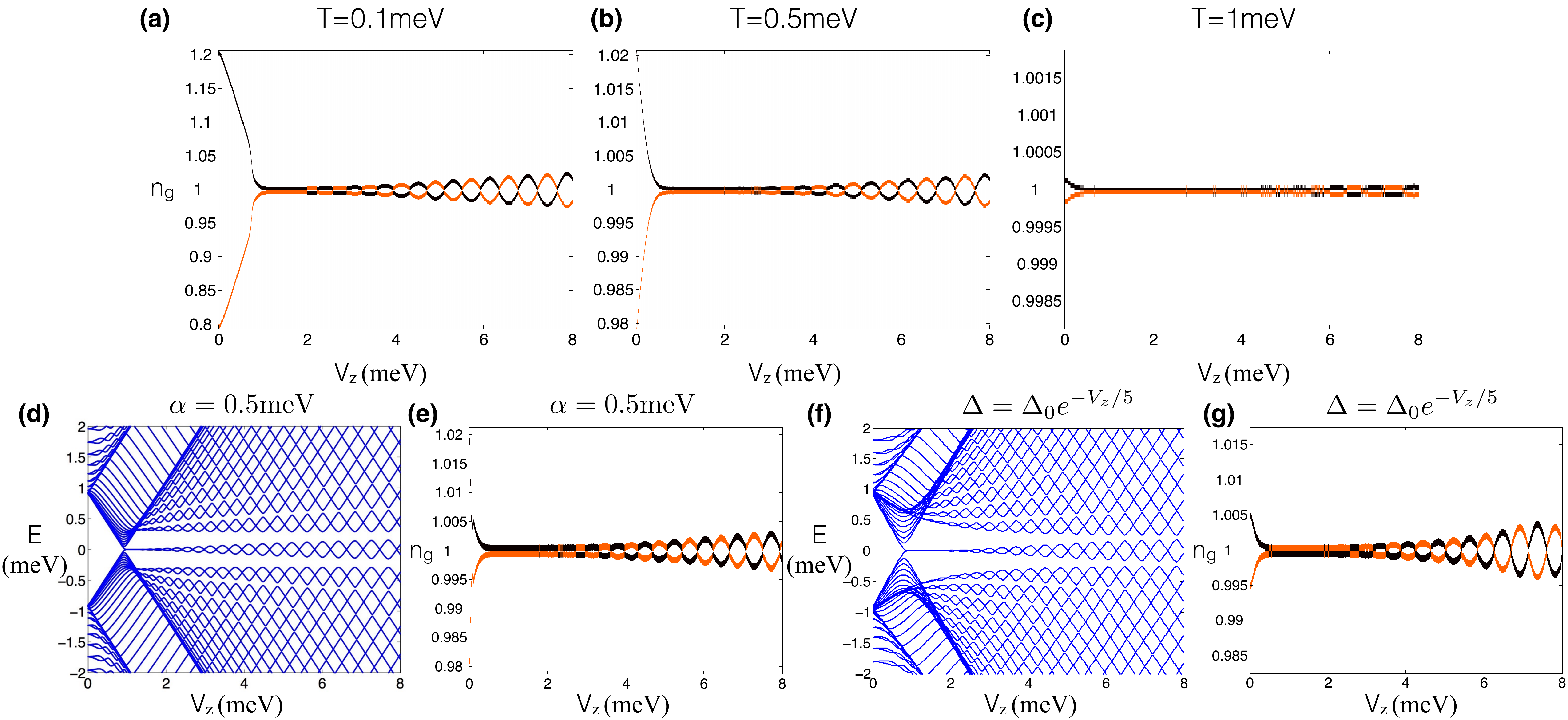}
\end{center}
  \caption{The even $S_e$(black) and odd $S_o$(orange) conductance peak spacings at difference temperatures (a-c) and at the same temperature ($T=0.5$meV) with different parameters (e,g). (a-c) show the spacings at three different temperatures ($T=0.1,\ 0.5,\ 1$meV). Although high temperature dramatically squeezes the OCPSs, the oscillation amplitude increases. (d) the energy spectrum of the nanowire as the spin-orbital coupling $\alpha$ is adjusted to $0.5$meV to shrink the bulk gap. (e) even when the bulk gap is smaller, the oscillation amplitude still increases. (f) as the order parameter has exponential decay $\Delta=\Delta_o e^{-V_z/5}$, the bulk gap also shrinks. (g) the oscillation becomes larger, except for the last pocket, as the magnetic field increases.  } \label{regular_finite_T_oscillation}
\end{figure*}

We numerically compute the \emph{magnetic-field-dependent} conductance of the superconducting nanowire (\ref{nanowire Hamiltonian}) hosting MBSs on the ends after passing through the TQPT. The nanowire described by the Hamiltonian (\ref{nanowire Hamiltonian}) \emph{accidentally} preserves reflection symmetry and all of the eigenstates are extended, except for the two localized Majoranas at the ends; hence, the tunneling rate on the left and right are equal ($\Gamma_i^r=\Gamma_i^l,\ \Lambda_i^r=\Lambda_i^l$ for all the energy levels). Due to the Majorana hybridization in the finite length nanowire, the tunneling ratio of the hybridized states should be $\Gamma_1^r=\Gamma_1^l,\ \Lambda_1^r=\Lambda_1^l$. With these fixed tunneling ratios, the conductance can be simply computed by using \cref{evenc,oddc}. The energy spectrum and the wavefunction on the ends can be obtained by solving the eigenvalue problem of the Hamiltonian (\ref{nanowire Hamiltonian}). For computing the conductance, we still need to know one more physical quantity, which is the magnitude of the electrostatic energy difference between the total numbers $N-1$ and $N$ arising from Coulomb blockade physics
\begin{align}
\Delta U  &= U(N)-U(N-1) \nonumber \\
&=E_c(1-2n_g)+E_c(2N+1)
\end{align}
When $E_c$ is less than the nanowire gap ($< \Delta =0.9$meV), the 2$e$ periodicity of the conductance peak, arising from pure Andreev process in the SC wire, dominates. Here our focus is only on the $1e$ periodicity region since we are interested in the MBS physics; therefore, we choose $E_c=2$meV greater than the superconducting nanowire gap. (We do not discuss the 2e-periodic physics \cite{conductance_coulomb_blockade_roman} in this work focusing entirely on the large charging energy, $E_c>\Delta$, regime since this is presumably not connected with MBSs.) Furthermore, the gate voltage ($V_g$) of the nanowire is an experimentally controllable physical parameter, which is proportional to $n_g$ since $n_g=CV_g/e$, where $C$ is the capacitance of the nanowire. In the following, the conductance of the nanowire will be computed in wide regions of normalized gate voltage $(n_g)$ and Zeeman splitting ($V_z$) from the magnetic field ($B$). Furthermore, the energy levels much higher than temperature can be neglected since they do not participate in the transport process. 
We keep the ten lowest energy levels for the conductance calculations carried out in the current work although it should be feasible to add a few more levels in the numerical work if there is good reason to do so.  We believe that 10 levels should suffice for qualitative conclusions as long as the temperature is not too high (which would kill the superconductivity any way).  We emphasize that at very low temperatures, the theoretical results for OCPSs disagree qualitatively with the experimental results of Ref.~\cite{Albrecht:2016aa} in terms of the magnetic field dependence as described already in the introduction (\cref{section I}) of this paper.

\subsection{Finite temperature}
We numerically compute the conductance of the superconducting nanowire at three different temperatures $T=0.1,\ 0.5,\ 1.0$meV as the magnetic field increases. 
We emphasize that these absolute temperature scales are somewhat arbitrary being totally dependent on the input parameters of our theory (the corresponding parameters for the actual wires in Ref.~\cite{Albrecht:2016aa} are not known), and the three cases should be qualitatively considered as the low, high, and very high temperatures respectively. The low temperature in this manuscript means the temperature is much smaller than the second lowest energy state and still $T\gg \Gamma_p^{l,r}, \Lambda_p^{l,r}$, while the high temperature means at least more than two energy levels are close to or smaller than the temperature scale.  
(Note that all of our chosen temperatures are on the high side since otherwise the experimental results simply cannot be understood as arising from Majorana physics at all; obviously, if all the energy scales are much lower than our input parameter values, then the same physics could emerge at lower temperatures -- we discuss later in this paper the relevance of the energy scale for the experimental nanowires; our choice of model parameters dictate these particular temperature scales given here, which are indeed higher than the quoted temperatures in Ref.~\cite{Albrecht:2016aa} perhaps because the experimental induced gap is lower than our theoretical value.)
 Fig. \ref{conductance_finite_T} shows the calculated nanowire conductance for different $V_z$ and $n_g$. The conductance peaks  ($n^o_g(N),\ n^e_g(N)$) for even and odd $N$'s can be clearly seen at the low temperature as shown in panel (a) and (d)  respectively. At the low temperature, keeping just the lowest energy level is enough to compute the conductance in the topological region (after the bulk gap closes) and OCPSs depend on the the hybridization energy of the MBSs. 
Thus, at the lowest temperature, the oscillation amplitude as a function of Zeeman splitting is indeed a reflection of the low-lying MBS energy spectrum, thus always manifesting an oscillatory amplitude increasing with increasing magnetic field in contradiction to the experimental finding in Ref.~\cite{Albrecht:2016aa}.
Since at higher temperature the conductance peaks, which are thermally broadened, are difficult to visualize, we present in \cref{regular_finite_T_oscillation}, the even and odd conductance peak spacings $S_e=n_g^e(N+1)-n_g^o(N)$ and $S_o=n_g^o(N+1)-n_g^e(N)$ (see the special case in \cref{SO,SE}). 
At the low temperature, as expected and as shown in \cref{regular_finite_T_oscillation} (a), the conductance can be described by \cref{lowT_conductance_odd,lowT_conductance_even}. The spacings should be identical to the two energy levels close to zero energy in \cref{spectrum_L80key}. The important feature of these two energy levels is that after the bulk gap closes, the OCPS amplitude becomes larger as the magnetic field increases due to the Majorana hybridization. This is the same as in the simpler theory~\cite{PhysRevB.86.220506} without Coulomb blockade and in disagreement with the experimental data in Ref.~\cite{Albrecht:2016aa}. Although high temperature suppresses the OCPS due to thermal damping, the amplitude of the OCPS still increases as the magnetic field increases in agreement with the low temperature results even if the actual increase is quantitatively damped by temperature. 
Therefore, the observed amplitude deceasing with increasing magnetic field in Ref.~\cite{Albrecht:2016aa} cannot be explained by temperature effects (and the associated multilevel occupancy) alone -- some other element of physics is still missing in the theory.

	To understand this thermal damping effect of the OCPS, we can consider the limit as the temperature is extremely high. Although the conductance peaks are broadened, the system becomes an effective non-SC metallic Coulomb blockade; the conductance peaks exhibit $1e$ periodicity. Hence, the conductance peak spacings are fixed as $V_z$ increases. In this limit, the OCPS completely vanishes.  In the other limit, at low temperature the conductance peak is reflected by the lowest energy of state in the nanowire. Then, the OCPS due to the Majorana hybridization is expected. Therefore, these two limits show that high temperature suppresses the MBS-induced OPCS without changing its qualitative behavior.

In the next two sub-sections we consider two possible physical effects, shrinking bulk gap with increasing field and contributions from ordinary Andreev bound states in the trivial regime, to see if we can qualitatively reproduce the puzzling results of Ref.~\cite{Albrecht:2016aa}. We note that the orbital effect~\cite{PhysRevB.93.235434}, stemming from parallel magnetic field through the nanowire with a finite cross-sectional diameter, does not explain the decreasing amplitude of the OCPS since the Majorana splitting energy still increases as the magnetic field increases.

%
%
%
%
%
%
%
%

%
%

\subsection{Shrinking bulk gap }
Although the conductance behavior from the standard 1D Majorana Hamiltonian (\ref{nanowire Hamiltonian}) is not consistent with the experimental observation~\cite{Albrecht:2016aa}, we might suspect that the conductance peak might be affected not only by the Majorana hybridization but also by the size of the bulk SC gap. The reason is that, when the bulk gap collapses, both $S_e$ and $S_o$ should be unity without any oscillation similar to an effective Coulomb blockaded normal quantum dot with $1e$ periodicity. We mention that indeed in Ref.~\cite{Albrecht:2016aa} there is experimental evidence supporting the collapse of the bulk gap near where the oscillations are suppressed with increasing field. There are several ways or mechanisms for the bulk gap to shrink with increasing magnetic field -- the most obvious one being that the gap of the parent SC producing the proximity effect itself shrinks with increasing magnetic field.  Some possible mechanisms could be:
a.\ reduce the strength of the spin-orbit coupling ($\alpha$) b.\ reduce the strength of the superconductor order parameter ($\Delta$) as the magnetic field increases. c.\ change the direction of the magnetic field. d. introduce sub-bands having bulk gap closing after the TQPT point. Since at low temperature the conductance peak spacings are almost identical to the two levels close to zero energy; we only need to compute the conductance spacing at high temperature ($T=0.5$meV unless specified) in the following. Clearly, at low temperatures, the gap closing does not affect the results of the last subsection.

\paragraph{$(\alpha)$} the spin-orbit coupling $\alpha$ is adjusted to $0.5$meV from $1.2$meV. As shown in \cref{regular_finite_T_oscillation} (f), the bulk gap near $V_z=8$meV is around $0.25$meV compared with $0.4$meV in \cref{spectrum_L80key} with $\alpha=1.2$meV. At low temperature, the OCPS amplitude still increases as the magnetic field increases based on the spectrum (\cref{regular_finite_T_oscillation} (d)). Thus, at any temperature, the oscillation amplitude still increases at higher magnetic field in spite of decreasing bulk SC gap. 

\paragraph{$(\Delta)$} keeping the strong spin-orbit coupling $\alpha=1.2$meV, we change the superconducting order parameter exhibiting arbitrarily exponential decay $\Delta=\Delta_o e^{-V_z/5}$meV, where $\Delta_o=0.9$meV. (There is no particular significance to this particular form for the bulk gap except that it captures its decay with increasing field in a quantitative manner.) Similarly, as shown in \cref{regular_finite_T_oscillation} (g) the oscillation amplitude still increases as the magnetic field increases, although the last peak (near the bulk gap) becomes smaller. Although real $\Delta$ might exhibit a different decay behavior as the magnetic field grows, the qualitative trend of increasing amplitude of the OCPS with increasing magnetic field should remain unchanged.

\paragraph{Direction of the magnetic field} To keep the bulk gap open after the TQPT, the direction of the magnetic field has to be perpendicular to the direction ($\hat{y}$) of the spin-orbit coupling. On the contrary, when the magnetic field is parallel to the the direction of the spin orbit coupling, the bulk nanowire is gapless after the TQPT point. We define an angle $\theta$ between the direction of the magnetic field in the $yz$-plane and the $z$ direction, which is perpendicular to the spin-orbital direction. That is, $V_z=V\cos \theta $ and  $V_y= V\sin \theta$. 

\begin{figure*}[htp!]
\begin{center}
\includegraphics[clip,width=2\columnwidth]{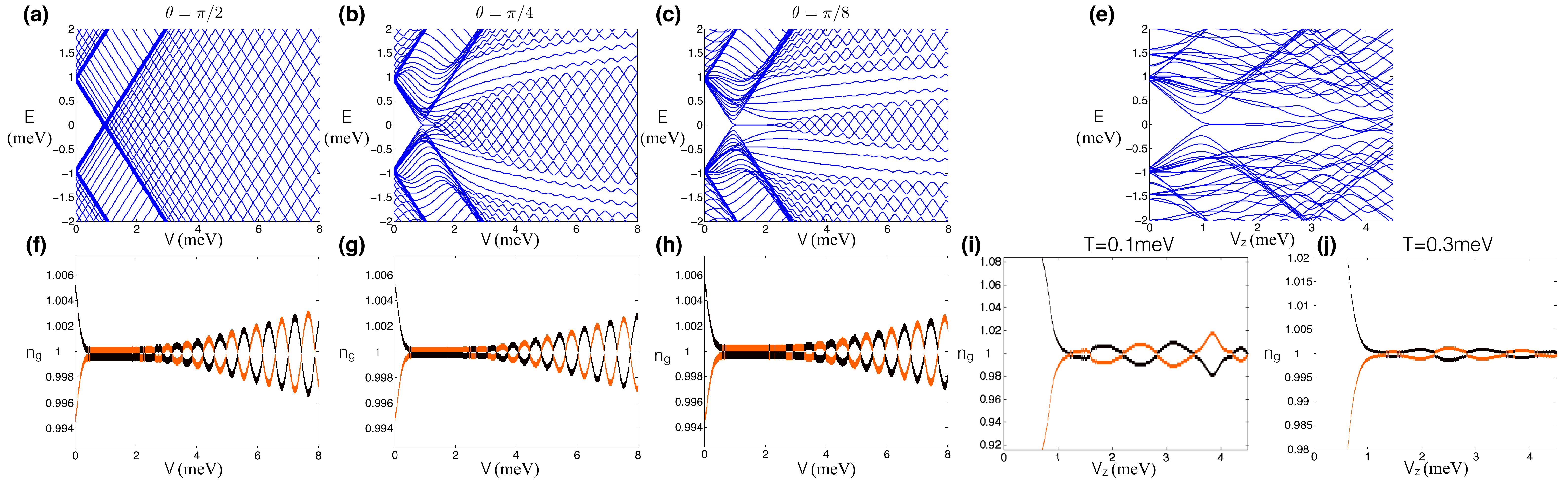}
\end{center}
  \caption{The energy spectra and the even $S_e$(black) and odd $S_o$(orange) conductance peak spacings for different directions of the magnetic field or in the presence of the bulk gap closing subbands. (a-c) the bulk gap region becomes smaller as the direction of the magnetic field moves toward to the spin-orbital direction. (e) we introduce the subbands, by including the additional Hamiltonian (\ref{nanowire Hamiltonian}) with $\mu=5$meV, has gap closing after the TQPT. (f-h) at $T=0.5$meV the peak spacings, which are independent of the magnetic field direction, have larger oscillation in the higher field. (i) at the low temperature ($T=0.1$meV), the peak spacings is roughly identical to the energy levels close to zero energy. (j) at high temperature ($T=0.3$meV), the last few oscillation packets become smaller as the magnetic field increases.} \label{B_angle}
\end{figure*}

For $\theta=0$, the direction of the spin-orbit coupling is perpendicular to the magnetic field, and the bulk spectrum (\cref{spectrum_L80key}) is gapped hosting stable MBSs. Now we slightly tilt the direction of the magnetic field. For $\theta=\pi/8$, after the transition point, the bulk gap is open until $V_z\sim 3$meV as shown in \cref{B_angle} (c). The MBSs hybridize and are no longer zero energy modes in the high magnetic field region. When the angle $\theta$ increases, the gap region becomes smaller. At $\theta=\pi/2$, the nanowire is completely gapless after the transition point as shown in \cref{B_angle} (a).  

Now we can consider the peak spacings at higher temperature ($T=0.5$meV) in this tilted magnetic field situation.
As shown in \cref{B_angle} (f-h), the peak spacings exhibit very similar oscillations as the $\theta=0$ case. It is quite surprising that even without the MBSs, the peak spacing can manifest oscillations, \cref{B_angle} (f), which are almost identical to the nanowire possessing MBSs. Therefore, even if this oscillation of the peak spacings is observed in experiment, it is difficult to definitely conclude the existence of MBSs.

\paragraph{$subbands$} To have the bulk subband gap closing as the magnetic field increases, we consider two copies of the nanowire Hamiltonians (\ref{nanowire Hamiltonian}). The chemical potential $\mu$ in one of the Hamiltonians is adjusted to $5$meV so that the bulk gap starts to decrease in the zero field and close near $V_z=5$meV~\cite{parameter_1DMajorana}. This model is obviously not able to describe the spectrum beyond $V_z=5$meV. In reality, after the subband gap closes, the density of states should be large at zero energy. Hence, the OCPS is suppressed since $S_{\rm{e}}=S_{\rm{o}}=1$ in normal quantum dots. 

To see a larger OCPS, the length of the nanowire is reduced to $L=40$ lattice sites from $L=80$. With the remaining parameters unchanged, the energy spectrum is shown in \cref{B_angle} (e). While the oscillation of the Majorana hybridization energy becomes larger, the bulk gap of the subbands becomes smaller. Consider the peak spacings at two different temperatures $T=0.1,\ 0.3$meV. At the lower temperature ($T=0.1$meV), the spacings are almost identical to the two energy levels close to the zero energy (i.e.~the OCPS is a map of the energy spectrum). Although the last oscillation is smaller, the oscillation magnitude is similar to the case without the subbands. At the higher temperature ($T=0.3$meV), the subbands close to zero energy squeeze the OCPSs and the oscillation amplitude decreases, but the conductance is no longer a map of the energy spectrum. \Cref{B_angle} (j) shows that for $V_z=3$meV, where the subbands come down at energy level $0.3$meV, the oscillation amplitude starts to become smaller as the magnetic field increases. Although this scenario seems to be partially consistent with the experimental observation, the experimental oscillation amplitude \emph{never} increases with magnetic field, which disagrees with \cref{B_angle}(j). In \cref{B_angle}(c) the second and third oscillations are still larger than the previous ones; thus, these 1D models cannot completely explain the oscillation observed in the experiment~\cite{Albrecht:2016aa} in spite of our incorporating many mechanisms affecting the SC gap. 
We need something more (i.e.~an additional mechanism beyond just gap collapse, finite temperature, and multi-subband occupancy) to make theory and experiment consistent.  We include Andreev bound state contribution to the conductance in the next subsection as this new mechanism.




\subsection{Contribution of end Andreev bound states}

\begin{figure}[t!]
\begin{center}
\includegraphics[clip,width=0.98\columnwidth]{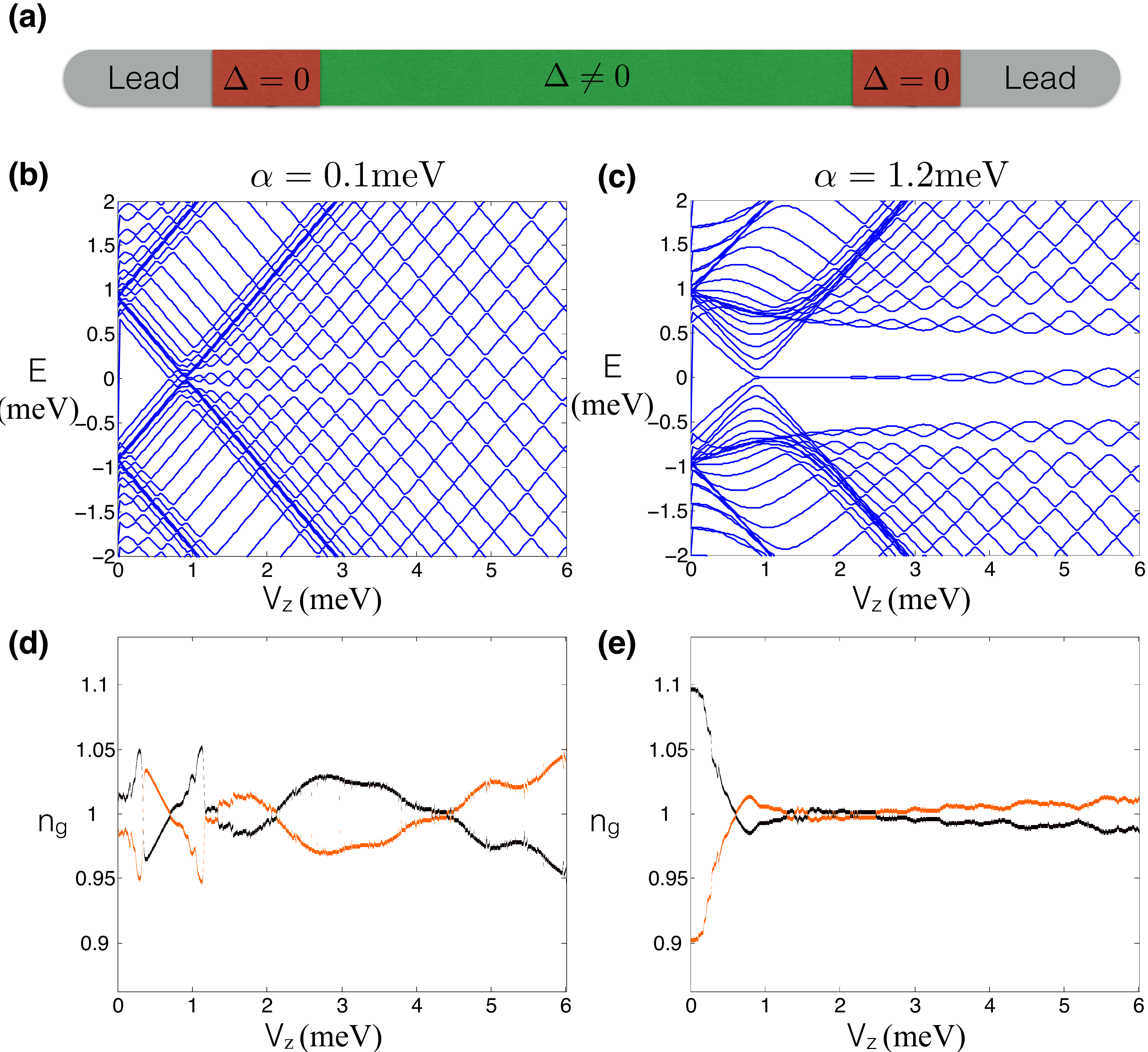}
\end{center}
  \caption{The energy spectra and the even $S_e$(black) and odd $S_o$(orange) conductance peak spacings for the nanowire having the superconducting order parameter vanishes near the ends of the nanowire as schematically illustrated in (a).  The red regions in panel (a) defining the quantum dots (and leading to Andreev bound states) are 3 and 4 sites respectively on the right and the left whereas the green region defining the wire is much longer. (b,c) the energy spectra describe the $L=60$ nanowire with different spin-orbital couplings $\alpha=0.1,\ 1.2$meV respectively. The energy levels of the Andreev bound states are closest to zero energy before the TQPT. 
(d) at $T=0.5$meV the presence of the Andreev bound states leads to the random oscillation. (e) at $T=0.5$meV the Majoranas stabilize the oscillation but the oscillation becomes larger in higher magnetic field.  } \label{Andreev_states}
\end{figure}

\begin{figure}[t!]
\begin{center}
\includegraphics[clip,width=0.98\columnwidth]{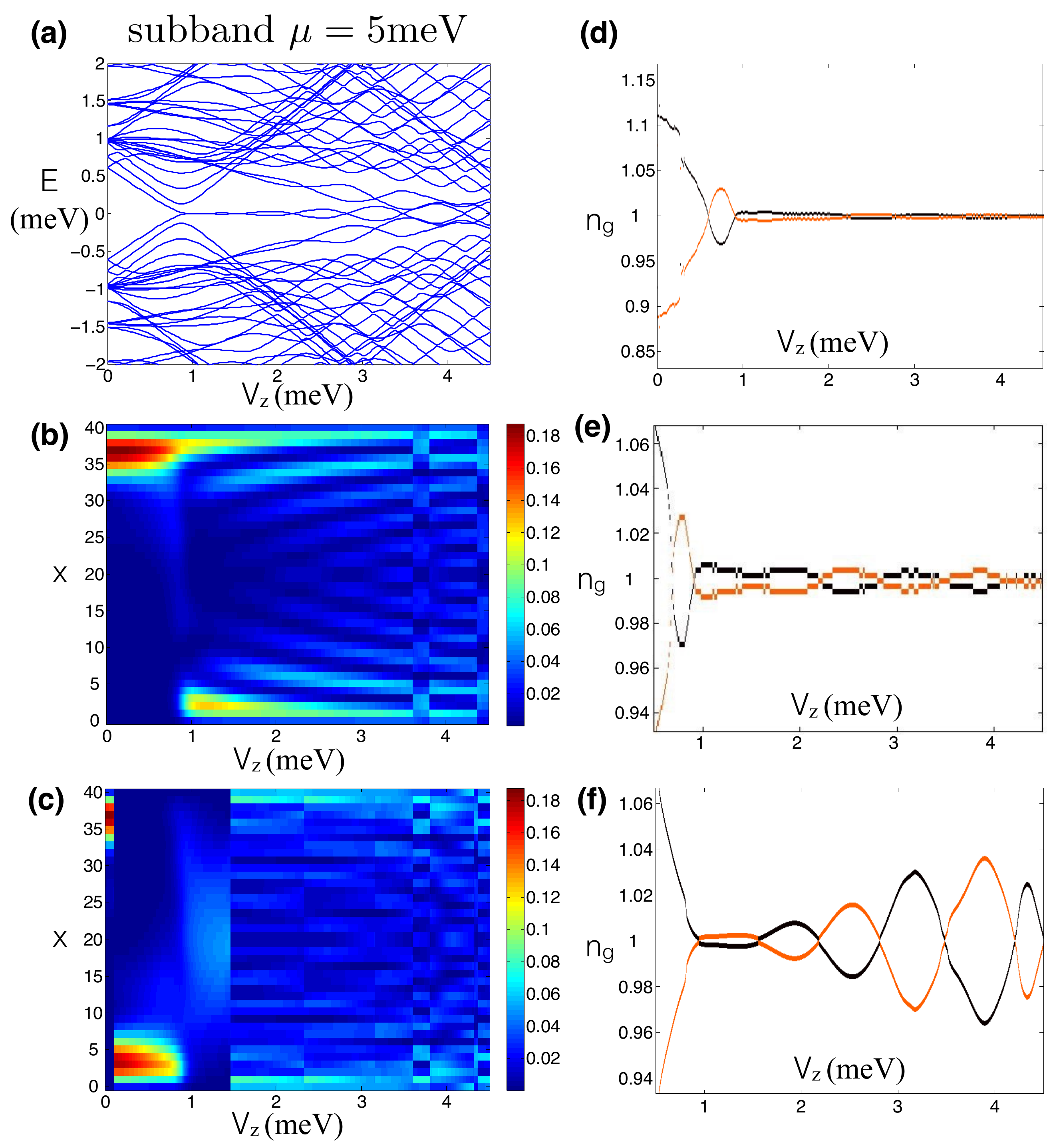}
\end{center}
  \caption{ The nanowire illustrated in \cref{Andreev_states} (a) with $L=40$ and $\alpha=1.2$meV hosts Andreev bound states in the trivial region and MBSs in the topological region and includes the additional Hamiltonian (\ref{nanowire Hamiltonian}) with $\mu=5$meV.  (a) shows the energy spectrum of the nanowire. (b,c) show the density of the first and second lowest energy states ($\sum_{\alpha} (|u_{1,x\alpha}|^2+|v_{1,x\alpha}|^2)$ and $\sum_{\alpha} (|u_{2,x\alpha}|^2+|v_{2,x\alpha}|^2)$ respectively. The Andreev bound states are separately localized on the two ends of the nanowire before the TQPT point ($V_z=0.92$meV). The first lowest energy state is localized near $x=L$ ($L=40$); (c) the second lowest energy state is localized near $x=1$ and its discontinuity of the wavefunction distribution near $V_z\sim 0.1$meV stems from the second and third level crossing. The reason that the energy of the Andeev bound state near $x=1$ is higher than the one near $x=L$ is that we have chose $\Delta=0$ region near $x=1$ is smaller than near $x=L$.
  (d,e) show the conductance peak spacings at $T=0.3, 0.2$meV respectively. The subbands play a final role to suppress the oscillation in the high magnetic field. Now the oscillation amplitude is monotonically decreasing as the function of the magnetic field. 
  (f) shows that at the low temperature ($T=0.02$meV) the oscillation is almost identical to the energy level closest to zero energy, although near the TQPT the presence of the Andreev bound states slightly affects the oscillation. 
    }\label{app_fig}
\end{figure}

The results obtained in the previous section for the ideal topological Coulomb blockaded nanowire with MBSs at the end of the nanowire do not appear
to produce
 conductance peak spacings in agreement with the experimental observation~\cite{Albrecht:2016aa}. 
Therefore, motivated by the experimental geometry and the conductance results, we consider a case where there are Andreev states 
at the wire end. Such Andreev bound states are generated by ensuring that the superconductivity vanishes near the ends of the nanowire as shown in \cref{Andreev_states} (a). This is a reasonable (perhaps even necessary) consideration since in the experiment~\cite{Albrecht:2016aa} the ends of the nanowire, which do not touch the parent superconductor, do not possess proximity-induced superconductivity in all likelihood (see \cref{Andreev_states} (a) for the experimental schematic). That is, between the leads and the superconducting nanowire there are small normal metallic regions, which could induce Andreev bound states in the system. There could also be unintentional (and therefore unknown) effective quantum dots inside the nanowire leading to Andreev bound in the system. 
Low energy Andreev bound states are difficult to distinguish from a split pair of MBSs, which have essentially the same energy spectrum -- in fact, one central concern in the Majorana nanowire field is how to distinguish the effects of MBS from those of regular non-topological subgap low-energy Andreev bound states. 
However, weakly split MBSs are, in principle, distinguishable from Andreev bound states via the localization properties of their wavefunctions. Suppose a 
low energy Andreev bound state is represented by a creation operator $a_{\epsilon_1}^\dagger$. Such an Andreev state can be thought of as a pair of weakly coupled MBSs if there exists a phase $\theta$ such that $(e^{i \theta}a_{\epsilon_1}+e^{-i\theta}a_{\epsilon_1}^\dagger)/\sqrt{2}$ is localized at one end and $(e^{i \theta}a_{\epsilon_1}-e^{-i\theta}a_{\epsilon_1}^\dagger)/\sqrt{2}$ is localized at the other end. In \cref{app_fig} (b), after the TQPT ($V_z \sim 0.92$meV), the lowest energy state satisfies this Majorana criterion. On the other hand, before the transition point  the low-energy states, which fail this MBS criterion, are the Andreev bound states in \cref{app_fig} (b,c). Hence, in this scenario the Andreev bound states and MBSs do not coexist at the same magnetic field since MBSs appear after the bulk gap closes. 
Thus, we are considering Andreev states and Majorana states on an equal footing because the experiment is carried out by sweeping the magnetic field, and since experimentally the topological transition point (i.e.~the critical field at TQPT) is not known, the possibility that  the low-field (high-field) behavior in the peak spacings arises from Andreev (Majorana) bound states cannot be ruled out.  This is the possibility we are investigating in this subsection. Of course, the criterion of localization is one of 
degree and therefore this does not precisely define a transition unless the system size goes to infinity where the MBSs can be infinitely far apart.
However, for the clean system we consider the critical value of the Zeeman field at the TQPT in agreement with the analytical result $V_z\sim\sqrt{\Delta^2+\mu^2}$~\cite{Sau_semiconductor_heterostructures}. 
Andreev (Majorana) states exist for magnetic field below (above) this critical field. One should think of the transition as a crossover in a finite system, and we need to take into account the possibility that the experimentally observed oscillations arise from a combination of Andreev and Majorana physics as the magnetic field is increased through the critical field near the first peak oscillation.


The localization of the Andreev bound states leads to different tunneling ratios ($\Gamma^l_p/\Gamma^r_p,\ \Lambda^l_p/\Lambda^r_p$) for different energy levels. This means that we must do considerable additional work in order to include the Andreev states in our theory at low magnetic field values below the TQPT. The conductance cannot be simply calculated using \cref{evenc,oddc}; we have to directly solve the master equations \cref{vPN,vPN1} for the numerical values of $\Phi_N$'s and $\Phi_{N+1}$'s and use the conductance equation (\ref{right current}) to obtain the conductance. 
Since this new calculation including Andreev states is much more involved and is computationally much more demanding than the MBS calculations just using \cref{evenc,oddc}, we consider only nine energy levels, instead of ten. 
To set up the nanowire hosting Andreev bound states for the simulation, we let $\Delta=0$ on the first four left and first three right lattice sites. The reason to choose the different numbers of the left and right lattice sites is to avoid any accidental energy level degeneracy, which leads to the incorrect and nongeneric conductance results from master equation numerical code.


	First, we assume that only Andreev bound states are present in the nanowire and MBSs are absent by adjusting the spin-orbit coupling to a low value $\alpha=0.1$meV. The spectrum (\cref{Andreev_states} (b)) shows that it is hard to distinguish the MBSs and the bulk states. By comparing with \cref{spectrum_L80key} without Andreev bound states and examining the wavefunctions of the low energy states, the energy levels of the Andreev bound states are the states closest to zero energy in \cref{Andreev_states} (b,c) before the bulk gap closing and these localized states become extended after the bulk gap closing point. Unfortunately, the random oscillation of the peak spacings in \cref{Andreev_states} (d) appears different from results reported in the experiment. Therefore, only Andreev bound states without any MBS do not appear to explain the experimental results in Ref.~\cite{Albrecht:2016aa}, although it gives a clue that the presence of Andreev bound states may indeed lead to a situation where peak spacings could sometime reflect a decreasing oscillatory amplitude with increasing magnetic field as observed experimentally. A word of caution is, however, in order here.  If chosen appropriately over different samples, it is certainly possible that some of these random Andreev peak spacings could lead to the observed experimental behavior except that other samples would manifest a different behavior. We cannot therefore decisively rule out the possibility that the physics described in Ref.~\cite{Albrecht:2016aa} arises from Andreev bound states. We believe that a recent paper~\cite{2017arXiv170502035L} gives a clue to why the Andreev bound states appear to produce random oscillatory behavior in our calculated OCPS results. It is shown~\cite{2017arXiv170502035L} that the Andreev bound states could actually come very close to zero energy accidentally and stay there for finite regimes of magnetic field (``zero-sticking property") because of the presence of SO coupling and Zeeman splitting.  Such accidental ``zero-sticking" of Andreev bound states is nonuniversal as a function of the magnetic field and depends on all the parameters of the system.  The corresponding OCPS arising from such zero-sticking of Andreev bound states would not manifest any systematic magnetic field dependence and would appear random in a sample to sample measurement.  We believe that this zero-sticking property of Andreev bound states is responsible for the OCPS behavior in \cref{Andreev_states}(d), which may very well be what is being observed in Ref.~\cite{Albrecht:2016aa}.

	Now we introduce MBSs in the nanowire by tuning the spin-orbit coupling back to $\alpha=1.2$meV. \cref{Andreev_states} (b) shows the energy of the MBSs to be close to zero and the Majoranas are protected by the large bulk gap. The Andreev bound states are the lowest energy states before the TQPT point. For $V_z$ less than $3$meV, as shown in \cref{Andreev_states} (e), the peak spacings are similar to the experimental observation since the oscillation becomes smaller as the magnetic field increases. However, in the high magnetic field region the peak spacings are affected by the Majorana hybridization, and the oscillation becomes larger. Of course, the magnetic field at which this change occurs is non-universal, and it is possible that Ref.~\cite{Albrecht:2016aa} probes only the low field regime where the Andreev states dominate leading to suppressed oscillations, but then at much higher field eventually oscillations with increasing amplitude as appropriate for MBSs should return. Furthermore, comparing \cref{Andreev_states} (c,e), we find the crossing of the conductance peak spacings ($V_z \sim 0.6$meV) to occur before the bulk closing point ($V_z \sim 0.92$meV). This example shows that the conductance peak spacings cannot be simply described by the lowest energy level (\cref{SO,SE}). The localized states play a subtle role changing the conductance and the possibility that the Andreev bound states is playing a role cannot be ruled out.
	
	Finally, we add the following ingredient to the physics of the nanowire keeping the Andreev states in the analysis:~the bulk gap closing subbands. Since at higher temperature the subbands lead to smaller oscillations at higher magnetic field as shown in \cref{B_angle} (j), inclusion of all the ingredients (the higher temperature, the Majoranas in high magnetic field, the Andreev bound states in low magnetic field, and the gap closing subbands) leads to a decreasing oscillation amplitude with increasing magnetic field except perhaps at very high magnetic field (which may be outside the experimental regime) as shown in \cref{app_fig} (d,e). These peak spacings are qualitatively similar to the experimental finding in Ref.~\cite{Albrecht:2016aa}. Thus, higher temperature and the presence of the closing subband gap as well as the MBSs in the topological region and the Andreev bound states in the trivial region (all of these mechanisms taken together) are currently our best explanation for the experimental observations. 
	Although we do not find a situation where MBSs without Andreev bound states produce results in agreement with experiment, the reverse is not, strictly speaking, true -- just Andreev bound states without MBSs give random peak spacings, which, in some situations, may mimic decreasing oscillations with increasing field in a narrow field region.
	We note that \cref{Andreev_states,app_fig} show that the same physics applies for long (L=60; \cref{Andreev_states}) and short (L=40; \cref{app_fig}) wires with the inclusion of Andreev bound states in the calculation always producing decreasing oscillations with increasing magnetic field as observed experimentally. Our findings are therefore generic.
	
\begin{figure}[t!]
\begin{center}
\includegraphics[clip,width=0.98\columnwidth]{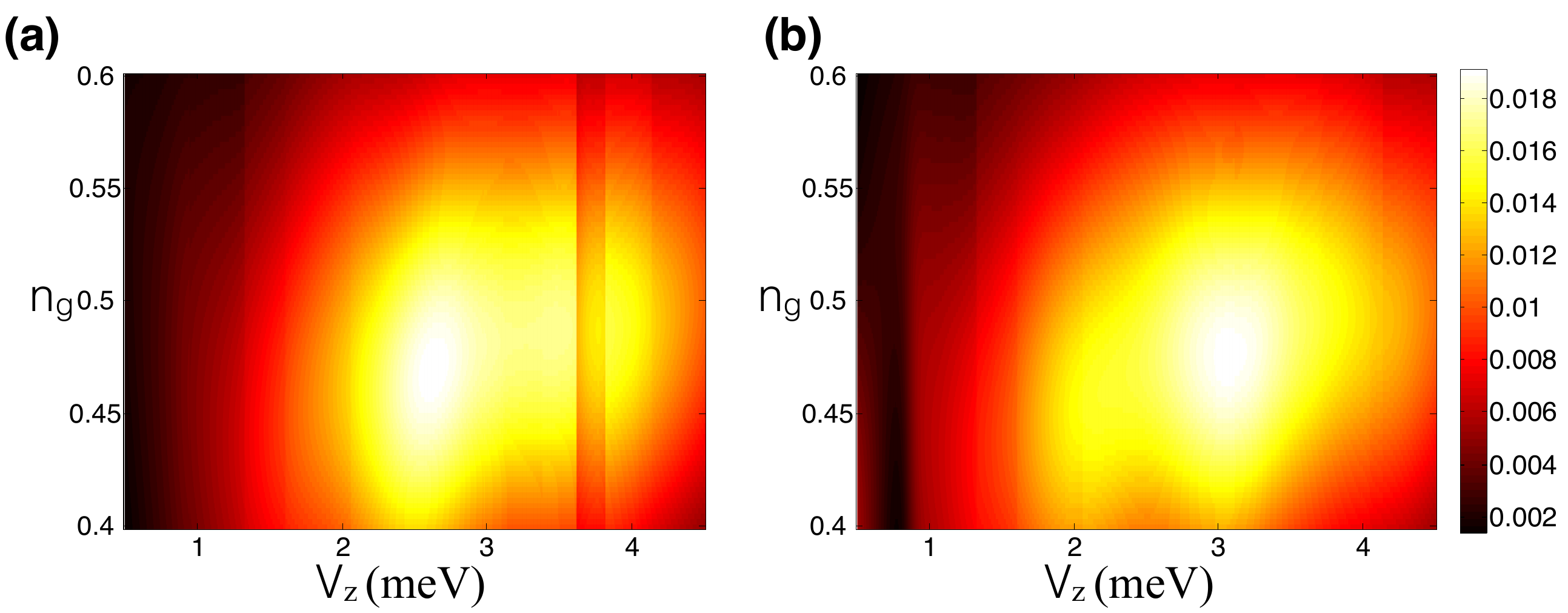}
\end{center}
  \caption{ 
The conductance (arbitrary unit)  of the nanowire with the parameters identical to \cref{app_fig} at $T=0.2$meV for odd $N$ and even $N$ respectively. After the first bulk gap closing, the conductance is greater than before the closing.  }  \label{conductance_andreev} 
\end{figure}

	The conductance including all effects at $T=0.2$meV is shown in \cref{conductance_andreev}. The conductance before the TQPT point is much smaller than the one after the TQPT. This result is in agreement with the experiment \cite{Albrecht:2016aa} and has been explained in Ref.~\cite{conductance_coulomb_blockade_roman}. 
	
	The discontinuity appears in the conductance plot in \cref{conductance_andreev} (a) near $V_z=3.6$meV. The conductance discontinuity occurs when the two lowest energy levels are degenerate $E_1=E_2$ as shown in the spectrum plot \cref{app_fig} (a). That is, the qausiparticles $a_{\epsilon_1}^\dagger$ and $a_{\epsilon_2}^\dagger$ exchange so that the BCS ground state in the odd parity $|\rm{BCS_o}\rangle=$$a_{\epsilon_1}^\dagger|\rm{BCS_e}\rangle$ changes sharply as $V_z$ passes through the degenerate points; this sharp change leads to the conductance discontinuity. In the reality, since $a_{\epsilon_1}^\dagger$ and $a_{\epsilon_2}^\dagger$ usually weakly couple, $E_1$ and $E_2$ are close but are never identical. The change of the $|\rm{BCS_o}\rangle$ is smoothed out and then the conductance discontinuity should not be expected. 

	To suppress the oscillation as the magnetic field increases (purely in the high-field topological regime where Andreev states cannot play any role) we estimate the lower bound of the temperature to be around $0.2$meV, which corresponds roughly to one fifth of the SC gap, (for the specific parameters used in our study) in \cref{app_fig} (e) since in high magnetic field the oscillation grows at $T=0.1$meV as shown in \cref{B_angle} (i). We emphasize that the Andreev bound states can only explain the decrease in the oscillation amplitude in the low-field region (near the first oscillation), and once the MBS oscillations set in, the theory predicts unequivocally that the oscillation amplitude must increase with increasing magnetic field, which is not seen experimentally.  We therefore need a mechanism to suppress the oscillations at higher field, and we use temperature as this damping mechanism. 
	
One may wonder what happens at much lower temperature in the presence of both Andreev and Majorana bound states as the magnetic field sweeps through the TQPT.  We show our calculated low-temperature results for this general situation in \cref{app_fig} (f) to be compared with the high-temperature results shown in  \cref{app_fig} (e).
	As shown in \cref{app_fig} (f), although near the TQPT the presence of the Andreev bound states does slightly change the oscillation at the low temperature, the oscillation amplitude always grows with increasing magnetic field in contrast to the experimental results of Ref.~\cite{Albrecht:2016aa}.  This is, in fact, expected since at low enough temperatures, the conductance should be a direct map of the low energy spectrum, and the MBS splitting always increases with increasing magnetic field.  Thus, even in the presence of Andreev (and Majorana) bound states, the OCPS amplitude increases with increasing magnetic field at low temperatures.
	 The damped oscillation should not be apparent in the experiment unless the electron temperature is pretty high. 
	 Of course, as mentioned earlier, what this ``pretty high'' temperature scale is in absolute units must depend crucially on the experimental parameters which are unknown. All we can claim is that experiments should see distinct behaviors at low and high temperatures with the high-temperature result manifesting decreasing oscillation amplitude with increasing magnetic field.

	 A serious note of caution is in order about the absolute temperatures used in our simulations, which are much higher than the quoted temperature ($T=50\sim 100 mK$) in Ref.~\cite{Albrecht:2016aa}.  We emphasize that our temperature scale is determined entirely by the parameters used in our model.  For example, if we use an increased effective mass (i.e. a lower nanowire hopping matrix element), the energy scale goes down, and such an adjustment can induce arbitrary lowering of the temperature in our model.  Another (perhaps even more important) point is that the precise chemical potential is not known in the experiment.  If the Fermi level is somehow near the bottom of the 1D subbands in the nanowire (not an unlikely scenario given the large density of states near the 1D subband bottom), again the effective energy scale is suppressed lowering the temperatures used in our simulations.  Similarly, if the typical subband energy spacing in the nanowire is small (our confinement model for the nanowire corresponds to a hard-wall infinite square well-confinement), that will again make our effective temperatures much smaller.  Since these energy scales in the experimental nanowires are unknown, not much significance should be attached to our absolute energy scales.
	 In addition, we assume that the nanowire possesses a hard gap~\cite{ChangW.:2015aa,PhysRevB.92.174511} after the TQPT point. In reality, the experimental gap is very soft after the TQPT where a zero bias peak emerges, and therefore some (unknown) low energy states are present in the gap. (These low-energy subgap states producing the soft gap most probably arise from disorder in the underlying parent superconductor due to the strong superconductor-semiconductor coupling as discussed in~\cite{PhysRevB.94.140505,PhysRevLett.110.186803,PhysRevB.92.174512}, an effect beyond the scope of the current work.) Therefore, even if the temperature is low, these low energy states in the soft gap might be able to suppress the oscillation spacings by acting like effective Andreev states as considered in our theory. 
	 Since the nature of these low energy states is unknown, they cannot be included in the theory, but it is quite possible that these low energy fermionic subgap states at high magnetic field act similar to the low magnetic field Andreev subgap states included in our theory, leading to suppressed oscillations with increasing magnetic field even at low temperatures.
	Since our current conductance numerical program can compute at most 10 energy levels, the numerical result in the manuscript is only for the hard gap. We leave the soft gap simulation, which must include many more energy levels as well as a physical model for the subgap states causing the soft gap, for future work. But, other than suppressing the energy scale at which oscillation amplitude dampens thermally (i.e. by providing many available low energy states already at low temperatures), such low energy states should not change any of our results qualitatively.  
More experimental and theoretical work will be necessary to decisively settle the question of which subgap states may actually be contributing to the finite temperature conductance causing the increasing oscillation with increasing field.  What we have accomplished is to show that a mechanism combining finite temperature and additional subgap states provides a possible explanation for the observed oscillation amplitude decreasing with increasing magnetic field.

\begin{figure}[t!]
\begin{center}
\includegraphics[clip,width=0.99\columnwidth]{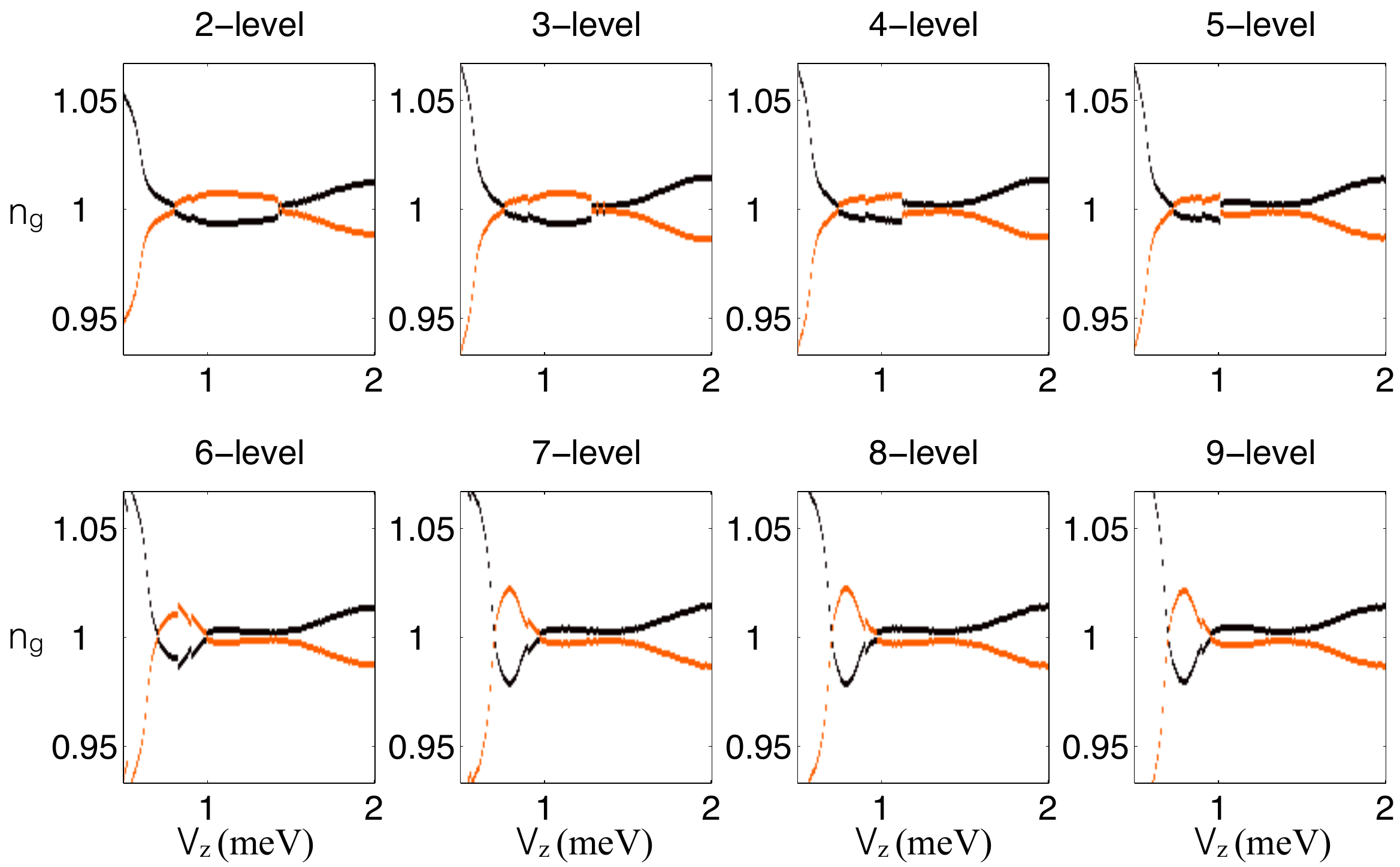}
\end{center}
  \caption{ The OCPS for the conductance master equations including different numbers of the energy levels for $L=30$ at $T=0.2$meV. The remaining parameters are identical to \cref{app_fig}. Since the OCPSs including few different energy levels are completely distinct, the multi-band physics does affect the OCPS dramatically. The mechanism of the Coulomb blockaded nanowire therefore becomes complicated at \emph{higher} temperatures where multi-band effects are important. The OCPSs including 7, 8, 9 energy levels, which are identical, lead to the correct estimation of the conductance for the input parameters used in our theory.
 }  \label{Energy level} 
\end{figure}

Finally, we show our calculated numerical results in \cref{Energy level} as a function of different number of energy levels included in the theory in order to establish that the 9-level calculation is adequate to obtain convergent results for the parameters used in our theory in the high-temperature regime.  We emphasize that the necessary number of levels is obviously a non-universal quantity depending crucially on the input parameters (and temperature) used in our simulation, but the point we make is that multilevel effects must play an important role in understanding the observations in Ref.~\cite{Albrecht:2016aa}.  It is possible (actually likely) that the renormalization by the superconductor makes the higher nanowire energy levels almost degenerate~\cite{Tudor_private}, leading to a participation by several energy levels even at much lower temperatures.  As discussed already, `low' and `high' temperatures in our calculation are nonuniversal quantities (in absolute terms) as they depend crucially on the unknown microscopic details of the hybrid semiconductor-superconductor structures.  The numbers used in our theory are for demonstrative purposes only.

\section{Length dependence}

So far all our results and discussions focused entirely on the magnetic field dependence of $1e$-tunneling OCPSs in the Coulomb blockaded nanowire experiment of Ref.~\cite{Albrecht:2016aa}, leaving out all considerations of the length dependence.  This is appropriate since the measurements are always carried out on a wire of fixed length ($L$) as a function of an applied magnetic field.  The experiment is NOT done as a function of length keeping all other system parameters (e.g. magnetic field, chemical potential, superconducting gap) fixed.  Thus, any conclusion about an estimated length dependence of splitting is subject to criticism since the other relevant system parameters certainly vary along with the wire length.  In fact, in Ref.~\cite{Albrecht:2016aa}, the length dependence is extracted from the measurement of peak splitting in five different wires with each splitting in each wire measured at different magnetic fields.  In fact, not only the magnitude, but even the direction of the magnetic field, is different in some of the wires used to obtain the length dependence.  It is expected that these wires of different lengths have different chemical potentials, confinement potentials, superconducting gaps, Coulomb energies,  and disorder as well since these are not controllable parameters in the experiment.  What was kept common is that the splitting measurement was always done at the first oscillation maxima of the peak spacing, which was often the only oscillation presenting a measurable amplitude since the oscillation amplitude typically decayed rapidly with increasing magnetic field (which is the main point of our theoretical analyses).  Some additional limitations of the experiment are that the measured splitting is around $1K$, which is much larger than even the parent SC gap in Al at that magnetic field, for two of the shorter wires ($< 0.5\mu m$), and the measured splitting is around $\sim 10 mK$ for the longest wire ($L\sim 1.6 \mu m$), which is almost an order of magnitude lower than the experimental temperature, leaving only two (or even three) data points in the experiment where the measured spacings can be reasonably construed to be meaningful for extracting an exponential dependence.  These two splittings are both $\sim 100 mK$ for two wires of $L \sim 1\mu m$. (Here by assuming the low temperature limit, the maximum oscillatory amplitude in the unit of eV is identical to the energy splitting. We refer the readers to Fig.~2 in Ref.~\cite{Albrecht:2016aa} for the details.)  Whether a decisive $\exp (-L)$ behavior can be meaningfully extracted on the basis of two data points of nearby $L$ values remains unclear. In addition, the fact that the experimental L-dependence in Ref.~\cite{Albrecht:2016aa} is extracted by using different samples of different lengths at different magnetic fields (which are sometimes oriented in different directions) casts some doubt in the accuracy of the exponential dependence conclusion reached in Ref.~\cite{Albrecht:2016aa}. Obviously, many more results are necessary for compellingly establishing an exponential dependence of the splitting in wire length.  Nevertheless, the question arises about our theoretical $L$-dependence, specifically for those results in our simulations (including both Andreev and Majorana bound states) which approximately mimic the experimental magnetic field dependence in the OCPSs (i.e.~oscillations decreasing with increasing magnetic field).  

\begin{figure}[t!]
\begin{center}
\includegraphics[clip,width=0.75\columnwidth]{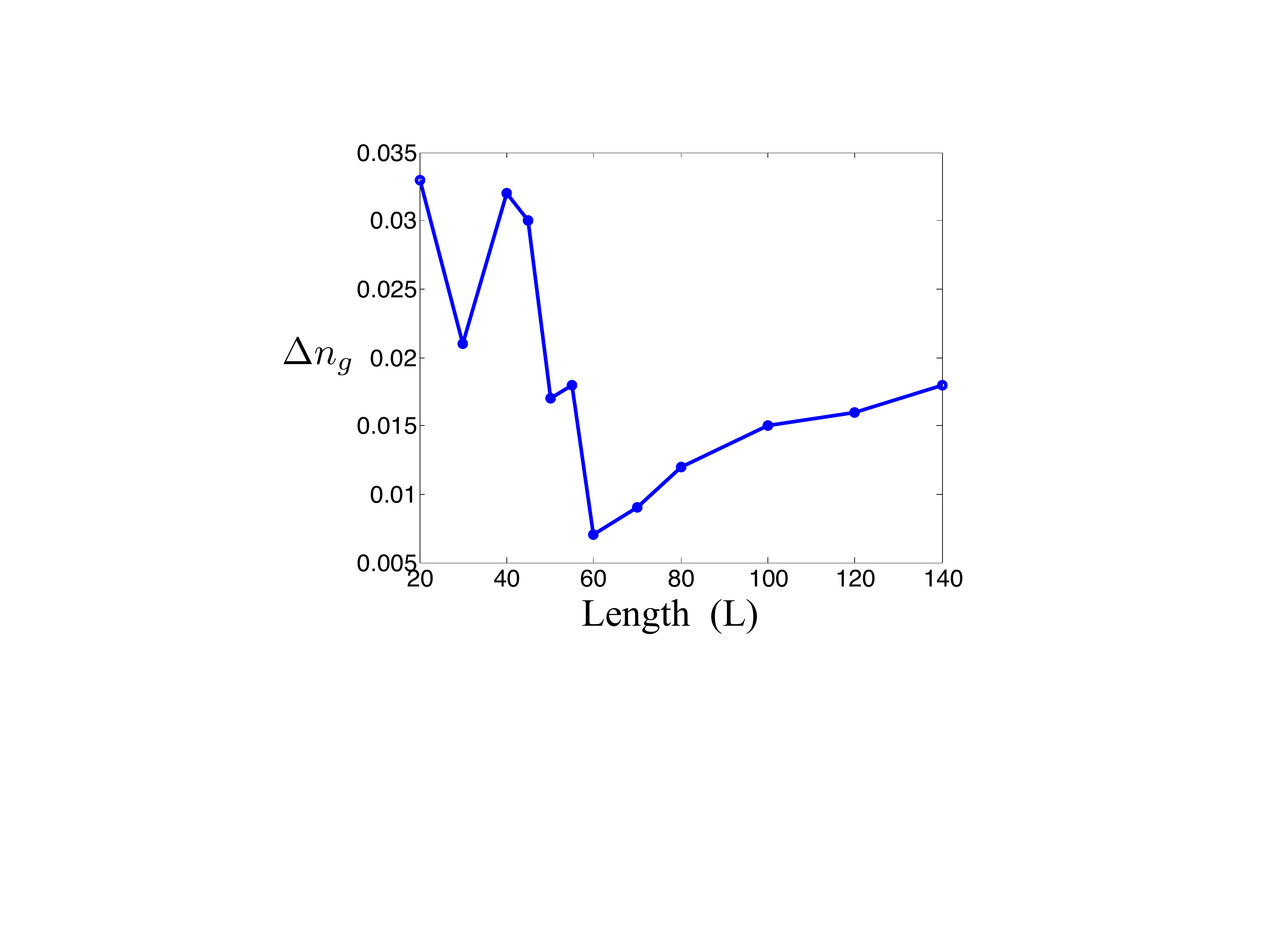}
\end{center}
  \caption{ The first oscillation maxima of the conductance peak spacing $(|S_o- S_e|/2)$ for different lengths of the nanowires for our high-temperature case at $T=0.2$meV. The remaining parameters are identical to \cref{app_fig}. The maximum oscillatory amplitude has a rapid oscillatory decrease with increasing length for shorter wires and a very slow increase with increasing length for longer wires.  This complex length dependence arises from Andreev physics in our theory. }  \label{L dependence} 
\end{figure}


\begin{figure*}[t!]
\begin{center}
\includegraphics[clip,width=1.98\columnwidth]{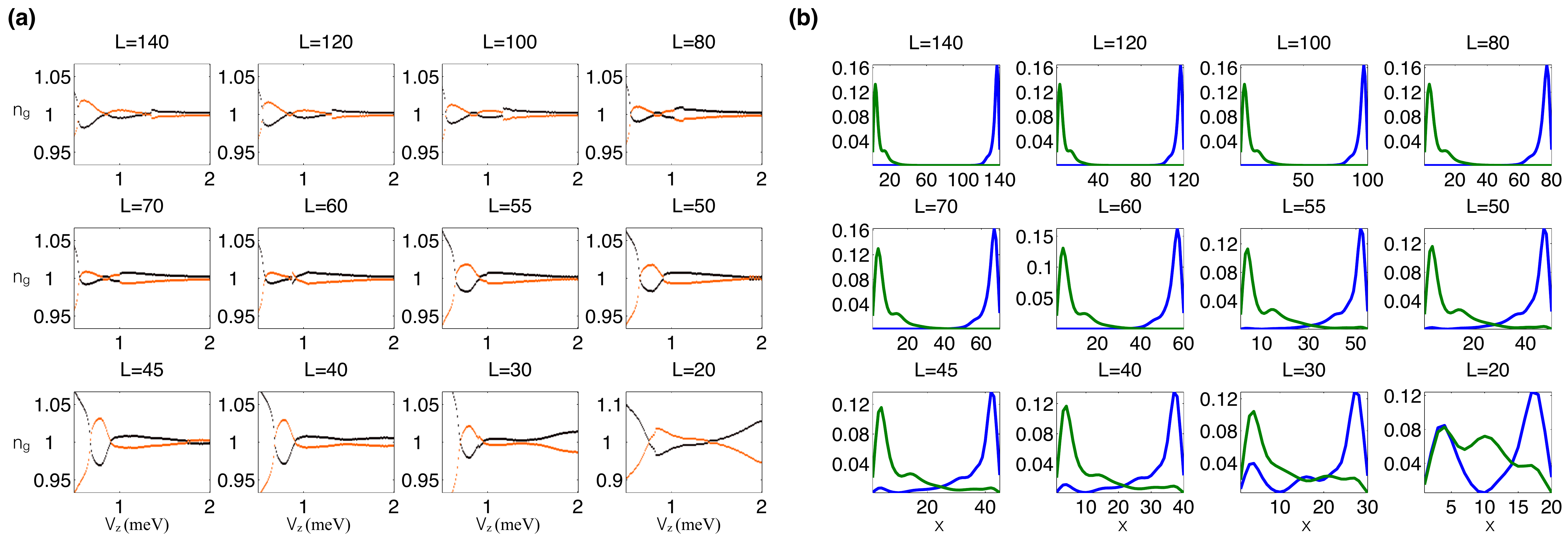}
\end{center}
  \caption{ (a) OCPSs for different lengths of the nanowire at higher temperature $T=0.2$meV. The remaining parameters are identical to \cref{app_fig}. The first oscillation maxima of the conductance peak spacing are obtained from this subfigure. For small $L$, the first oscillation peak is very close to TQPT. As $L$ increases, the peak of the first oscillation moves toward smaller magnetic field. (b) probability density spatial distribution of the lowest (blue) and second lowest (green) energy states, which are Andreev bound states, for different lengths of the wire at the magnetic field corresponding to the first oscillation peak. Since for small $L$ the corresponding magnetic field is close to TQPT, the sizes of the low energy states are almost extended in the whole wire. The overlap of the states vanishes at $L=60$, which is the characteristic length of the first splitting defining the crossover point (i.e. the minimum) in \cref{L dependence}. }  \label{different L} 
\end{figure*}

Although we do not expect any well-defined generic length dependence in our simulated peak spacings since the results include effects of both Andreev and Majorana states as well as finite temperature and multilevel contributions, we show our calculated length dependence in \cref{L dependence} for the peak spacings obtained operationally in the same way as in Ref.~\cite{Albrecht:2016aa}.  We simply plot the amplitude of the first peak spacing oscillation as a function of $L$ in \cref{L dependence} making sure that for each value of $L$, the corresponding magnetic field dependence manifests qualitatively similar behavior as observed in Ref.~\cite{Albrecht:2016aa} (i.e. amplitude deceasing with increasing field). We keep all other parameters fixed except for the spin splitting, which must change since wires of different lengths manifest the first oscillations at different magnetic field values (precisely as happens in the experiment also).  As can be seen in \cref{L dependence}, the theoretically extracted $L$-dependence is nonmonotonic -- it shows a strong decrease with increasing length for shorter wires and then a very slow increase with increasing length for longer wires.  
We have no precise explanation for the nonmonotonic length dependence in \cref{L dependence} except that this is what we get from the simulations (and the behavior is most certainly non-universal here as there is no reason to expect a universal behavior).  Obviously, the decrease in the oscillation for shorter wires, mimicking the findings of Ref.~\cite{Albrecht:2016aa}, is not a manifestation of the exponential $\exp (-L/ \xi ) $ behavior of Majorana splitting, since such a behavior should be more prominent for $L\gg\xi$, i.e.~for longer wires.  In fact, our estimated $\xi$ values for our results are around $100\ (10)$ for high magnetic field (TQPT) values, and since the first oscillation always occurs just below the TQPT in our simulations, it is unclear that the $L$-dependence can have much to do with MBS properties.
It is possible, but by no means certain, that the experimentally observed $L$-dependence in Ref.~\cite{Albrecht:2016aa} is related to the $L$-dependence in our \cref{L dependence} for smaller values of $L$, but we emphasize that this short-length behavior has nothing whatsoever to do with topological protection as should be obvious from the larger-$L$ behavior in \cref{L dependence}.
While we cannot comment on the significance of the experimentally observed length dependence in Ref.~\cite{Albrecht:2016aa}, our theoretical length dependence in \cref{L dependence} is likely to be nongeneric arising from the complex interplay between finite temperature and Andreev bound states and not much significance should be attached to our theoretical $L$-dependence in \cref{L dependence}.

To understand our calculated $L$-dependence in \cref{L dependence} better, we show in \cref{different L} two sets of numerical results (in panels (a) and (b), respectively) for the theoretical energy splittings and low-lying Andreev bound state wave functions (actually, the squared amplitude, i.e., the probability density) for different values of the wire length $L$ (keeping the quantum dot size always 3 and 4 $\ll L$).  The basic picture that emerges is that for small $L$, the first oscillation peak is always close to the TQPT point, and the Andreev bound state is very extended at this low magnetic field, covering much of the wire whereas at higher values of $L$, the first oscillation is associated with Andreev states which are mostly localized near the wire ends.  Thus, the physics of the first oscillation is different for small and large values of $L$ with the crossover occurring at some nonuniversal characteristic $L$-value ($\sim 60$ in  \cref{L dependence} for our parameters) which is determined by the system parameters including the size of the quantum dot regimes (see \cref{Andreev_states}(a)).  Typically this characteristic $L$-value defines the approximate crossover for the Andreev bound states being mostly extended over the wire (small-$L$) to being mostly localized near the wire ends (large-$L$).  Thus, the decrease in the energy splitting with increasing $L$ in our  \cref{L dependence} has nothing to do with Majorana protection. It arises from the overlap of the Andreev bound states at the two ends rather than the Majorana bound states at the two ends. At this stage, our results, therefore, disagree with an exponential protection conclusion for the nanowire OCPS.

\section{Conclusion}

Our focus in this work is on the transport properties of the superconducting Coulomb blockaded nanowire in the weak tunneling situation with small applied voltage between the two leads. The generic transport equations in the $1e$ periodicity Coulomb blockaded superconducting region are derived including Coulomb blockade and superconductivity on an equal footing. These transport master equations capture various physical effects, such as multi-band tunneling, finite temperature, tunneling from localized Andreev states, and possible gap closing. The goal is to qualitatively understand in depth the recent experimental report of topological protection in Coulomb blockaded nanowires~\cite{Albrecht:2016aa}, which is warranted since the direct interpretation of the experiment in terms of the existence of Majorana zero modes shows inconsistency with the theoretical expectation that in a given wire the Majorana oscillations must increase in amplitude with increasing magnetic field (because the SC gap decreases with increasing magnetic field leading to an increasing coherence length). 
We obtain analytical solutions to our equations to calculate the low temperature conductance arising from the resonant tunneling through MBSs in several simple situations. These low temperature Majorana conductance results cannot explain the experimental observation~\cite{Albrecht:2016aa} of decreasing oscillation amplitude with increasing magnetic field.
In the presence of Andreev bound states (at low fields) and/or at higher temperatures where many levels contribute to transport, we solve our master equations numerically, and find that the presence of both finite temperature and Andreev bound states are essential in obtaining results which are in qualitative agreement with experiment.  We also find that eventually with increasing field, the oscillation amplitude must always increase, but it is possible that the experiment would not work in this high field region because of the complete collapse of the SC gap. 
We do not find any way to explain the experimental behavior in the low-temperature regime, but we emphasize that the temperature scale in the problem is nonuniversal and is affected by numerous unknown system parameters (effective mass, hopping amplitude, SC gap, chemical potential, SO coupling, confinement, soft gap, subgap states, etc.).  We cannot rule out the possibility that Ref.~\cite{Albrecht:2016aa} indeed observes our ``high-temperature" behavior reflecting the anomalous decrease in the oscillation amplitude with increasing field because of the combined contributions from Andreev states and Majorana states.

When the temperature is less than or close to the lowest energy level of the superconducting nanowire and the next excited energy level is much higher, the OCPSs (as studied experimentally in Ref.~\cite{Albrecht:2016aa} and theoretically in Ref.~\cite{conductance_coulomb_blockade_roman}) is proportional to the lowest energy level. Therefore, for the simplest Majorana nanowire model \cite{Tudor_oscillation_damp}, the OCPS always grows as the magnetic field increases and the thermal broadening of the conductance peak is  proportional to $T$. To explain the damped oscillation (i.e.~decreasing with increasing field) observed in the puzzling experiment, we consider several physical mechanisms. Unfortunately, higher temperature, small spin-orbit coupling, and changing of the magnetic field direction always lead to growing oscillations with increasing magnetic field in conflict with the experimental report in Ref.~\cite{Albrecht:2016aa}. We further introduce the SC gap closing of the subbands after the TQPT point in order to see if gap closing could explain the anomalous behavior. At high temperature in high magnetic field (away from the TQPT point) the oscillation becomes smaller in the presence of gap closing (and in qualitative agreement with experiment); however, the oscillation still grows with increasing magnetic field \emph{right} after the TQPT point in the low field region. Finally, when we introduce Andreev bound states in the trivial region, the theory becomes much more complicated and the conductance has to be computed by numerically solving the master equations. The calculation time grows exponentially with the number of the included energy levels in the theory. Although the oscillation is suppressed near the first transition point, only the presence of the Andreev bound states by itself does not lead to the consistency of the oscillation behavior with the experiment. 
In particular, the oscillations in the trivial phase arising purely from Andreev bound states (and no Majorana states at any magnetic field) appear to have random amplitudes although some samples in some limited regimes of magnetic field (for fine-tuned gate voltage values) may very well manifest OCPSs in qualitative agreement with the results in Ref.~\cite{Albrecht:2016aa}, but this agreement is nongeneric because of the random nature of the theoretical results.
Including the high temperature, the Majorana bound states (in the topological regime), the subband gap closing, and the Andreev bound states (in the trivial regime), we find that the OCPS amplitude \emph{always} decreases as the magnetic field increases. Obviously in this scenario, which is in excellent qualitative agreement with the experimental data presented in Ref.~\cite{Albrecht:2016aa}, the first oscillations in peak spacing as a function of magnetic field are taking the system from the trivial phase (with Andreev states) to the topological phase (with Majorana states), and as such, no concept of topological protection applies to these first oscillations. Therefore, the Majorana scenario at high field values along with Andreev states at lower magnetic fields can explain the recent puzzling experimental results. We emphasize, however, that in this scenario no significance can be attached to the length dependence of the peak size of the first oscillation since the first oscillation happens below TQPT and is dominated by Andreev, and not Majorana, physics.  Although our theoretical length dependence for the peak size of this first oscillation mimics the experimental finding for shorter wire lengths, this behavior is not directly connected with the Majorana splitting. 
We in fact establish through direct numerical simulations that the decreasing splitting with increasing length (in the shorter wire regime) arises entirely from the physics of Andreev bound states which tend to be more extended through the wire for shorter wires leading to this decrease.  For longer wires, where the Andreev states from the two ends are localized and no longer overlap with each other, we find the splitting to be essentially independent of length (actually increasing slowly with increasing length).

	Our work shows that it is difficult to understand the important findings in Ref.~\cite{Albrecht:2016aa} by invoking only Majorana bound states in the nanowire.  At the minimum one also needs finite temperature and Andreev bound states playing crucial roles.  These additional mechanisms together can not only explain the puzzling observed decrease in the OCPS amplitude with increasing magnetic field, but also implies that the concept of an exponential protection in length may not apply to the situation since the system is transitioning from the trivial to the topological phase in its oscillatory regime.  The only way we see to experimentally establish the topological protection is to use wires of variable length (imposed for example by applying suitable gate voltages) and then study oscillations as a function of length at a fixed magnetic field.  These oscillations at a fixed field (in the topological regime) should eventually fall off exponentially as the length increases, signifying topological protection~\cite{PhysRevB.86.220506}.  
	Our work showing that the first oscillation may arise from the physics of Andreev bound states makes the situation very complex with the conclusion that great care is necessary to establish the purely isolated Majorana topological protection regime experimentally.
	
	If our interpretation of Ref.~\cite{Albrecht:2016aa} is qualitatively correct (i.e.~both Andreev and Majorana bound states contributing to OCPSs), then there is an immediate significant (and highly encouraging) implication.  Since the interpretation most likely (but not absolutely certainly) requires the presence of both Andreev and Majorana bound states in the system, it is clear that going to higher magnetic field should confine the physics entirely to MBSs since the Andreev bound states are operational only near the low magnetic field part involving the first oscillation.  This means that the higher magnetic field regime of these nanowires, if accessible, should enable a study of only MBS without any complications arising from Andreev bound states.  Unfortunately, this may not be possible in hybrid core-shell epitaxial structures where Al is used as the superconductor since the Al SC gap collapses precisely around where MBS physics becomes operational.  But the data of Ref.~\cite{Albrecht:2016aa} may be indicating that the true topological Majorana nanowire regime may very well be very close to the parameter values used in Ref.~\cite{Albrecht:2016aa} except that one needs to somehow suppress the soft gap behavior so that the SC gap persists to somewhat higher magnetic field. We believe that at magnetic field values higher than that used in Ref.~\cite{Albrecht:2016aa} Majorana physics will indeed manifest itself if the SC gap can be kept finite in this high-field MBS regime.
	
	Before concluding, we make two salient comments.  First, our conclusion regarding the key role that Andreev bound states may be playing in the experimental results of Ref.~\cite{Albrecht:2016aa} is completely consistent with the recent work by Liu et al.\cite{2017arXiv170502035L}, who shows that Andreev bound states have an interesting generic zero-sticking tendency in Majorana nanowires, thus necessarily conflating the physics of Majorana bound states and Andreev bound states.  More work is therefore necessary in understanding the role of Andreev bound states vis a vis Majorana properties in nanowire experiments. Second, our results, albeit including many effects, are still obtained within the minimal Majorana nanowire model, and we have no way of ruling out a more complicated model going beyond the minimal model providing a more compelling explanation for the data of Ref.~\cite{Albrecht:2016aa} although the interpretation provided in Ref.~\cite{Albrecht:2016aa} also uses a minimal model leaving out all the complications (e.g. Andreev bound states, finite temperature, several energy levels, gap collapse, etc.) considered in our work.


\section{Acknowledgements}

The authors are indebted to J. Alicea, W. S. Cole, A. P. Higginbotham, C.-X. Liu, F. Kuemmeth, R. M. Lutchyn, F. Setiawan, and T. D. Stanescu for discussions. 
This work is supported by Microsoft Q and LPS-MPO-CMTC.


%
%
%
%
%
%


\bibliographystyle{apsrev4-1}
\bibliography{TOPO}

\begin{thebibliography}{58}%
\makeatletter
\providecommand \@ifxundefined [1]{%
 \@ifx{#1\undefined}
}%
\providecommand \@ifnum [1]{%
 \ifnum #1\expandafter \@firstoftwo
 \else \expandafter \@secondoftwo
 \fi
}%
\providecommand \@ifx [1]{%
 \ifx #1\expandafter \@firstoftwo
 \else \expandafter \@secondoftwo
 \fi
}%
\providecommand \natexlab [1]{#1}%
\providecommand \enquote  [1]{``#1''}%
\providecommand \bibnamefont  [1]{#1}%
\providecommand \bibfnamefont [1]{#1}%
\providecommand \citenamefont [1]{#1}%
\providecommand \href@noop [0]{\@secondoftwo}%
\providecommand \href [0]{\begingroup \@sanitize@url \@href}%
\providecommand \@href[1]{\@@startlink{#1}\@@href}%
\providecommand \@@href[1]{\endgroup#1\@@endlink}%
\providecommand \@sanitize@url [0]{\catcode `\\12\catcode `\$12\catcode
  `\&12\catcode `\#12\catcode `\^12\catcode `\_12\catcode `\%12\relax}%
\providecommand \@@startlink[1]{}%
\providecommand \@@endlink[0]{}%
\providecommand \url  [0]{\begingroup\@sanitize@url \@url }%
\providecommand \@url [1]{\endgroup\@href {#1}{\urlprefix }}%
\providecommand \urlprefix  [0]{URL }%
\providecommand \Eprint [0]{\href }%
\providecommand \doibase [0]{http://dx.doi.org/}%
\providecommand \selectlanguage [0]{\@gobble}%
\providecommand \bibinfo  [0]{\@secondoftwo}%
\providecommand \bibfield  [0]{\@secondoftwo}%
\providecommand \translation [1]{[#1]}%
\providecommand \BibitemOpen [0]{}%
\providecommand \bibitemStop [0]{}%
\providecommand \bibitemNoStop [0]{.\EOS\space}%
\providecommand \EOS [0]{\spacefactor3000\relax}%
\providecommand \BibitemShut  [1]{\csname bibitem#1\endcsname}%
\let\auto@bib@innerbib\@empty
\bibitem [{\citenamefont {Klitzing}\ \emph {et~al.}(1980)\citenamefont
  {Klitzing}, \citenamefont {Dorda},\ and\ \citenamefont {Pepper}}]{Klitzing}%
  \BibitemOpen
  \bibfield  {author} {\bibinfo {author} {\bibfnamefont {K.~v.}\ \bibnamefont
  {Klitzing}}, \bibinfo {author} {\bibfnamefont {G.}~\bibnamefont {Dorda}}, \
  and\ \bibinfo {author} {\bibfnamefont {M.}~\bibnamefont {Pepper}},\
  }\href@noop {} {\bibfield  {journal} {\bibinfo  {journal} {Phys. Rev. Lett.}\
  }\textbf {\bibinfo {volume} {45}},\ \bibinfo {pages} {494} (\bibinfo {year}
  {1980})}\BibitemShut {NoStop}%
\bibitem [{\citenamefont {Thouless}\ \emph {et~al.}(1982)\citenamefont
  {Thouless}, \citenamefont {Kohmoto}, \citenamefont {Nightingale},\ and\
  \citenamefont {den Nijs}}]{Thouless:1982rz}%
  \BibitemOpen
  \bibfield  {author} {\bibinfo {author} {\bibfnamefont {D.~J.}\ \bibnamefont
  {Thouless}}, \bibinfo {author} {\bibfnamefont {M.}~\bibnamefont {Kohmoto}},
  \bibinfo {author} {\bibfnamefont {M.~P.}\ \bibnamefont {Nightingale}}, \ and\
  \bibinfo {author} {\bibfnamefont {M.}~\bibnamefont {den Nijs}},\ }\href@noop
  {} {\bibfield  {journal} {\bibinfo  {journal} {Phys. Rev. Lett.}\ }\textbf
  {\bibinfo {volume} {49}},\ \bibinfo {pages} {405} (\bibinfo {year}
  {1982})}\BibitemShut {NoStop}%
\bibitem [{\citenamefont {Haldane}(1988)}]{Haldane1988}%
  \BibitemOpen
  \bibfield  {author} {\bibinfo {author} {\bibfnamefont {F.~D.~M.}\
  \bibnamefont {Haldane}},\ }\href@noop {} {\bibfield  {journal} {\bibinfo
  {journal} {Phys. Rev. Lett.}\ }\textbf {\bibinfo {volume} {61}},\ \bibinfo
  {pages} {2015} (\bibinfo {year} {1988})}\BibitemShut {NoStop}%
\bibitem [{\citenamefont {Kitaev}(2009)}]{Kitaev2009}%
  \BibitemOpen
  \bibfield  {author} {\bibinfo {author} {\bibfnamefont {A.}~\bibnamefont
  {Kitaev}},\ }\href@noop {} {\bibfield  {journal} {\bibinfo  {journal} {AIP
  Conf. Proc.}\ }\textbf {\bibinfo {volume} {1134}},\ \bibinfo {pages} {22}
  (\bibinfo {year} {2009})}\BibitemShut {NoStop}%
\bibitem [{\citenamefont {{Schnyder}}\ \emph {et~al.}(2008)\citenamefont
  {{Schnyder}}, \citenamefont {{Ryu}}, \citenamefont {{Furusaki}},\ and\
  \citenamefont {{Ludwig}}}]{Schnyder2008}%
  \BibitemOpen
  \bibfield  {author} {\bibinfo {author} {\bibfnamefont {A.~P.}\ \bibnamefont
  {{Schnyder}}}, \bibinfo {author} {\bibfnamefont {S.}~\bibnamefont {{Ryu}}},
  \bibinfo {author} {\bibfnamefont {A.}~\bibnamefont {{Furusaki}}}, \ and\
  \bibinfo {author} {\bibfnamefont {A.~W.~W.}\ \bibnamefont {{Ludwig}}},\
  }\href {\doibase 10.1103/PhysRevB.78.195125} {\bibfield  {journal} {\bibinfo
  {journal} {\prb}\ }\textbf {\bibinfo {volume} {78}},\ \bibinfo {eid} {195125}
  (\bibinfo {year} {2008})}\BibitemShut {NoStop}%
\bibitem [{\citenamefont {Hasan}\ and\ \citenamefont {Kane}(2010)}]{hasan:rmp}%
  \BibitemOpen
  \bibfield  {author} {\bibinfo {author} {\bibfnamefont {M.~Z.}\ \bibnamefont
  {Hasan}}\ and\ \bibinfo {author} {\bibfnamefont {C.~L.}\ \bibnamefont
  {Kane}},\ }\href@noop {} {\bibfield  {journal} {\bibinfo  {journal} {Rev.
  Mod. Phys.}\ }\textbf {\bibinfo {volume} {82}},\ \bibinfo {pages} {3045}
  (\bibinfo {year} {2010})}\BibitemShut {NoStop}%
\bibitem [{\citenamefont {Qi}\ and\ \citenamefont {Zhang}(2010)}]{review_TIb}%
  \BibitemOpen
  \bibfield  {author} {\bibinfo {author} {\bibfnamefont {X.-L.}\ \bibnamefont
  {Qi}}\ and\ \bibinfo {author} {\bibfnamefont {S.-C.}\ \bibnamefont {Zhang}},\
  }\href {\doibase 10.1103/RevModPhys.83.1057} {\bibfield  {journal} {\bibinfo
  {journal} {Rev. Mod. Phys.}\ }\textbf {\bibinfo {volume} {83}},\ \bibinfo
  {pages} {1057} (\bibinfo {year} {2010})}\BibitemShut {NoStop}%
\bibitem [{\citenamefont {Chiu}\ \emph {et~al.}(2016)\citenamefont {Chiu},
  \citenamefont {Teo}, \citenamefont {Schnyder},\ and\ \citenamefont
  {Ryu}}]{RevModPhys.88.035005}%
  \BibitemOpen
  \bibfield  {author} {\bibinfo {author} {\bibfnamefont {C.-K.}\ \bibnamefont
  {Chiu}}, \bibinfo {author} {\bibfnamefont {J.~C.~Y.}\ \bibnamefont {Teo}},
  \bibinfo {author} {\bibfnamefont {A.~P.}\ \bibnamefont {Schnyder}}, \ and\
  \bibinfo {author} {\bibfnamefont {S.}~\bibnamefont {Ryu}},\ }\href {\doibase
  10.1103/RevModPhys.88.035005} {\bibfield  {journal} {\bibinfo  {journal}
  {Rev. Mod. Phys.}\ }\textbf {\bibinfo {volume} {88}},\ \bibinfo {pages}
  {035005} (\bibinfo {year} {2016})}\BibitemShut {NoStop}%
\bibitem [{\citenamefont {Elliott}\ and\ \citenamefont
  {Franz}(2015)}]{elliott_franz_review}%
  \BibitemOpen
  \bibfield  {author} {\bibinfo {author} {\bibfnamefont {S.~R.}\ \bibnamefont
  {Elliott}}\ and\ \bibinfo {author} {\bibfnamefont {M.}~\bibnamefont
  {Franz}},\ }\href {\doibase 10.1103/RevModPhys.87.137} {\bibfield  {journal}
  {\bibinfo  {journal} {Rev. Mod. Phys.}\ }\textbf {\bibinfo {volume} {87}},\
  \bibinfo {pages} {137} (\bibinfo {year} {2015})}\BibitemShut {NoStop}%
\bibitem [{\citenamefont {{Kitaev}}(2001)}]{Kitaev2001}%
  \BibitemOpen
  \bibfield  {author} {\bibinfo {author} {\bibfnamefont {A.}~\bibnamefont
  {{Kitaev}}},\ }\href {\doibase 10.1070/1063-7869/44/10S/S29} {\bibfield
  {journal} {\bibinfo  {journal} {Physics Uspekhi}\ }\textbf {\bibinfo {volume}
  {44}},\ \bibinfo {pages} {131} (\bibinfo {year} {2001})}\BibitemShut
  {NoStop}%
\bibitem [{\citenamefont {Sarma}\ \emph {et~al.}(2015)\citenamefont {Sarma},
  \citenamefont {Freedman},\ and\ \citenamefont {Nayak}}]{Sarma:2015aa}%
  \BibitemOpen
  \bibfield  {author} {\bibinfo {author} {\bibfnamefont {S.~D.}\ \bibnamefont
  {Sarma}}, \bibinfo {author} {\bibfnamefont {M.}~\bibnamefont {Freedman}}, \
  and\ \bibinfo {author} {\bibfnamefont {C.}~\bibnamefont {Nayak}},\ }\href
  {http://dx.doi.org/10.1038/npjqi.2015.1} {\bibfield  {journal} {\bibinfo
  {journal} {Npj Quantum Information}\ }\textbf {\bibinfo {volume} {1}},\
  \bibinfo {pages} {15001 EP } (\bibinfo {year} {2015})}\BibitemShut {NoStop}%
\bibitem [{\citenamefont {Nayak}\ \emph {et~al.}(2008)\citenamefont {Nayak},
  \citenamefont {Simon}, \citenamefont {Stern}, \citenamefont {Freedman},\ and\
  \citenamefont {Das~Sarma}}]{RMP_braiding}%
  \BibitemOpen
  \bibfield  {author} {\bibinfo {author} {\bibfnamefont {C.}~\bibnamefont
  {Nayak}}, \bibinfo {author} {\bibfnamefont {S.~H.}\ \bibnamefont {Simon}},
  \bibinfo {author} {\bibfnamefont {A.}~\bibnamefont {Stern}}, \bibinfo
  {author} {\bibfnamefont {M.}~\bibnamefont {Freedman}}, \ and\ \bibinfo
  {author} {\bibfnamefont {S.}~\bibnamefont {Das~Sarma}},\ }\href {\doibase
  10.1103/RevModPhys.80.1083} {\bibfield  {journal} {\bibinfo  {journal} {Rev.
  Mod. Phys.}\ }\textbf {\bibinfo {volume} {80}},\ \bibinfo {pages} {1083}
  (\bibinfo {year} {2008})}\BibitemShut {NoStop}%
\bibitem [{\citenamefont {Stanescu}(2016)}]{Ivanov_braiding}%
  \BibitemOpen
  \bibfield  {author} {\bibinfo {author} {\bibfnamefont {T.}~\bibnamefont
  {Stanescu}},\ }\href@noop {} {\emph {\bibinfo {title} {Introduction to
  Topological Quantum Matter and Quantum Computation}}}\ (\bibinfo  {publisher}
  {CRC Press},\ \bibinfo {year} {2016})\BibitemShut {NoStop}%
\bibitem [{\citenamefont {Fu}\ and\ \citenamefont {Kane}(2008)}]{Fu:2008fk}%
  \BibitemOpen
  \bibfield  {author} {\bibinfo {author} {\bibfnamefont {L.}~\bibnamefont
  {Fu}}\ and\ \bibinfo {author} {\bibfnamefont {C.~L.}\ \bibnamefont {Kane}},\
  }\href {http://link.aps.org/doi/10.1103/PhysRevLett.100.096407} {\bibfield
  {journal} {\bibinfo  {journal} {Phys. Rev. Lett.}\ }\textbf {\bibinfo
  {volume} {100}},\ \bibinfo {pages} {096407} (\bibinfo {year}
  {2008})}\BibitemShut {NoStop}%
\bibitem [{\citenamefont {Lutchyn}\ \emph {et~al.}(2010)\citenamefont
  {Lutchyn}, \citenamefont {Sau},\ and\ \citenamefont
  {Das~Sarma}}]{Roman_SC_semi}%
  \BibitemOpen
  \bibfield  {author} {\bibinfo {author} {\bibfnamefont {R.~M.}\ \bibnamefont
  {Lutchyn}}, \bibinfo {author} {\bibfnamefont {J.~D.}\ \bibnamefont {Sau}}, \
  and\ \bibinfo {author} {\bibfnamefont {S.}~\bibnamefont {Das~Sarma}},\ }\href
  {\doibase 10.1103/PhysRevLett.105.077001} {\bibfield  {journal} {\bibinfo
  {journal} {Phys. Rev. Lett.}\ }\textbf {\bibinfo {volume} {105}},\ \bibinfo
  {pages} {077001} (\bibinfo {year} {2010})}\BibitemShut {NoStop}%
\bibitem [{\citenamefont {Oreg}\ \emph {et~al.}(2010)\citenamefont {Oreg},
  \citenamefont {Refael},\ and\ \citenamefont {von Oppen}}]{Gil_Majorana_wire}%
  \BibitemOpen
  \bibfield  {author} {\bibinfo {author} {\bibfnamefont {Y.}~\bibnamefont
  {Oreg}}, \bibinfo {author} {\bibfnamefont {G.}~\bibnamefont {Refael}}, \ and\
  \bibinfo {author} {\bibfnamefont {F.}~\bibnamefont {von Oppen}},\ }\href
  {\doibase 10.1103/PhysRevLett.105.177002} {\bibfield  {journal} {\bibinfo
  {journal} {Phys. Rev. Lett.}\ }\textbf {\bibinfo {volume} {105}},\ \bibinfo
  {pages} {177002} (\bibinfo {year} {2010})}\BibitemShut {NoStop}%
\bibitem [{\citenamefont {Sau}\ \emph {et~al.}(2010{\natexlab{a}})\citenamefont
  {Sau}, \citenamefont {Tewari}, \citenamefont {Lutchyn}, \citenamefont
  {Stanescu},\ and\ \citenamefont {Das~Sarma}}]{PhysRevB.82.214509}%
  \BibitemOpen
  \bibfield  {author} {\bibinfo {author} {\bibfnamefont {J.~D.}\ \bibnamefont
  {Sau}}, \bibinfo {author} {\bibfnamefont {S.}~\bibnamefont {Tewari}},
  \bibinfo {author} {\bibfnamefont {R.~M.}\ \bibnamefont {Lutchyn}}, \bibinfo
  {author} {\bibfnamefont {T.~D.}\ \bibnamefont {Stanescu}}, \ and\ \bibinfo
  {author} {\bibfnamefont {S.}~\bibnamefont {Das~Sarma}},\ }\href {\doibase
  10.1103/PhysRevB.82.214509} {\bibfield  {journal} {\bibinfo  {journal} {Phys.
  Rev. B}\ }\textbf {\bibinfo {volume} {82}},\ \bibinfo {pages} {214509}
  (\bibinfo {year} {2010}{\natexlab{a}})}\BibitemShut {NoStop}%
\bibitem [{\citenamefont {Sengupta}\ \emph {et~al.}(2001)\citenamefont
  {Sengupta}, \citenamefont {\ifmmode \check{Z}\else
  \v{Z}\fi{}uti\ifmmode~\acute{c}\else \'{c}\fi{}}, \citenamefont {Kwon},
  \citenamefont {Yakovenko},\ and\ \citenamefont
  {Das~Sarma}}]{PhysRevB.63.144531}%
  \BibitemOpen
  \bibfield  {author} {\bibinfo {author} {\bibfnamefont {K.}~\bibnamefont
  {Sengupta}}, \bibinfo {author} {\bibfnamefont {I.}~\bibnamefont {\ifmmode
  \check{Z}\else \v{Z}\fi{}uti\ifmmode~\acute{c}\else \'{c}\fi{}}}, \bibinfo
  {author} {\bibfnamefont {H.-J.}\ \bibnamefont {Kwon}}, \bibinfo {author}
  {\bibfnamefont {V.~M.}\ \bibnamefont {Yakovenko}}, \ and\ \bibinfo {author}
  {\bibfnamefont {S.}~\bibnamefont {Das~Sarma}},\ }\href {\doibase
  10.1103/PhysRevB.63.144531} {\bibfield  {journal} {\bibinfo  {journal} {Phys.
  Rev. B}\ }\textbf {\bibinfo {volume} {63}},\ \bibinfo {pages} {144531}
  (\bibinfo {year} {2001})}\BibitemShut {NoStop}%
\bibitem [{\citenamefont {Mourik}\ \emph {et~al.}(2012)\citenamefont {Mourik},
  \citenamefont {Zuo}, \citenamefont {Frolov}, \citenamefont {Plissard},
  \citenamefont {Bakkers},\ and\ \citenamefont
  {Kouwenhoven}}]{Mourik_zero_bias}%
  \BibitemOpen
  \bibfield  {author} {\bibinfo {author} {\bibfnamefont {V.}~\bibnamefont
  {Mourik}}, \bibinfo {author} {\bibfnamefont {K.}~\bibnamefont {Zuo}},
  \bibinfo {author} {\bibfnamefont {S.~M.}\ \bibnamefont {Frolov}}, \bibinfo
  {author} {\bibfnamefont {S.~R.}\ \bibnamefont {Plissard}}, \bibinfo {author}
  {\bibfnamefont {E.~P. A.~M.}\ \bibnamefont {Bakkers}}, \ and\ \bibinfo
  {author} {\bibfnamefont {L.~P.}\ \bibnamefont {Kouwenhoven}},\ }\href
  {\doibase 10.1126/science.1222360} {\bibfield  {journal} {\bibinfo  {journal}
  {Science}\ }\textbf {\bibinfo {volume} {336}},\ \bibinfo {pages} {1003}
  (\bibinfo {year} {2012})}\BibitemShut {NoStop}%
\bibitem [{\citenamefont {Das}\ \emph {et~al.}(2012)\citenamefont {Das},
  \citenamefont {Ronen}, \citenamefont {Most}, \citenamefont {Oreg},
  \citenamefont {Heiblum},\ and\ \citenamefont {Shtrikman}}]{Das:2012aa}%
  \BibitemOpen
  \bibfield  {author} {\bibinfo {author} {\bibfnamefont {A.}~\bibnamefont
  {Das}}, \bibinfo {author} {\bibfnamefont {Y.}~\bibnamefont {Ronen}}, \bibinfo
  {author} {\bibfnamefont {Y.}~\bibnamefont {Most}}, \bibinfo {author}
  {\bibfnamefont {Y.}~\bibnamefont {Oreg}}, \bibinfo {author} {\bibfnamefont
  {M.}~\bibnamefont {Heiblum}}, \ and\ \bibinfo {author} {\bibfnamefont
  {H.}~\bibnamefont {Shtrikman}},\ }\href {http://dx.doi.org/10.1038/nphys2479}
  {\bibfield  {journal} {\bibinfo  {journal} {Nat Phys}\ }\textbf {\bibinfo
  {volume} {8}},\ \bibinfo {pages} {887} (\bibinfo {year} {2012})}\BibitemShut
  {NoStop}%
\bibitem [{\citenamefont {{Zhang}}\ \emph {et~al.}(2016)\citenamefont
  {{Zhang}}, \citenamefont {{G{\"u}l}}, \citenamefont {{Conesa-Boj}},
  \citenamefont {{Zuo}}, \citenamefont {{Mourik}}, \citenamefont {{de Vries}},
  \citenamefont {{van Veen}}, \citenamefont {{van Woerkom}}, \citenamefont
  {{Nowak}}, \citenamefont {{Wimmer}}, \citenamefont {{Car}}, \citenamefont
  {{Plissard}}, \citenamefont {{Bakkers}}, \citenamefont
  {{Quintero-P{\'e}rez}}, \citenamefont {{Goswami}}, \citenamefont
  {{Watanabe}}, \citenamefont {{Taniguchi}},\ and\ \citenamefont
  {{Kouwenhoven}}}]{ballistic_M}%
  \BibitemOpen
  \bibfield  {author} {\bibinfo {author} {\bibfnamefont {H.}~\bibnamefont
  {{Zhang}}}, \bibinfo {author} {\bibfnamefont {{\"O}.}~\bibnamefont
  {{G{\"u}l}}}, \bibinfo {author} {\bibfnamefont {S.}~\bibnamefont
  {{Conesa-Boj}}}, \bibinfo {author} {\bibfnamefont {K.}~\bibnamefont {{Zuo}}},
  \bibinfo {author} {\bibfnamefont {V.}~\bibnamefont {{Mourik}}}, \bibinfo
  {author} {\bibfnamefont {F.~K.}\ \bibnamefont {{de Vries}}}, \bibinfo
  {author} {\bibfnamefont {J.}~\bibnamefont {{van Veen}}}, \bibinfo {author}
  {\bibfnamefont {D.~J.}\ \bibnamefont {{van Woerkom}}}, \bibinfo {author}
  {\bibfnamefont {M.~P.}\ \bibnamefont {{Nowak}}}, \bibinfo {author}
  {\bibfnamefont {M.}~\bibnamefont {{Wimmer}}}, \bibinfo {author}
  {\bibfnamefont {D.}~\bibnamefont {{Car}}}, \bibinfo {author} {\bibfnamefont
  {S.}~\bibnamefont {{Plissard}}}, \bibinfo {author} {\bibfnamefont
  {E.~P.~A.~M.}\ \bibnamefont {{Bakkers}}}, \bibinfo {author} {\bibfnamefont
  {M.}~\bibnamefont {{Quintero-P{\'e}rez}}}, \bibinfo {author} {\bibfnamefont
  {S.}~\bibnamefont {{Goswami}}}, \bibinfo {author} {\bibfnamefont
  {K.}~\bibnamefont {{Watanabe}}}, \bibinfo {author} {\bibfnamefont
  {T.}~\bibnamefont {{Taniguchi}}}, \ and\ \bibinfo {author} {\bibfnamefont
  {L.~P.}\ \bibnamefont {{Kouwenhoven}}},\ }\href@noop {} {\bibfield  {journal}
  {\bibinfo  {journal} {ArXiv e-prints}\ } (\bibinfo {year} {2016})},\ \Eprint
  {http://arxiv.org/abs/1603.04069} {arXiv:1603.04069} \BibitemShut {NoStop}%
\bibitem [{\citenamefont {Deng}\ \emph {et~al.}(2012)\citenamefont {Deng},
  \citenamefont {Yu}, \citenamefont {Huang}, \citenamefont {Larsson},
  \citenamefont {Caroff},\ and\ \citenamefont {Xu}}]{doi:10.1021/nl303758w}%
  \BibitemOpen
  \bibfield  {author} {\bibinfo {author} {\bibfnamefont {M.~T.}\ \bibnamefont
  {Deng}}, \bibinfo {author} {\bibfnamefont {C.~L.}\ \bibnamefont {Yu}},
  \bibinfo {author} {\bibfnamefont {G.~Y.}\ \bibnamefont {Huang}}, \bibinfo
  {author} {\bibfnamefont {M.}~\bibnamefont {Larsson}}, \bibinfo {author}
  {\bibfnamefont {P.}~\bibnamefont {Caroff}}, \ and\ \bibinfo {author}
  {\bibfnamefont {H.~Q.}\ \bibnamefont {Xu}},\ }\href {\doibase
  10.1021/nl303758w} {\bibfield  {journal} {\bibinfo  {journal} {Nano Letters}\
  }\textbf {\bibinfo {volume} {12}},\ \bibinfo {pages} {6414} (\bibinfo {year}
  {2012})}\BibitemShut {NoStop}%
\bibitem [{\citenamefont {Churchill}\ \emph {et~al.}(2013)\citenamefont
  {Churchill}, \citenamefont {Fatemi}, \citenamefont {Grove-Rasmussen},
  \citenamefont {Deng}, \citenamefont {Caroff}, \citenamefont {Xu},\ and\
  \citenamefont {Marcus}}]{PhysRevB.87.241401}%
  \BibitemOpen
  \bibfield  {author} {\bibinfo {author} {\bibfnamefont {H.~O.~H.}\
  \bibnamefont {Churchill}}, \bibinfo {author} {\bibfnamefont {V.}~\bibnamefont
  {Fatemi}}, \bibinfo {author} {\bibfnamefont {K.}~\bibnamefont
  {Grove-Rasmussen}}, \bibinfo {author} {\bibfnamefont {M.~T.}\ \bibnamefont
  {Deng}}, \bibinfo {author} {\bibfnamefont {P.}~\bibnamefont {Caroff}},
  \bibinfo {author} {\bibfnamefont {H.~Q.}\ \bibnamefont {Xu}}, \ and\ \bibinfo
  {author} {\bibfnamefont {C.~M.}\ \bibnamefont {Marcus}},\ }\href {\doibase
  10.1103/PhysRevB.87.241401} {\bibfield  {journal} {\bibinfo  {journal} {Phys.
  Rev. B}\ }\textbf {\bibinfo {volume} {87}},\ \bibinfo {pages} {241401}
  (\bibinfo {year} {2013})}\BibitemShut {NoStop}%
\bibitem [{\citenamefont {Finck}\ \emph {et~al.}(2013)\citenamefont {Finck},
  \citenamefont {Van~Harlingen}, \citenamefont {Mohseni}, \citenamefont
  {Jung},\ and\ \citenamefont {Li}}]{PhysRevLett.110.126406}%
  \BibitemOpen
  \bibfield  {author} {\bibinfo {author} {\bibfnamefont {A.~D.~K.}\
  \bibnamefont {Finck}}, \bibinfo {author} {\bibfnamefont {D.~J.}\ \bibnamefont
  {Van~Harlingen}}, \bibinfo {author} {\bibfnamefont {P.~K.}\ \bibnamefont
  {Mohseni}}, \bibinfo {author} {\bibfnamefont {K.}~\bibnamefont {Jung}}, \
  and\ \bibinfo {author} {\bibfnamefont {X.}~\bibnamefont {Li}},\ }\href
  {\doibase 10.1103/PhysRevLett.110.126406} {\bibfield  {journal} {\bibinfo
  {journal} {Phys. Rev. Lett.}\ }\textbf {\bibinfo {volume} {110}},\ \bibinfo
  {pages} {126406} (\bibinfo {year} {2013})}\BibitemShut {NoStop}%
\bibitem [{\citenamefont {{Chen}}\ \emph {et~al.}(2016)\citenamefont {{Chen}},
  \citenamefont {{Yu}}, \citenamefont {{Stenger}}, \citenamefont {{Hocevar}},
  \citenamefont {{Car}}, \citenamefont {{Plissard}}, \citenamefont {{Bakkers}},
  \citenamefont {{Stanescu}},\ and\ \citenamefont
  {{Frolov}}}]{2016arXiv161004555C}%
  \BibitemOpen
  \bibfield  {author} {\bibinfo {author} {\bibfnamefont {J.}~\bibnamefont
  {{Chen}}}, \bibinfo {author} {\bibfnamefont {P.}~\bibnamefont {{Yu}}},
  \bibinfo {author} {\bibfnamefont {J.}~\bibnamefont {{Stenger}}}, \bibinfo
  {author} {\bibfnamefont {M.}~\bibnamefont {{Hocevar}}}, \bibinfo {author}
  {\bibfnamefont {D.}~\bibnamefont {{Car}}}, \bibinfo {author} {\bibfnamefont
  {S.~R.}\ \bibnamefont {{Plissard}}}, \bibinfo {author} {\bibfnamefont
  {E.~P.~A.~M.}\ \bibnamefont {{Bakkers}}}, \bibinfo {author} {\bibfnamefont
  {T.~D.}\ \bibnamefont {{Stanescu}}}, \ and\ \bibinfo {author} {\bibfnamefont
  {S.~M.}\ \bibnamefont {{Frolov}}},\ }\href@noop {} {\bibfield  {journal}
  {\bibinfo  {journal} {ArXiv e-prints}\ } (\bibinfo {year} {2016})},\ \Eprint
  {http://arxiv.org/abs/1610.04555} {arXiv:1610.04555} \BibitemShut {NoStop}%
\bibitem [{\citenamefont {Albrecht}\ \emph {et~al.}(2016)\citenamefont
  {Albrecht}, \citenamefont {Higginbotham}, \citenamefont {Madsen},
  \citenamefont {Kuemmeth}, \citenamefont {Jespersen}, \citenamefont
  {Nyg{\aa}rd}, \citenamefont {Krogstrup},\ and\ \citenamefont
  {Marcus}}]{Albrecht:2016aa}%
  \BibitemOpen
  \bibfield  {author} {\bibinfo {author} {\bibfnamefont {S.~M.}\ \bibnamefont
  {Albrecht}}, \bibinfo {author} {\bibfnamefont {A.~P.}\ \bibnamefont
  {Higginbotham}}, \bibinfo {author} {\bibfnamefont {M.}~\bibnamefont
  {Madsen}}, \bibinfo {author} {\bibfnamefont {F.}~\bibnamefont {Kuemmeth}},
  \bibinfo {author} {\bibfnamefont {T.~S.}\ \bibnamefont {Jespersen}}, \bibinfo
  {author} {\bibfnamefont {J.}~\bibnamefont {Nyg{\aa}rd}}, \bibinfo {author}
  {\bibfnamefont {P.}~\bibnamefont {Krogstrup}}, \ and\ \bibinfo {author}
  {\bibfnamefont {C.~M.}\ \bibnamefont {Marcus}},\ }\href
  {http://dx.doi.org/10.1038/nature17162} {\bibfield  {journal} {\bibinfo
  {journal} {Nature}\ }\textbf {\bibinfo {volume} {531}},\ \bibinfo {pages}
  {206} (\bibinfo {year} {2016})}\BibitemShut {NoStop}%
\bibitem [{\citenamefont {Cheng}\ \emph {et~al.}(2009)\citenamefont {Cheng},
  \citenamefont {Lutchyn}, \citenamefont {Galitski},\ and\ \citenamefont
  {Das~Sarma}}]{PhysRevLett.103.107001}%
  \BibitemOpen
  \bibfield  {author} {\bibinfo {author} {\bibfnamefont {M.}~\bibnamefont
  {Cheng}}, \bibinfo {author} {\bibfnamefont {R.~M.}\ \bibnamefont {Lutchyn}},
  \bibinfo {author} {\bibfnamefont {V.}~\bibnamefont {Galitski}}, \ and\
  \bibinfo {author} {\bibfnamefont {S.}~\bibnamefont {Das~Sarma}},\ }\href
  {\doibase 10.1103/PhysRevLett.103.107001} {\bibfield  {journal} {\bibinfo
  {journal} {Phys. Rev. Lett.}\ }\textbf {\bibinfo {volume} {103}},\ \bibinfo
  {pages} {107001} (\bibinfo {year} {2009})}\BibitemShut {NoStop}%
\bibitem [{\citenamefont {Cheng}\ \emph {et~al.}(2012)\citenamefont {Cheng},
  \citenamefont {Lutchyn},\ and\ \citenamefont
  {Das~Sarma}}]{PhysRevB.85.165124}%
  \BibitemOpen
  \bibfield  {author} {\bibinfo {author} {\bibfnamefont {M.}~\bibnamefont
  {Cheng}}, \bibinfo {author} {\bibfnamefont {R.~M.}\ \bibnamefont {Lutchyn}},
  \ and\ \bibinfo {author} {\bibfnamefont {S.}~\bibnamefont {Das~Sarma}},\
  }\href {\doibase 10.1103/PhysRevB.85.165124} {\bibfield  {journal} {\bibinfo
  {journal} {Phys. Rev. B}\ }\textbf {\bibinfo {volume} {85}},\ \bibinfo
  {pages} {165124} (\bibinfo {year} {2012})}\BibitemShut {NoStop}%
\bibitem [{\citenamefont {Das~Sarma}\ \emph {et~al.}(2012)\citenamefont
  {Das~Sarma}, \citenamefont {Sau},\ and\ \citenamefont
  {Stanescu}}]{PhysRevB.86.220506}%
  \BibitemOpen
  \bibfield  {author} {\bibinfo {author} {\bibfnamefont {S.}~\bibnamefont
  {Das~Sarma}}, \bibinfo {author} {\bibfnamefont {J.~D.}\ \bibnamefont {Sau}},
  \ and\ \bibinfo {author} {\bibfnamefont {T.~D.}\ \bibnamefont {Stanescu}},\
  }\href {\doibase 10.1103/PhysRevB.86.220506} {\bibfield  {journal} {\bibinfo
  {journal} {Phys. Rev. B}\ }\textbf {\bibinfo {volume} {86}},\ \bibinfo
  {pages} {220506} (\bibinfo {year} {2012})}\BibitemShut {NoStop}%
\bibitem [{\citenamefont {H\"utzen}\ \emph {et~al.}(2012)\citenamefont
  {H\"utzen}, \citenamefont {Zazunov}, \citenamefont {Braunecker},
  \citenamefont {Yeyati},\ and\ \citenamefont
  {Egger}}]{PhysRevLett.109.166403}%
  \BibitemOpen
  \bibfield  {author} {\bibinfo {author} {\bibfnamefont {R.}~\bibnamefont
  {H\"utzen}}, \bibinfo {author} {\bibfnamefont {A.}~\bibnamefont {Zazunov}},
  \bibinfo {author} {\bibfnamefont {B.}~\bibnamefont {Braunecker}}, \bibinfo
  {author} {\bibfnamefont {A.~L.}\ \bibnamefont {Yeyati}}, \ and\ \bibinfo
  {author} {\bibfnamefont {R.}~\bibnamefont {Egger}},\ }\href {\doibase
  10.1103/PhysRevLett.109.166403} {\bibfield  {journal} {\bibinfo  {journal}
  {Phys. Rev. Lett.}\ }\textbf {\bibinfo {volume} {109}},\ \bibinfo {pages}
  {166403} (\bibinfo {year} {2012})}\BibitemShut {NoStop}%
\bibitem [{\citenamefont {Stanescu}\ \emph {et~al.}(2013)\citenamefont
  {Stanescu}, \citenamefont {Lutchyn},\ and\ \citenamefont
  {Das~Sarma}}]{Tudor_oscillation_damp}%
  \BibitemOpen
  \bibfield  {author} {\bibinfo {author} {\bibfnamefont {T.~D.}\ \bibnamefont
  {Stanescu}}, \bibinfo {author} {\bibfnamefont {R.~M.}\ \bibnamefont
  {Lutchyn}}, \ and\ \bibinfo {author} {\bibfnamefont {S.}~\bibnamefont
  {Das~Sarma}},\ }\href {\doibase 10.1103/PhysRevB.87.094518} {\bibfield
  {journal} {\bibinfo  {journal} {Phys. Rev. B}\ }\textbf {\bibinfo {volume}
  {87}},\ \bibinfo {pages} {094518} (\bibinfo {year} {2013})}\BibitemShut
  {NoStop}%
\bibitem [{\citenamefont {Kastner}(1992)}]{RevModPhys.64.849}%
  \BibitemOpen
  \bibfield  {author} {\bibinfo {author} {\bibfnamefont {M.~A.}\ \bibnamefont
  {Kastner}},\ }\href {\doibase 10.1103/RevModPhys.64.849} {\bibfield
  {journal} {\bibinfo  {journal} {Rev. Mod. Phys.}\ }\textbf {\bibinfo {volume}
  {64}},\ \bibinfo {pages} {849} (\bibinfo {year} {1992})}\BibitemShut
  {NoStop}%
\bibitem [{\citenamefont {Fulton}\ and\ \citenamefont
  {Dolan}(1987)}]{PhysRevLett.59.109}%
  \BibitemOpen
  \bibfield  {author} {\bibinfo {author} {\bibfnamefont {T.~A.}\ \bibnamefont
  {Fulton}}\ and\ \bibinfo {author} {\bibfnamefont {G.~J.}\ \bibnamefont
  {Dolan}},\ }\href {\doibase 10.1103/PhysRevLett.59.109} {\bibfield  {journal}
  {\bibinfo  {journal} {Phys. Rev. Lett.}\ }\textbf {\bibinfo {volume} {59}},\
  \bibinfo {pages} {109} (\bibinfo {year} {1987})}\BibitemShut {NoStop}%
\bibitem [{\citenamefont {Fulton}\ \emph {et~al.}(1989)\citenamefont {Fulton},
  \citenamefont {Gammel}, \citenamefont {Bishop}, \citenamefont
  {Dunkleberger},\ and\ \citenamefont {Dolan}}]{PhysRevLett.63.1307}%
  \BibitemOpen
  \bibfield  {author} {\bibinfo {author} {\bibfnamefont {T.~A.}\ \bibnamefont
  {Fulton}}, \bibinfo {author} {\bibfnamefont {P.~L.}\ \bibnamefont {Gammel}},
  \bibinfo {author} {\bibfnamefont {D.~J.}\ \bibnamefont {Bishop}}, \bibinfo
  {author} {\bibfnamefont {L.~N.}\ \bibnamefont {Dunkleberger}}, \ and\
  \bibinfo {author} {\bibfnamefont {G.~J.}\ \bibnamefont {Dolan}},\ }\href
  {\doibase 10.1103/PhysRevLett.63.1307} {\bibfield  {journal} {\bibinfo
  {journal} {Phys. Rev. Lett.}\ }\textbf {\bibinfo {volume} {63}},\ \bibinfo
  {pages} {1307} (\bibinfo {year} {1989})}\BibitemShut {NoStop}%
\bibitem [{\citenamefont {Scott-Thomas}\ \emph {et~al.}(1989)\citenamefont
  {Scott-Thomas}, \citenamefont {Field}, \citenamefont {Kastner}, \citenamefont
  {Smith},\ and\ \citenamefont {Antoniadis}}]{PhysRevLett.62.583}%
  \BibitemOpen
  \bibfield  {author} {\bibinfo {author} {\bibfnamefont {J.~H.~F.}\
  \bibnamefont {Scott-Thomas}}, \bibinfo {author} {\bibfnamefont {S.~B.}\
  \bibnamefont {Field}}, \bibinfo {author} {\bibfnamefont {M.~A.}\ \bibnamefont
  {Kastner}}, \bibinfo {author} {\bibfnamefont {H.~I.}\ \bibnamefont {Smith}},
  \ and\ \bibinfo {author} {\bibfnamefont {D.~A.}\ \bibnamefont {Antoniadis}},\
  }\href {\doibase 10.1103/PhysRevLett.62.583} {\bibfield  {journal} {\bibinfo
  {journal} {Phys. Rev. Lett.}\ }\textbf {\bibinfo {volume} {62}},\ \bibinfo
  {pages} {583} (\bibinfo {year} {1989})}\BibitemShut {NoStop}%
\bibitem [{\citenamefont {Meirav}\ \emph {et~al.}(1989)\citenamefont {Meirav},
  \citenamefont {Kastner}, \citenamefont {Heiblum},\ and\ \citenamefont
  {Wind}}]{PhysRevB.40.5871}%
  \BibitemOpen
  \bibfield  {author} {\bibinfo {author} {\bibfnamefont {U.}~\bibnamefont
  {Meirav}}, \bibinfo {author} {\bibfnamefont {M.~A.}\ \bibnamefont {Kastner}},
  \bibinfo {author} {\bibfnamefont {M.}~\bibnamefont {Heiblum}}, \ and\
  \bibinfo {author} {\bibfnamefont {S.~J.}\ \bibnamefont {Wind}},\ }\href
  {\doibase 10.1103/PhysRevB.40.5871} {\bibfield  {journal} {\bibinfo
  {journal} {Phys. Rev. B}\ }\textbf {\bibinfo {volume} {40}},\ \bibinfo
  {pages} {5871} (\bibinfo {year} {1989})}\BibitemShut {NoStop}%
\bibitem [{\citenamefont {Meirav}\ \emph {et~al.}(1990)\citenamefont {Meirav},
  \citenamefont {Kastner},\ and\ \citenamefont {Wind}}]{PhysRevLett.65.771}%
  \BibitemOpen
  \bibfield  {author} {\bibinfo {author} {\bibfnamefont {U.}~\bibnamefont
  {Meirav}}, \bibinfo {author} {\bibfnamefont {M.~A.}\ \bibnamefont {Kastner}},
  \ and\ \bibinfo {author} {\bibfnamefont {S.~J.}\ \bibnamefont {Wind}},\
  }\href {\doibase 10.1103/PhysRevLett.65.771} {\bibfield  {journal} {\bibinfo
  {journal} {Phys. Rev. Lett.}\ }\textbf {\bibinfo {volume} {65}},\ \bibinfo
  {pages} {771} (\bibinfo {year} {1990})}\BibitemShut {NoStop}%
\bibitem [{\citenamefont {Glazman}\ and\ \citenamefont
  {Shekhter}(1989)}]{early_Blockade_glazman}%
  \BibitemOpen
  \bibfield  {author} {\bibinfo {author} {\bibfnamefont {L.~I.}\ \bibnamefont
  {Glazman}}\ and\ \bibinfo {author} {\bibfnamefont {R.~I.}\ \bibnamefont
  {Shekhter}},\ }\href {http://stacks.iop.org/0953-8984/1/i=33/a=027}
  {\bibfield  {journal} {\bibinfo  {journal} {Journal of Physics: Condensed
  Matter}\ }\textbf {\bibinfo {volume} {1}},\ \bibinfo {pages} {5811} (\bibinfo
  {year} {1989})}\BibitemShut {NoStop}%
\bibitem [{\citenamefont {Beenakker}(1991)}]{generic_blockade}%
  \BibitemOpen
  \bibfield  {author} {\bibinfo {author} {\bibfnamefont {C.~W.~J.}\
  \bibnamefont {Beenakker}},\ }\href {\doibase 10.1103/PhysRevB.44.1646}
  {\bibfield  {journal} {\bibinfo  {journal} {Phys. Rev. B}\ }\textbf {\bibinfo
  {volume} {44}},\ \bibinfo {pages} {1646} (\bibinfo {year}
  {1991})}\BibitemShut {NoStop}%
\bibitem [{\citenamefont {van Houten}\ and\ \citenamefont
  {Beenakker}(1989)}]{PhysRevLett.63.1893}%
  \BibitemOpen
  \bibfield  {author} {\bibinfo {author} {\bibfnamefont {H.}~\bibnamefont {van
  Houten}}\ and\ \bibinfo {author} {\bibfnamefont {C.~W.~J.}\ \bibnamefont
  {Beenakker}},\ }\href {\doibase 10.1103/PhysRevLett.63.1893} {\bibfield
  {journal} {\bibinfo  {journal} {Phys. Rev. Lett.}\ }\textbf {\bibinfo
  {volume} {63}},\ \bibinfo {pages} {1893} (\bibinfo {year}
  {1989})}\BibitemShut {NoStop}%
\bibitem [{\citenamefont {Hekking}\ \emph {et~al.}(1993)\citenamefont
  {Hekking}, \citenamefont {Glazman}, \citenamefont {Matveev},\ and\
  \citenamefont {Shekhter}}]{PhysRevLett.70.4138}%
  \BibitemOpen
  \bibfield  {author} {\bibinfo {author} {\bibfnamefont {F.~W.~J.}\
  \bibnamefont {Hekking}}, \bibinfo {author} {\bibfnamefont {L.~I.}\
  \bibnamefont {Glazman}}, \bibinfo {author} {\bibfnamefont {K.~A.}\
  \bibnamefont {Matveev}}, \ and\ \bibinfo {author} {\bibfnamefont {R.~I.}\
  \bibnamefont {Shekhter}},\ }\href {\doibase 10.1103/PhysRevLett.70.4138}
  {\bibfield  {journal} {\bibinfo  {journal} {Phys. Rev. Lett.}\ }\textbf
  {\bibinfo {volume} {70}},\ \bibinfo {pages} {4138} (\bibinfo {year}
  {1993})}\BibitemShut {NoStop}%
\bibitem [{\citenamefont {Eiles}\ \emph {et~al.}(1993)\citenamefont {Eiles},
  \citenamefont {Martinis},\ and\ \citenamefont
  {Devoret}}]{PhysRevLett.70.1862}%
  \BibitemOpen
  \bibfield  {author} {\bibinfo {author} {\bibfnamefont {T.~M.}\ \bibnamefont
  {Eiles}}, \bibinfo {author} {\bibfnamefont {J.~M.}\ \bibnamefont {Martinis}},
  \ and\ \bibinfo {author} {\bibfnamefont {M.~H.}\ \bibnamefont {Devoret}},\
  }\href {\doibase 10.1103/PhysRevLett.70.1862} {\bibfield  {journal} {\bibinfo
   {journal} {Phys. Rev. Lett.}\ }\textbf {\bibinfo {volume} {70}},\ \bibinfo
  {pages} {1862} (\bibinfo {year} {1993})}\BibitemShut {NoStop}%
\bibitem [{\citenamefont {Tuominen}\ \emph {et~al.}(1992)\citenamefont
  {Tuominen}, \citenamefont {Hergenrother}, \citenamefont {Tighe},\ and\
  \citenamefont {Tinkham}}]{PhysRevLett.69.1997}%
  \BibitemOpen
  \bibfield  {author} {\bibinfo {author} {\bibfnamefont {M.~T.}\ \bibnamefont
  {Tuominen}}, \bibinfo {author} {\bibfnamefont {J.~M.}\ \bibnamefont
  {Hergenrother}}, \bibinfo {author} {\bibfnamefont {T.~S.}\ \bibnamefont
  {Tighe}}, \ and\ \bibinfo {author} {\bibfnamefont {M.}~\bibnamefont
  {Tinkham}},\ }\href {\doibase 10.1103/PhysRevLett.69.1997} {\bibfield
  {journal} {\bibinfo  {journal} {Phys. Rev. Lett.}\ }\textbf {\bibinfo
  {volume} {69}},\ \bibinfo {pages} {1997} (\bibinfo {year}
  {1992})}\BibitemShut {NoStop}%
\bibitem [{\citenamefont {Amar}\ \emph {et~al.}(1994)\citenamefont {Amar},
  \citenamefont {Song}, \citenamefont {Lobb},\ and\ \citenamefont
  {Wellstood}}]{PhysRevLett.72.3234}%
  \BibitemOpen
  \bibfield  {author} {\bibinfo {author} {\bibfnamefont {A.}~\bibnamefont
  {Amar}}, \bibinfo {author} {\bibfnamefont {D.}~\bibnamefont {Song}}, \bibinfo
  {author} {\bibfnamefont {C.~J.}\ \bibnamefont {Lobb}}, \ and\ \bibinfo
  {author} {\bibfnamefont {F.~C.}\ \bibnamefont {Wellstood}},\ }\href {\doibase
  10.1103/PhysRevLett.72.3234} {\bibfield  {journal} {\bibinfo  {journal}
  {Phys. Rev. Lett.}\ }\textbf {\bibinfo {volume} {72}},\ \bibinfo {pages}
  {3234} (\bibinfo {year} {1994})}\BibitemShut {NoStop}%
\bibitem [{\citenamefont {Zazunov}\ \emph {et~al.}(2011)\citenamefont
  {Zazunov}, \citenamefont {Yeyati},\ and\ \citenamefont
  {Egger}}]{PhysRevB.84.165440}%
  \BibitemOpen
  \bibfield  {author} {\bibinfo {author} {\bibfnamefont {A.}~\bibnamefont
  {Zazunov}}, \bibinfo {author} {\bibfnamefont {A.~L.}\ \bibnamefont {Yeyati}},
  \ and\ \bibinfo {author} {\bibfnamefont {R.}~\bibnamefont {Egger}},\ }\href
  {\doibase 10.1103/PhysRevB.84.165440} {\bibfield  {journal} {\bibinfo
  {journal} {Phys. Rev. B}\ }\textbf {\bibinfo {volume} {84}},\ \bibinfo
  {pages} {165440} (\bibinfo {year} {2011})}\BibitemShut {NoStop}%
\bibitem [{\citenamefont {Sau}\ \emph {et~al.}(2015)\citenamefont {Sau},
  \citenamefont {Swingle},\ and\ \citenamefont {Tewari}}]{PhysRevB.92.020511}%
  \BibitemOpen
  \bibfield  {author} {\bibinfo {author} {\bibfnamefont {J.~D.}\ \bibnamefont
  {Sau}}, \bibinfo {author} {\bibfnamefont {B.}~\bibnamefont {Swingle}}, \ and\
  \bibinfo {author} {\bibfnamefont {S.}~\bibnamefont {Tewari}},\ }\href
  {\doibase 10.1103/PhysRevB.92.020511} {\bibfield  {journal} {\bibinfo
  {journal} {Phys. Rev. B}\ }\textbf {\bibinfo {volume} {92}},\ \bibinfo
  {pages} {020511} (\bibinfo {year} {2015})}\BibitemShut {NoStop}%
\bibitem [{\citenamefont {Fu}(2010)}]{PhysRevLett.104.056402}%
  \BibitemOpen
  \bibfield  {author} {\bibinfo {author} {\bibfnamefont {L.}~\bibnamefont
  {Fu}},\ }\href {\doibase 10.1103/PhysRevLett.104.056402} {\bibfield
  {journal} {\bibinfo  {journal} {Phys. Rev. Lett.}\ }\textbf {\bibinfo
  {volume} {104}},\ \bibinfo {pages} {056402} (\bibinfo {year}
  {2010})}\BibitemShut {NoStop}%
\bibitem [{\citenamefont {van Heck}\ \emph {et~al.}(2016)\citenamefont {van
  Heck}, \citenamefont {Lutchyn},\ and\ \citenamefont
  {Glazman}}]{conductance_coulomb_blockade_roman}%
  \BibitemOpen
  \bibfield  {author} {\bibinfo {author} {\bibfnamefont {B.}~\bibnamefont {van
  Heck}}, \bibinfo {author} {\bibfnamefont {R.~M.}\ \bibnamefont {Lutchyn}}, \
  and\ \bibinfo {author} {\bibfnamefont {L.~I.}\ \bibnamefont {Glazman}},\
  }\href {\doibase 10.1103/PhysRevB.93.235431} {\bibfield  {journal} {\bibinfo
  {journal} {Phys. Rev. B}\ }\textbf {\bibinfo {volume} {93}},\ \bibinfo
  {pages} {235431} (\bibinfo {year} {2016})}\BibitemShut {NoStop}%
\bibitem [{\citenamefont {Liu}\ \emph {et~al.}(2017)\citenamefont {Liu},
  \citenamefont {Sau},\ and\ \citenamefont {Das~Sarma}}]{parameter_1DMajorana}%
  \BibitemOpen
  \bibfield  {author} {\bibinfo {author} {\bibfnamefont {C.-X.}\ \bibnamefont
  {Liu}}, \bibinfo {author} {\bibfnamefont {J.~D.}\ \bibnamefont {Sau}}, \ and\
  \bibinfo {author} {\bibfnamefont {S.}~\bibnamefont {Das~Sarma}},\ }\href
  {\doibase 10.1103/PhysRevB.95.054502} {\bibfield  {journal} {\bibinfo
  {journal} {Phys. Rev. B}\ }\textbf {\bibinfo {volume} {95}},\ \bibinfo
  {pages} {054502} (\bibinfo {year} {2017})}\BibitemShut {NoStop}%
\bibitem [{\citenamefont {Nijholt}\ and\ \citenamefont
  {Akhmerov}(2016)}]{PhysRevB.93.235434}%
  \BibitemOpen
  \bibfield  {author} {\bibinfo {author} {\bibfnamefont {B.}~\bibnamefont
  {Nijholt}}\ and\ \bibinfo {author} {\bibfnamefont {A.~R.}\ \bibnamefont
  {Akhmerov}},\ }\href {\doibase 10.1103/PhysRevB.93.235434} {\bibfield
  {journal} {\bibinfo  {journal} {Phys. Rev. B}\ }\textbf {\bibinfo {volume}
  {93}},\ \bibinfo {pages} {235434} (\bibinfo {year} {2016})}\BibitemShut
  {NoStop}%
\bibitem [{\citenamefont {Sau}\ \emph {et~al.}(2010{\natexlab{b}})\citenamefont
  {Sau}, \citenamefont {Lutchyn}, \citenamefont {Tewari},\ and\ \citenamefont
  {Das~Sarma}}]{Sau_semiconductor_heterostructures}%
  \BibitemOpen
  \bibfield  {author} {\bibinfo {author} {\bibfnamefont {J.~D.}\ \bibnamefont
  {Sau}}, \bibinfo {author} {\bibfnamefont {R.~M.}\ \bibnamefont {Lutchyn}},
  \bibinfo {author} {\bibfnamefont {S.}~\bibnamefont {Tewari}}, \ and\ \bibinfo
  {author} {\bibfnamefont {S.}~\bibnamefont {Das~Sarma}},\ }\href {\doibase
  10.1103/PhysRevLett.104.040502} {\bibfield  {journal} {\bibinfo  {journal}
  {Phys. Rev. Lett.}\ }\textbf {\bibinfo {volume} {104}},\ \bibinfo {pages}
  {040502} (\bibinfo {year} {2010}{\natexlab{b}})}\BibitemShut {NoStop}%
\bibitem [{\citenamefont {{Liu}}\ \emph {et~al.}(2017)\citenamefont {{Liu}},
  \citenamefont {{Sau}}, \citenamefont {{Stanescu}},\ and\ \citenamefont {{Das
  Sarma}}}]{2017arXiv170502035L}%
  \BibitemOpen
  \bibfield  {author} {\bibinfo {author} {\bibfnamefont {C.-X.}\ \bibnamefont
  {{Liu}}}, \bibinfo {author} {\bibfnamefont {J.~D.}\ \bibnamefont {{Sau}}},
  \bibinfo {author} {\bibfnamefont {T.~D.}\ \bibnamefont {{Stanescu}}}, \ and\
  \bibinfo {author} {\bibfnamefont {S.}~\bibnamefont {{Das Sarma}}},\
  }\href@noop {} {\bibfield  {journal} {\bibinfo  {journal} {ArXiv e-prints}\ }
  (\bibinfo {year} {2017})},\ \Eprint {http://arxiv.org/abs/1705.02035}
  {arXiv:1705.02035} \BibitemShut {NoStop}%
\bibitem [{\citenamefont {Chang}\ \emph {et~al.}(2015)\citenamefont {Chang},
  \citenamefont {Albrecht}, \citenamefont {Jespersen}, \citenamefont
  {Kuemmeth}, \citenamefont {Krogstrup}, \citenamefont {Nyg{\aa}rd},\ and\
  \citenamefont {Marcus}}]{ChangW.:2015aa}%
  \BibitemOpen
  \bibfield  {author} {\bibinfo {author} {\bibfnamefont {W.}~\bibnamefont
  {Chang}}, \bibinfo {author} {\bibfnamefont {S.~M.}\ \bibnamefont {Albrecht}},
  \bibinfo {author} {\bibfnamefont {T.~S.}\ \bibnamefont {Jespersen}}, \bibinfo
  {author} {\bibfnamefont {F.}~\bibnamefont {Kuemmeth}}, \bibinfo {author}
  {\bibfnamefont {P.}~\bibnamefont {Krogstrup}}, \bibinfo {author}
  {\bibfnamefont {J.}~\bibnamefont {Nyg{\aa}rd}}, \ and\ \bibinfo {author}
  {\bibfnamefont {C.~M.}\ \bibnamefont {Marcus}},\ }\href
  {http://dx.doi.org/10.1038/nnano.2014.306} {\bibfield  {journal} {\bibinfo
  {journal} {Nat Nano}\ }\textbf {\bibinfo {volume} {10}},\ \bibinfo {pages}
  {232} (\bibinfo {year} {2015})}\BibitemShut {NoStop}%
\bibitem [{\citenamefont {Cole}\ \emph {et~al.}(2015)\citenamefont {Cole},
  \citenamefont {Das~Sarma},\ and\ \citenamefont
  {Stanescu}}]{PhysRevB.92.174511}%
  \BibitemOpen
  \bibfield  {author} {\bibinfo {author} {\bibfnamefont {W.~S.}\ \bibnamefont
  {Cole}}, \bibinfo {author} {\bibfnamefont {S.}~\bibnamefont {Das~Sarma}}, \
  and\ \bibinfo {author} {\bibfnamefont {T.~D.}\ \bibnamefont {Stanescu}},\
  }\href {\doibase 10.1103/PhysRevB.92.174511} {\bibfield  {journal} {\bibinfo
  {journal} {Phys. Rev. B}\ }\textbf {\bibinfo {volume} {92}},\ \bibinfo
  {pages} {174511} (\bibinfo {year} {2015})}\BibitemShut {NoStop}%
\bibitem [{\citenamefont {Cole}\ \emph {et~al.}(2016)\citenamefont {Cole},
  \citenamefont {Sau},\ and\ \citenamefont {Das~Sarma}}]{PhysRevB.94.140505}%
  \BibitemOpen
  \bibfield  {author} {\bibinfo {author} {\bibfnamefont {W.~S.}\ \bibnamefont
  {Cole}}, \bibinfo {author} {\bibfnamefont {J.~D.}\ \bibnamefont {Sau}}, \
  and\ \bibinfo {author} {\bibfnamefont {S.}~\bibnamefont {Das~Sarma}},\ }\href
  {\doibase 10.1103/PhysRevB.94.140505} {\bibfield  {journal} {\bibinfo
  {journal} {Phys. Rev. B}\ }\textbf {\bibinfo {volume} {94}},\ \bibinfo
  {pages} {140505} (\bibinfo {year} {2016})}\BibitemShut {NoStop}%
\bibitem [{\citenamefont {Takei}\ \emph {et~al.}(2013)\citenamefont {Takei},
  \citenamefont {Fregoso}, \citenamefont {Hui}, \citenamefont {Lobos},\ and\
  \citenamefont {Das~Sarma}}]{PhysRevLett.110.186803}%
  \BibitemOpen
  \bibfield  {author} {\bibinfo {author} {\bibfnamefont {S.}~\bibnamefont
  {Takei}}, \bibinfo {author} {\bibfnamefont {B.~M.}\ \bibnamefont {Fregoso}},
  \bibinfo {author} {\bibfnamefont {H.-Y.}\ \bibnamefont {Hui}}, \bibinfo
  {author} {\bibfnamefont {A.~M.}\ \bibnamefont {Lobos}}, \ and\ \bibinfo
  {author} {\bibfnamefont {S.}~\bibnamefont {Das~Sarma}},\ }\href {\doibase
  10.1103/PhysRevLett.110.186803} {\bibfield  {journal} {\bibinfo  {journal}
  {Phys. Rev. Lett.}\ }\textbf {\bibinfo {volume} {110}},\ \bibinfo {pages}
  {186803} (\bibinfo {year} {2013})}\BibitemShut {NoStop}%
\bibitem [{\citenamefont {Hui}\ \emph {et~al.}(2015)\citenamefont {Hui},
  \citenamefont {Sau},\ and\ \citenamefont {Das~Sarma}}]{PhysRevB.92.174512}%
  \BibitemOpen
  \bibfield  {author} {\bibinfo {author} {\bibfnamefont {H.-Y.}\ \bibnamefont
  {Hui}}, \bibinfo {author} {\bibfnamefont {J.~D.}\ \bibnamefont {Sau}}, \ and\
  \bibinfo {author} {\bibfnamefont {S.}~\bibnamefont {Das~Sarma}},\ }\href
  {\doibase 10.1103/PhysRevB.92.174512} {\bibfield  {journal} {\bibinfo
  {journal} {Phys. Rev. B}\ }\textbf {\bibinfo {volume} {92}},\ \bibinfo
  {pages} {174512} (\bibinfo {year} {2015})}\BibitemShut {NoStop}%
\bibitem [{\citenamefont {Stanescu}()}]{Tudor_private}%
  \BibitemOpen
  \bibfield  {author} {\bibinfo {author} {\bibfnamefont {T.}~\bibnamefont
  {Stanescu}},\ }\href@noop {} {}\bibinfo {howpublished} {private
  communication}\BibitemShut {NoStop}%
\end{thebibliography}%

\end{document}